\documentclass[11pt]{report}
\usepackage{amsmath} 
\usepackage[dvips]{graphics}
\usepackage{latexsym}
\usepackage{psfig}
\usepackage{amscd}     \usepackage{amsxtra}
\usepackage{upref}     \usepackage{amsthm}
\usepackage{amssymb}   

\usepackage{amsfonts}
\begin{document}

\title{{\bf From Maxwell Stresses to Photon-like Objects through
Frobenius Curvature Geometrization of Local Physical Interaction}}

\author{{\bf Stoil Donev}\footnote{e-mail:
 sdonev@inrne.bas.bg}, {\bf Maria Tashkova}, \\ Institute for Nuclear
Research and Nuclear Energy,\\ Bulg.Acad.Sci., 1784 Sofia,
blvd.Tzarigradsko chaussee 72\\ Bulgaria\\}

\date{}

\maketitle

\begin{abstract} This paper aims to review our recent results on exploring the
capabilities of nonquantum field theory as a possible tool for describing
single photon-like objects, considered as massless time-stable spatially finite
physical entities with compatible translational-rotational dynamical structure.
It consists of five chapters and includes 22 sections and subsections.

In Chapter 1 ({\bf Introduction}) we present briefly some remarks on Maxwell
vacuum equations and our notions concerning the two basic concepts in physics:
{\it physical object} and {\it interaction}, with an accent on the view that
physical interaction necessarily implies energy-momentum exchange, and on the
suggestion that the energy-momentum exchanging subsystems of a general field
should be described rather by $(F,*F)$, than by $(\mathbf{E},\mathbf{B})$.

Chapter 2 ({\bf Nonrelativistic considerations}) begins with a reasoning on the
status, right understanding and appropriate use of the Coulomb force law in
electrostatics, leading to the conclusion that the usual way of introducing
static electric field as local object directly from the Coulomb force law
violates in definite sense the local conservation laws in the frame
of Maxwell electrodynamics. Then we concentrate on the role, significance and
eigen properties of the Maxwell stress tensor in nonrelativistic terms. This
part of the paper culminates in writing down nonlinear field equations for the
vacuum electromagnetic fields paying due respect of the Newton view on the
sense of dynamical equations as local balance relations of conserved
quantities carried by two appropriately defined subsystems of the general
electromagnetic field, and presenting some important properties of the
nonlinear solutions. Finally, we discuss briefly the contents of this part of
the paper.

Chapter 3 ({\bf Relativistic considerations}) of the paper makes use of the
relativistic formalism. We begin with presenting the notion of photon-like
object(s) (PhLO). The existing corresponding relativistic description of PhLO
in the frame of Extended Electrodynamics is briefly recalled. In terms of
integrability and nonintegrability properties of distributions (differential
systems) on a manifold a principle of geometric interaction between two
nonintegrable distributions is formulated and a corresponding physical
interpretation is given. This principle is further substantially used in
building mathematical description of single PhLO, considered as composed of two
individualized and interacting subsystems, and realizing a special kind of
dynamical equilibrium. The mathematical model is built on the assumptions that
Frobenius integrability can be made to correspond to physical time-stability,
and the nonintegrability of subdistributions of integrable distribution to
correspond to local physical interaction between subsystems of a time-stable
continuous physical system. Two approaches are considered: direct use of the
Frobenius theorem and the corresponding curvature being  a measure of
nonintegrability, and the recently developed approach known as "non-linear
connections". Both these approaches make use of the corresponding curvatures
for  generating appropriate quantities describing local physical interaction,
i.e. local energy-momentum exchange

A corresponding concept of electromagnetic strain is defined and the basic
stress-energy-momentum relations, obtained before, are represented in
electromagnetic strain terms.

Chapter 4 ({\bf Equations of motion for PhLO.Solutions}) deals with various
views on dynamical equations for PhLO and gives illustrative examples of
appropriate solutions.

Finally, Chapter 5 ({\bf Retrospect}) gives a retrospective discussion of the
contents of the paper.

\end{abstract}
\tableofcontents


\pagenumbering{arabic}
\setcounter{page}{1}

\chapter{Introduction}
\section{Preliminary remarks}
As it is well known the vacuum Maxwell equations (zero charge density:
$\rho=0$) do not admit spatially finite time-stable solutions of photon-like
type, i.e. solutions, having compatible translational-rotational dynamical
structure and propagating as a whole along straight lines in the space with the
fundamental velocity "c" - the speed of light in vacuum and without dispersion.
This is in corressponding degree due to the fact that in the frame of Maxwell
vacuum equations every component $U(x,y,z,t)$ of the electric {\bf E} and
magnetic {\bf B} fields, satisfying corresponding smoothness conditions,
necessarily satisfies the D'Alembert wave equation $\square U=0$, and according
to the Poisson's theorem for this equation, every spatially finite initial
condition $U(x,y,z,0)=\varphi(x,y,z);\ \frac{\partial U}{\partial
t}(x,y,z,0)=\psi(x,y,z)$, where $\varphi$ and $\psi$ are finite functions
satisfying definite differentiability conditions, blows up radially and goes to
infinity with the speed of light [1,2,3]. So, in such a case, through every
spatial point outside such an initial condition pass fore-front and back-front,
and after this the corresponding point forgets about what has happened. Hence,
photon-like objects require new kind of description.

On the other hand the more than a century successful usage of Maxwell equations
in describing various electromagnetic physical systems and processes undoubtedly
suggests that their adequateness to reality is hardly understood and used
fully. For example, one of the crucially important relations that follows from
Maxwell equations, and that has been successfully used throughout all these
years, is the Poynting energy-momentum balance equation in vacuum $$
\frac{\partial}{\partial t}\left(\frac{\mathbf{E}^2+\mathbf{B}^2}{2}\right)=
-c\,\mathrm{div}(\mathbf{E}\times\mathbf{B}),
$$
where $c$ is the velocity of light in
vacuum. Considered from the above mentioned point of view it turns out that
this Poynting equation admits time-stable, spatially finite and propagating
along straight lines solutions of the following kind
$$
\mathbf{E}=[u(x,y,ct+\varepsilon z),\,p\,(x,y,ct+\varepsilon z),0],\ \ \
\mathbf{B}=
[\varepsilon p\,(x,y,ct+\varepsilon z),\, -\varepsilon u(x,y,ct+\varepsilon
z),0],  \ \ \varepsilon=\pm1\ ,
$$
where $u$ and $p$ are {\it arbitrary} functions, so they can be chosen {\it
finite}. This observation suggests to look deeper and more carefully into the
structures and assumptions used for mathematical interpretation of the
experimental electric-magnetic induction discoveries made in the 19th century.
In other words, which relations, and on what grounds, should be defined as
basic, so that the further deduced equations and relations to give reasonable
and physically acceptable results when viewed from the viewpoint for {\it
spatial finiteness} and {\it time stability}. Finding the right way to choosing
adequate mathematical objects and corresponding equations seems specially
important when we try to describe the intrinsic dynamical properties of such
spatially finite and time stable field objects. Therefore, it seems very
important to have the right notion about concepts like physical object,
intrinsic property, dynamical property, identifying characteristics, dynamical
structure, admissible changes, field equations, etc.

The idea to extend the vacuum Maxwell equations in such a way that spatially
finite time stable and straight-line propagating solutions to be incorporated
is, of course, not new [4]. Moreover, a general principle concerning all
theoretical physics was formulated by Born and Infeld [5], stating: {\it a
satisfactory theory should avoid letting physical quantities become infinite}.
Other recent efforts in this direction in the frame of electrodynamics may be
found in [6,7,8,9]. We are not going to analyze here all these various
approaches, what however deserves to be noted is that the new equations offered
therein have not usually direct physical sense of energy-momentum balance
relations as the Newton law in mechanics has. And this is true also for the
very Maxwell equations. The above mentioned example with the Poynting relation
clearly suggests a more serious and physically motivated respect to be paid to
this aspect of the problem when some basic relation in a physical theory is
postulated. For example,  should we consider the very electric and magnetic
fields $(\mathbf{E,B})$ as force fields in case of no charges present, or the
divergence $\nabla_iM^{ij}$ of Maxwell's stress tensor $M^{ij}$ [10] should be
accepted in the pure field case as force field? Each of these three vector
fields generates integral lines, so, which integral lines should be considered
as force-lines, and what is a force-line in case of absence of charged
particles? Clearly, since the general concept of force, considered as local
object, is {\it local energy-momentum exchange}, more reliable seems to be the
divergence $\nabla_iM^{ij}$, so, the Newton-like equations of motion in
the vacuum case should read $\nabla_iM^{ij}=0$. In other words, the implied
energy-momentum exchange between the electric and magnetic components of the
field should be represented by relations having direct energy-momentum exchange
sense, and not by the local versions of the so called induction laws
\[
\mathrm{rot}\,\mathbf{E}+\frac{\partial\mathbf{B}}{\partial \xi}=0,\quad
\mathrm{rot}\,\mathbf{B}-\frac{\partial\mathbf{E}}{\partial \xi}=0,
\]
where $\xi=ct$. In fact, we can NOT observe and verify {\bf directly} these
relations since we have not corresponding devices, we verify them indirectly
through observing corresponding energy-momentum changes, i.e. state-behavior
changes, of charged particles. In other words, we have a system having two
subsystems: field and charged particles, and, assuming the energy-momentum
local conservation law, we make energy-momentum change conclusions about the
field through the corresponding mechanical energy-momentum changes of the
charged particles. Therefore, when we want to understand and describe
intra-field energy-momentum exchanges, i.e. in case of no charged objects
presented, we must have in mind some preliminary pictures about the structure
and possible more or less individualized and time-stable substructures among
which this energy-momentum inter-exchange should take place. Assuming the
electric and magnetic components as such substructures we find that the absence
of well defined local interaction energy between these two components (recall
that the energy density of the field is $\frac12(\mathbf{E}^2+\mathbf{B}^2)$)
seriously complicates this problem: the two Maxwell equations above imply some
energy exchange between $\mathbf{E}$ and $\mathbf{B}$, so how it is performed
if there is NO local interaction energy, moreover, neither $\mathbf{E}$ nor
$\mathbf{B}$ are capable to carry separately momentum and angular momentum, so,
which are the corresponding energy-momentum exchanging substructures of
the field during propagation?

The relativistic development of classical electrodynamics introduced new point
of view: adequate mathematical objects that represent such two substructures of
the general vacuum field are not $\mathbf{E}$ and $\mathbf{B}$, but two
differential 2-forms $F_{(\mathbf{E},\mathbf{B})}$ and
$*F_{(-\mathbf{B},\mathbf{E})}$ on Minkowski space-time, so, from the new point
of view, any internal energy-momentum exchange should take place between $F$
and $*F$. However, the "new" field equations $\mathbf{d}F=0, \mathbf{d}*F=0$,
although in terms of $F$ and $*F$, keep the old viewpoint, and in a definite
sense they forbid such internal exchange (see Sec.3.1). This study is directed
to find corresponding new equations.

The paper is organized as follows. In Sec.1.2 we present some general
considerations concerning the mathematical description of a physical system.
Sections 2.1-2.5 are devoted to defining the model in pre-relativistic terms.
Sections 3.1-3.6 give relativistic approach, making use of Frobenius curvature,
non-linear connections and relativistic strain. Sections 4.1-4.4 present
various views on dynamical equations and give solutions of photon-like nature.
Finally, Ch. 5 gives a retrospect of this study.

\section{Physical Objects and Interactions}

From a definite point of view every physical system is characterized by two
kinds of properties.  The {\bf first} kind of properties we call {\it
identifying}, they identify the system throughout its existence in time, so,
the corresponding physical quantities/relations must show definite
conservation/constancy properties (with respect to the identification procedure
assumed). Without such experimentally established properties we could not talk
about physical objects/systems at all.  The {\bf second} kind of properties
(which may be called {\it kinematical}) characterize the time-evolution of the
system, the corresponding quantities are time-dependent, and the corresponding
evolution is {\it consistent} with the conservative/constant character of the
identifying properties/quantities. In this sense, the equations of motion of a
physical system can be considered as relations determining the admissible
time-changes of these quantities. For example, the mass $m$ of a classical
particle is an identifying quantity, while the velocity $\mathbf v$ is a
kinematical one. This view implies, of course, that the external world acts on
the system under consideration also in an admissible way, i.e. an assumption is
made that the system survives, the interaction with the outside world does not
lead to its destruction.

In theoretical physics we usually make use of quantities which are functions of
the identifying and of the kinematical characteristics of the system and call
them {\it dynamical} quantities. A well known example is the momentum
$\mathbf{p}$ of a particle: $\mathbf{p}=m\mathbf{v}$. Of crucial importance for
the description of admissible changes are the conservative dynamical
quantities, i.e.  those which may pass from one physical system to another with
NO LOSS.  For example {\bf energy} and {\bf momentum} are such quantities,
moreover, they are universal in the sense that every physical object carries
non-zero energy-momentum and, vice versa, every quantity of energy-momentum is
carried by some physical object. So, if a definite quantity of energy-momentum
passes from one object to another, this same quantity of energy-momentum can be
expressed in terms of the characteristics of the two objects/systems, and the
two expressions to be equalized. This allows to describe interaction between,
or among, physical objects. Thus we have a consistent with the requirement for
"identification through conservation"-way to write equations of motion, and
this is the way used by Newton to write down his famous equations $\dot{\mathbf
p} =\mathbf F $, where $\mathbf F$ carries information about where the momentum
change of the particle has gone, or has come from.  This also clarifies {\it
the physical sense of the concept of force as a change of momentum}, or as a
change of energy-momentum in relativistic terms. Paying due respect to Newton
we shall call some equations of motion of {\it Newton type} if on the two sides
of "=" stay physical quantities of energy-momentum change, or energy-momentum
density change in the case of continuous systems. Note that, written down for
the vector field $\mathbf{p}$, i.e. in terms of partial derivatives, the above
Newton equation looks like $\nabla_{\mathbf{p}}\mathbf{p}=m\mathbf{F}$, where
the left hand side means performing two steps: first, determining the "change
quantity" $\nabla{\mathbf{p}}$, second, projecting $\nabla{\mathbf{p}}$ on
$\mathbf{p}$, and the right hand side may be expressed as a function of the
characteristics of both: the particle and the external physical environment.

If there is no energy-momentum (or energy-momentum density) change, then
putting the corresponding expression equal to zero, e.g.
$\nabla_{\mathbf{p}}{\mathbf{p}}=0$, we obtain the "free particle" or "free
field" equations. In such a case we just declare that only those changes are
admissible which are consistent with the (local and integral) energy-momentum
conservation.

We note that an initial extent of knowledge about the system we are going to
describe mathematically is presupposed to be available, so that the assumptions
made to be, more or less, well grounded. This knowledge is the base that
generates corresponding insight and directs our attention to the appropriate
mathematical structures. This is exclusively important when we deal with
continuous, or field, physical objects/systems.

To illustrate our idea, let's consider a many-component continuous system,
i.e. such that each component is assumed to be time-stable and recognizable
during the system's existence. Accordingly, the {\bf
wholeness+structural integrity} of the system should be mathematically
represented by some complex $\Phi$ of interconnected fields:
$\Phi=\{\Phi_{a}\}, a=1,2,\dots$, where each $\Phi_{a}$ represents some
individualized subsystem and may be also many-component one:
$\Phi_{a}=\{\Phi_{a}^1,\Phi_{a}^2,\dots\}$. In view of the above
considerations, if our system is free, the steps to follow
are:
\vskip 0.3cm
1. Specify and consider the mathematical model-object
$\Phi=\{\Phi_{a}^1,\Phi_{a}^2,\dots\}, a=1,2,\dots$ which is chosen to
represent the {\bf wholeness+structural integrity} of the physical system
considered;

2. Define the change-objects $D(\Phi_{a})$, considered as due to internal
interactions;

3. "Project" $D(\Phi_{a})$ on $\Phi_{a}/\Phi_{b}, b\neq a$, by means of
some (in most cases bilinear) map $\mathfrak{P}$;

4. The projections $\mathfrak{P}(D(\Phi_{a}),\Phi_{b})$ and
$\mathfrak{P}(D(\Phi_{b}),\Phi_{a})$
obtained, we interpret
physically as local energy-momentum exchange between the individualized
subsystems described by $\Phi_{a}$ and $\Phi_{b}$:
$\mathfrak{P}(D(\Phi_{a}),\Phi_{b})$ is the energy-momentum that $\Phi_{a}$
transfers to $\Phi_{b}$, and $\mathfrak{P}(D(\Phi_{b}),\Phi_{a})$ is the
energy-momentum that $\Phi_{b}$ transfers to $\Phi_{a}$.

5. The subsystem described by $\Phi_a$ may, or may not, keep its
energy-momentum unchanged during the internal interaction.
Correspondingly, in the first case we'll have
$\mathfrak{P}(D(\Phi_{a}),\Phi_{a})=0$,
and in the second case we shall have $\mathfrak{P}(D(\Phi_{a}),\Phi_{a})\neq 0$.

6. The sum $\Sigma_{a}\mathfrak{P}(D(\Phi_{a}),\Phi_{a})$ should be equal to
zero, meaning that our system $\Phi=\{\Phi_{a}\}$
conserves locally its energy-momentum.

 \vskip 0.3cm The zero value of the projection
$\mathfrak{P}(D(\Phi_{a}),\Phi_{a})$ is interpreted in the sense that the
identifying characteristics of $\Phi_{a}$ have not been disturbed,
or, the change $D(\Phi_{a})$ is qualified as {\it admissible}. This
consideration shows the importance of knowing how much and in what way(s) a
given physical system is {\it potentially able} to lose, or gain
energy-momentum (locally or globally), without losing its identity.

It is always very important to take care of the physical sense of the
quantities that we put on the two sides of the relation $A=B$. Mathematically,
from set theory point of view [13], $A$ and $B$ denote {\it the same} element,
which element may be expressed in different terms, e.g. the real number 2 can
be expressed as
$3-1=6/3=\frac{d}{dx}(2x+const)$ and also in
many other ways. From physical point of view, however, we must be
preliminary sure that $A$ and $B$ denote the same thing {\it qualitatively} and
{\it quantitatively}, i.e. $A$ and $B$ must denote the same {\it physical}
quantity. This is specially important when the equation we want to write down
constitutes some basic relation. And the point is not the physical dimension of
the two sides to be the same: any two quantities by means of an appropriate
constant can be made of the same physical dimension, but this is a formal step.
The point is that {\it the physical nature of the physical quantity on the two
sides must be the same}, and this should be well understood and correspondingly
guaranteed beforehand.

For example, it is quite clear that on the two sides of the Newton's law
$\dot{\mathbf p} =\mathbf F $ stays the well defined for any physical system
quantity "change of momentum" since the {\it momentum} quantity is a universal
one. For a counterexample, which physical quantity stays on the two sides of
the Poisson equation $\Delta U=k\rho, k=const$?  On one hand, such a quantity is
expressed through $\Delta U$ and, since $grad\,U$ is usually interpreted as
force, $\Delta U$ appears as a "change of force" characteristic of the field
$U$ since it is essentially defined by the second derivatives of $U$. On the
other hand, the same quantity is expressed through $k\rho$ and appears as a
characteristic of the mass particles, so, do we know such a quantity? The same
question can be raised for one of the Maxwell equations:
$\mathrm{rot}\,\mathbf{B}-\frac1c\dot{\mathbf{E}}=\frac{4\pi}{c}\mathbf{j}$.

In the case of classical particles momentum is always represented as the
product $m\mathbf v$ and this is carried to fluid mechanics (continuous mass
distribution) as $\mu(x,y,z;t).\mathbf v(x,y,z;t)$, where $\mu $ is the
invariant mass density. A similar quantity is introduced in electrodynamics as
electric current density $\mathbf{j}=\rho(x,y,z;t).\mathbf v(x,y,z;t)$, where
$\rho$ is the electric charge density. The energy-momentum exchange between the
field energy-momentum and the mechanical energy-momentum of the available
charged particles is described by the force field $\mathbf{F}=\rho\mathbf{E} +
\frac{1}{c}\mathbf{j}\times\mathbf{B}$. So, the corresponding Faraday-Maxwell
{\it force lines} should be the integral lines of the vector field
$\mathbf{F}$. Clearly, in the {\it charge-free} case we get $\mathbf{F}=0$, so
{the concept of force-lines defined by $\mathbf{F}$ does not work}. Hence, if
we would like to use this concept appropriately in the charge-free case, we
have to introduce it appropriately. The simplest way seems to consider the
integral lines of $\mathbf{E}$ and $\mathbf{B}$ as force lines also in the
charge free case, but we do not share this view: if $\rho=0$ then
$\mathbf{j}=0$, the force-vector is zero and NO integral force lines exist. The
two vectors $\mathbf{E}$ and $\mathbf{B}$ generate, of course, integral lines,
but these integral lines are NOT force lines in the vacuum case since the
vacuum $\mathbf{E}$ and $\mathbf{B}$ are NOT force fields, and such an
interpretation of the integral lines of $\mathbf{E}$ and $\mathbf{B}$ would be
misleading. In fact, if in case of $\mathbf{E}$ some compromise could be made
since $\mathbf{E}$ and $\rho\mathbf{E}$ are colinear, in the case of
$\mathbf{B}$ this is impossible: the corresponding force lines are generated by
$\mathbf{j}\times\mathbf{B}$, so, at each point they are orthogonal to the
integral lines of $\mathbf{B}$. We note that these problems arise in
connection with passing from discrete (point-like) quantities to continuous
quantities: mass and charge distributions, currents, etc., so such a transition
must be carefully performed in any theory in view of the above remarks. The rule
we are going to follow is: the sense of quantities used must always be quite
clear and must not be misleading.

As we already mentioned, there exists a {\it sufficiently good} force field
defined by Maxwell in terms of the divergence of his stress tensor $M^{ij}$,
which definition {\bf works quite well also out of and away from any media
built of, or containing, charged mass particles}. So, in the frame of the
theory at the end of 19th century if we ask the question: {\it if there are NO
charged particles and the time-dependent EM-field cannot transfer
energy-momentum to them by means of the force field $\mathbf{F}$, and the
propagation of the free EM-field is available, so that energy-momentum internal
exchanges should necessarily take place, how these processes and the entire
propagational behaviour of the field could be understood and modeled}?, the
right answer in our view should be: {\bf turn to $M^{ij}$ and consider
carefully the divergence $\nabla_iM^{ij}$ terms as possible force fields
generating corresponding force lines along which energy-momentum is internally
and locally transported between/among subsystems}. As will be seen further in
the paper, such a look on the issue would necessarily lead Maxwell and his
followers to the prediction that real, free, spatially finite and time-stable
formations of electromagnetic field nature having compatible
translational-rotational dynamical structure should exist, a result that has
been proved in studying the photoeffect phenomena about 30 years after
Maxwell's death.

We consider as a remarkable achievement of Maxwell the determination of the
correct expressions for the energy density of the electromagnetic field through
the concept of {\it stress} [10]. His electromagnetic {\it stress tensor}
$M^{ij}$ still plays an essential role in modern electromagnetic theory as a
part of the modern relativistic stress-energy-momentum tensor. However, by some
reasons, Maxwell did not make further use of the computed by him divergence
$\nabla_iM^{ij}$ of the stress tensor (and called by him "force field" [10]) for
writing down Newton type equations of motion for a free electromagnetic field
through {\it equalizing different expressions for the same momentum change}.
Probably, he had missed an appropriate interpretation of the vector
$c\,\mathbf{E}\times\mathbf{B}$ (introduced by Poynting 5 years after his death
and called "electromagnetic energy flux" [14]).

In connection with the above considerations the following more general question
arises: does theoretical physics make the right step {\it allowing static
force fields} to be written on the right hand side of Newton equation of motion
in mechanics? Every {\it static} field, whatever is its nature and origin,
necessarily conserves locally and globally all its physically meaningful
characteristics, so its energy is also conserved and is {\it not at disposal to
other physical systems}. Moreover, static fields do NOT propagate, so they do
NOT carry momentum, therefore, there is NO WAY other physical systems to lose
or gain momentum at the expense of externel static fields. Hence, trying to pay
respect to the momentum conservation through writing down Newton type dynamical
equations $\dot{\mathbf {p}} =\mathbf{F}$  with static force field $\mathbf{F}$,
we, in fact, {\bf violate} it: due respect requires due usage, so, in our
view, no selfrespecting theory should allow dynamical equations with static
$\mathbf{F}$!

This concerns not only mechanics, and in fact, not only static fields: {\it
propagating composite time-stable physical systems are supposed to consist of
individualized time-stable subsystems capable to interact, i.e. to carry and
exchange energy-momentum during propagation}. Do we respect always this natural
principle in physical theories? Consider for example the two vector vacuum
Maxwell equations: $\dot{\mathbf{B}}=-c\,\mathrm{rot}\,\mathbf{E}, \
\dot{\mathbf{E}}=c\,\mathrm{rot}\,\mathbf{B}$. These equations imply that there
is mutual physical influence between $\mathbf{E}$ and $\mathbf{B}$, which we
understand physically as energy-momentum exchange between the individualized
$\mathbf{E}$ and $\mathbf{B}$ components: $\mathbf{E}$ acts upon $\mathbf{B}$
and $\mathbf{B}$ acts upon $\mathbf{E}$. Now, according to the equations,
each of these two components propagates and keeps its individualization during
propagation, so, it should be able to carry momentum. However, the energy
concept of the theory excludes nonzero interaction energy between these two
components to exist since the energy density $w$ is given by the sum of the
energies carried by $\mathbf{E}$ and $\mathbf{B}$:
$w=\frac12(\mathbf{E}^2+\mathbf{B}^2)$, so, how does energy exchange take
place? Further, the momentum concept in the theory is defined and
experimentally proved quantitatively to be given by
$\frac1c(\mathbf{E}\times\mathbf{B})$, so neither of the assumed in the theory
electric and magnetic components is allowed to carry momentum separately. In
view of this, how e.g. in the plane wave solutions, where the relations
$\mathbf{E}^2=\mathbf{B}^2$ and
$\mathbf{E}^2+\mathbf{B}^2=2|\mathbf{E}\times\mathbf{B}|$ always hold, the
implied by the above equations internal energy-momentum exchange between the
two subsystems mathematically identified as $\mathbf{E}$ and $\mathbf{B}$ is
performed? May be we have not made the right mathematical identification of
the subsystems?

The suggestion that we should come to is that we must very carefully make
significant conclusions and assumptions about the dynamical structure of the
physical system under consideration, especially in case of continuous and
spatially propagating composite systems. In this work we shall try to follow the
rule that {\it an adequate choice of mathematical structures in the theory must
correspond to a sufficiantly well recognized, identified and interacting
physical structures}. In the spirit of this, our hope is that, in general, an
isolated time-stable nonstatic continuous physical system should correspond to
a completely integrable distribution $\Delta$ on an appropriately defined
manifold, its interacting subsystems $\Sigma_1, \Sigma_2, ...$
should correspond to NONintegrable
subdistributions $\Delta_1, \Delta_2, ...$
of $\Delta$, and the very interaction, i.e. local energy-momentum exchange,
between any two subsystems $\Sigma_i\rightleftarrows\Sigma_j$ can be described
mathematically in terms of the corresponding Frobenius curvatures
$\Omega_{ij}$.

\chapter{Nonrelativistic Considerations}
\section{How to understand the Coulomb force ?}
Usually, the
Coulomb force field is introduced starting with the Coulomb force law:
$f=\frac{qQ}{r^2}$, where $q$ and $Q$ are the charges of two small bodies
(usually considered as point-particles) and $r$ is the euclidean distance
between them. Two fields $\mathbf{E}_Q$ and $\mathbf{E}_q$, considered as
generated correspondingly by any of the two charges $Q$ and $q$, are defined by
the relations $$ \mathbf{E}_Q=lim_{q\rightarrow 0}\left(\frac{f}{q}\right) \ \
\ \text{and} \ \ \ \mathbf{E}_q=lim_{Q\rightarrow 0}\left(\frac{f}{Q}\right).
$$
Now the so defined quantities $\mathbf{E}_Q$ and $\mathbf{E}_q$ are considered
as vector fields, i.e. local objects, defined outside the regions occupied by the
source objects of charge magnitudes $Q$ and $q$, and are interpreted as
force-fields acting on other unit charges, hence, the force acting on the
$q$-charge is $q\mathbf{E}_Q$ and the force acting on the $Q$-charge is
$Q\mathbf{E}_q$.

The mechanical behaviour of the $q-$particle in the
reference frame connected with the $Q$-particle is defined by the Newton law
$\dot{\mathbf{p}_{q}}=q\,\mathbf{E}_{Q}$, where $\mathbf{p}_{q}$ is the
mechanical momentum of the $q$-particle in this reference frame (clearly,
$\mathbf{p}_{Q}=0$). This
"dynamical" equation presupposes that the change of the mechanical momentum of
the $q$-particle comes from (or goes to) the corresponding change of the
momentum carried by the field $\mathbf{E}_{Q}$ in accordance with the universal
momentum conservation law. However, such a justification assumes that the field
$\mathbf{E}_{Q}$ carries non-zero energy and momentum and is capable to
exchange them with other physical objects. How much is, for example, the field
momentum? This question requires corresponding definition of the field
momentum. This directs our attention to the theory based on Maxwell vacuum
equations. Maxwell theory, however, gives three objections to this
understanding of the physical situation: \vskip0.2cm -the field
$\mathbf{E}_{Q}(x,y,z)$, considered as local object, i.e. vector field,
satisfies the vacuum equations $\mathbf{rot}\,\mathbf{E}_{Q}=0,\
\mathbf{div}\,\mathbf{E}_{Q}=0$ outside its source, and according to the
theory, every such solution field conserves its energy, momentum and angular
momentum;

-the static nature of the field requires no
time-changes of any field characteristic;

-the field momentum density in the theory is proportional to the Poynting
vector, so, neither the electric field $\mathbf{E}$ nor the
magnetic field $\mathbf{B}$ are allowed to carry momentum separately.
\vskip 0.2cm
\noindent
In general, every vacuum solution of Maxwell equations
conserves its energy, momentum and angular momentum, so, {\bf NO vacuum
solution $(\mathbf{E},\mathbf{B})$ should be allowed to participate directly as
force generating agent in the expression
$q\mathbf{E}+\frac{q}{c}\mathbf{v}\times\mathbf{B}$}.

We see that for vacuum fields the usual setting "charged particle in external
field" does not work: the field can not afford any chance to the "test
particle" to gain locally momentum from the field. In view of this, how to
understand the Coulomb force law from theoretical point of view in the frame of
Maxwell theory?

In order to answer this question we make the following considerations. First,
some clarifications concerning the structure and admissible changes of the
physical situation. We have two mass particles carrying electric charges $q$ and
$Q$. The two masses "generate" two gravitational fields which are further
neglected as physical factors. The two charges "generate" two electric fields:
$\omega_q$ - denoted further just by $\omega$, and $\Omega_Q$ - denoted further
just by $\Omega$. The whole system is static and time stable, so, the two fields
and the two particles considered as mechanical objects,
exist consistently with each other. \vskip 0.3cm \noindent {\bf Remark}.\hskip
0.1cm We put the term "generate" in commas intentionally,
because in this case the charge-field configuration we consider as the real
one, i.e. no charged particle can exist without such a field, and no such a
field can exist without charged particle. In the theory the opposite of the
usually stated idea that charges generate fields is realized: not the charged
particle "generates" field, but the field "generates" charge through the Gauss
theorem, so, both the charge and the field aspects of the situation should be
paid equal respect. \vskip 0.3cm \noindent Since we consider electrostatic
situation, no magnetic fields are assumed to be present. The admissible
changes, by assumption, do NOT lead to destruction of any of the objects.
Paying now due respect to the Gauss theorem we have to assume that each of the
two fields is NOT defined inside the small region that its source occupies.
Therefore, the two fields $\omega$ and $\Omega$ are defined on the
topologically non-trivial space $\Sigma=\mathbb{R}^3\diagdown(W_q\cup W_Q)$,
where $W_q$ and $W_Q$ are the two small nonintersecting regions, treated
further as two balls with boundaries $S^2_q$ and $S^2_Q$, occupied by the two
particles. How to specify the mathematical nature of $\omega$ and $\Omega$ ?

The topology of $\Sigma$, which {\bf must be kept unchanged}, and the assumed
spherical symmetry of each of the two fields with respect to (the centers of)
$W_q$ and $W_Q$ suggest to choose the fields
$\Omega$ and $\omega$ as spherically symmetric representatives of the
2-dimensional cohomology group of $\Sigma$. We introduce two spherical
coordinate systems $(r,\theta,\varphi)$ and
$(\bar{r},\bar{\theta},\bar{\varphi})$,  originating at the centers of
$W_Q$ and $W_q$ respectively, so, any two spherically symmetric with respect to
the centers of $W_Q$ and $W_q$ 2-forms will look as follows:
 $$
\Omega(r,\theta,\varphi)=h(r)\sin\theta\mathbf{d}\theta\wedge\mathbf{d}\varphi,
\ \ \ \omega(\bar{r},\bar{\theta},\bar\varphi)=
\bar{h}(\bar{r})\sin\bar{\theta}\mathbf{d}\bar{\theta}\wedge\mathbf{d}\bar{\varphi}.
$$
Being representatives of corresponding cohomology classes $\Omega$ and $\omega$
must satisfy $\mathbf{d}\Omega=0$ and $\mathbf{d}\omega=0$, so,
$h(r)=Const$ and $\bar{h}(\bar{r})=const$. We denote $Const=Q$ and $const=q$.
Now, the euclidean Hodge star operator $*$ and the euclidean identification of
vectors and covectors give
$$
\mathbf{E}_{Q}=*\Omega(r,\theta,\varphi)=\frac{Q}{r^2}dr,\ \ \
\mathbf{E}_{q}=*\omega(\bar{r},\bar{\theta},\bar\varphi)=\frac{q}{\bar{r}^2}d\bar{r}.
$$

Going further we note that at every point of $\Sigma$ the real field
configuration is built of two physical fields of the {\it same physical
nature}, therefore, the resulted stress should depend on the local mutual
interference/interaction between the two stress generating fields $\Omega$ and
$\omega$. The point is how to model mathematically this local interaction of
the two fields? At this moment the Maxwell stress tensor
$M^{j}_{i}=\mathbf{Z}_i\mathbf{Z}^j-\frac12\mathbf{Z}^2\delta_i^j
=\frac14\Omega(\mathbf{Z})_{mn}\,\Omega(\mathbf{Z})^{mn}\delta_i^j-
\Omega(\mathbf{Z})_{im}\,\Omega(\mathbf{Z})^{jm},\ \
\Omega(\mathbf{Z})=*(\mathbf{Z})$, which is defined by any vector field
$\mathbf{Z}$, could help us as follows.

Mathematically, the tensor $M$ can be considered as a quadratic map from the
vector fields on $\Sigma$ to $(1,1)$-tensors, i.e. to the linear maps in the
linear space of vector fields. Each of our two fields generates such
$(1,1)$-tensor field:  $M(\omega)$ and $M(\Omega)$. Recall
now that every quadratic map $\Phi$ between two linear spaces generates a
bilinear map $T_{\Phi}$ according to $T_{\Phi}(x,y)=\Phi(x+y)-\Phi(x)-\Phi(y)$,
where $(x,y)$ are corresponding variables. So, in our case we can define
corresponding bilinear map.

Identifying the vector fields and 1-forms on $\Sigma$ by means of the
euclidean metric $e$: $\mathbf{\bar{E}}^i=e^{ij}\mathbf{E}_{j}$,
for the two Maxwell stress tensors, expressing here stresses of topological
origin, we have
$$
M_q\equiv M(\mathbf{\bar{E}_{q}})=\mathbf{E_{q}}\otimes
\mathbf{\bar{E}_{q}}-\frac12\,\mathbf{\bar{E}^2_{q}}\,id_{T\Sigma}, \
\ \ M_Q\equiv M(\mathbf{\bar{E}_{Q}})=\mathbf{E}_{Q}\otimes
\mathbf{\bar{E}_{Q}}-\frac12\,\mathbf{\bar{E}}^2_{Q}\,id_{T\Sigma}.
$$
The corresponding bilinear map $\mathbb{T}$ will be
\begin{equation}
\mathbb{T}(\mathbf{E}_{q},\mathbf{E}_{Q})=\mathbf{E}_{q}\otimes\mathbf{\bar{E}}_{Q}+
\mathbf{E}_{Q}\otimes\mathbf{\bar{E}}_{q}-\mathbf{\bar{E}}_{Q}.\mathbf{\bar{E}}_{q}\,id_{T\Sigma}.
\end{equation}
In components we have correspondingly
$$
(M_{q})_i^j=(E_{q})_i(\bar{E}_{q})^j-\frac12(\mathbf{\bar{E}}_{q})^2\delta_i^j=
\frac14\omega_{mn}\,\omega^{mn}\delta_i^j-\omega_{im}\,\omega^{jm},
$$
$$
(M_{Q})_i^j=(E_{Q})_i(\bar{E}_{Q})^j-
\frac12(\mathbf{\bar{E}}_{Q})^2\delta_i^j=
\frac14\Omega_{mn}\,\Omega^{mn}\delta_i^j-\Omega_{im}\,\Omega^{jm},
$$
$$
\mathbb{T}_i^j(\mathbf{E}_{q},\mathbf{E}_{Q})=
(E_{q})_i(\bar{E}_{Q})^j+(E_{Q})_i(\bar{E}_{q})^j-\bar{\mathbf{E}}_{q}.\bar{\mathbf{E}}_{Q}\,\delta_i^j.
$$
The tensor field $(-\frac{1}{4\pi}\mathbb{T})$ may be called {\it mutual stress
tensor}, or {\it interaction stress tensor}. In $T^*\Sigma\otimes T\Sigma$
we have the {\it trace} form $tr$, and on $\Sigma$ we have the
standard volume form $\omega_o=dx\wedge dy\wedge dz$. So we can form the object
$tr\otimes\omega_o$. By definition, the quantities
\begin{equation}
w=(tr\otimes\omega_o)(-\frac{1}{4\pi}\mathbb{T})=
-\frac{1}{4\pi}<tr,\mathbb{T}>\omega_o,
\ \ \ \text{and} \ \ \
U=\int_{\Sigma}w
\end{equation}
will be called {\it interaction energy density} and {\it interaction energy} for
$\omega$ and $\Omega$. Clearly, $w$ and $U$ may be positive, zero, or negative.
Further we shall follow the rule that an isolated (quasistatic) physical system
of this kind tends to configurations with minimal value $U_{min}$ of the
integral interaction energy $U$, hence, an intrinsically induced drifting
between two allowed static configurations should satisfy the relation $\delta
U<0$ .

In order to compute $U$ we compute first $w$ and obtain
$$
w=\frac{1}{4\pi}\bar{\mathbf{E}}.\mathbf{E}\,\omega_o=
\frac{1}{8\pi}\,(\Omega\wedge *\omega+ \omega\wedge *\Omega)=
-\frac{1}{8\pi}\left[\mathbf{d}\left(\frac{q}{\bar{r}}\Omega\right)+
\mathbf{d}\left(\frac{Q}{r}\,\omega\right)\right].
$$
Making use of the Stokes theorem the integral of $w$ over $\Sigma$ is
transformed to 2-dimensional surface integral over the boundary
$\partial\Sigma$ of $\Sigma$: $\partial\Sigma=S^2_{r,\bar{r}=\infty}\cup
S^2_q\cup S^2_{Q}$. On $S^2_{\infty}$ the corresponding integral has zero
value. So, in the induced on $\partial \Sigma$ orientation, and denoting by
$R_q$ and $R_Q$ the radiuses of $S^2_q$ and $S^2_Q$ respectively, we have $$
U=\frac{q}{2}\frac{1}{4\pi R^2_Q}\int_{S^2_q\cup S^2_Q} \frac{R^2_Q
\,Q\,\mathrm{sin}\theta\,d\theta\wedge\,d\varphi}{\bar{r}}+
\frac{Q}{2}\frac{1}{4\pi R^2_q}\int_{S^2_q\cup S^2_Q}
\frac{R^2_q\,q\,\mathrm{sin}\bar{\theta}\,d\bar{\theta}\wedge
d\bar{\varphi}}{r}\ .
$$
On
$S^2_q$ we have $\bar{r}=const$ and $\int_{S^2_q}\Omega=0$. Similarly, on
$S^2_Q$ we have $r=const$ and $\int_{S^2_Q}\omega=0$. Notice further that
$\frac{1}{r}$ is a harmonic function, so, at every point $p\in \Sigma$ it can
be represented by its average value on the corresponding 2-sphere centered at
$p$. Now, the first integral reduces to integral over the 2-sphere $S^2_Q$ and
it is equal to $\frac{qQ}{2R}$, similarly, the second integral
reduces to integral over the 2-sphere $S^2_q$ and has the same value,
$\frac{qQ}{2R}$, where $R$ is the euclidean distance between the centers of the
two spheres. Thus, the computation gives finally $U=\frac{qQ}{R}$.

Now, according to the above mentioned rule that $\delta U<0$, and that $q$
and $Q$ do not change, for the  case $q.Q>0$ we obtain
$\delta U=-\frac{qQ}{R^2}\delta R<0$, so $\delta R>0$, i.e. repulsion should be
expected; and for the case $q.Q<0$ we obtain
$\delta U=-\frac{qQ}{R^2}\delta R<0$, so $\delta R<0$, i.e. atraction should be
expected.

The above consideration clearly suggests the idea that the Coulomb force law
 originates from available interaction
between the two fields $\mathbf{E}_{Q}$ and $\mathbf{E}_{q}$ under quasistatic
changes of the integral interaction energy $U$ leading to
minimization of $U$. In fact, if $U$ changes then the change
$\delta U=-\frac{qQ}{R^2}\delta R$ must be carried away mechanically by the
$(q,m)$-particle: $\delta\frac{\mathbf{p^2}}{2m}=\delta U$,
 since there is no other physical
factor in the system considered. So, the Coulomb force
can be understood as an integral characteristic of the system, therefore its
field, i.e local, interpretation may be reconsidered. On the other hand, in the
corresponding spherical coordinates, $(Q*\omega)$ and $(q*\Omega)$ look very
much as $\delta U$, but this first-sight resemblance should not
 mislead us. The difference is quite serious:
 $(Q*\omega)$ and $(q*\Omega)$ are 1-forms, local objects by definition, while
$\frac{qQ}{R^2}\delta R$ is not local object, $R$ is not the coordinate $r$ and,
contrary to $dr$, $\delta R$ is not 1-form on $\Sigma$: we should not try to
obtain local objects just through noting some possible change tendencies of
integral characteristics of a system. We may allow ourselves to call $(\omega)$
and $(\Omega)$, or $(*\omega)$ and $(*\Omega)$, Coulomb fields but NOT Coulomb
force fields because {\bf they can NOT generate any direct local change of
momentum}, since as we mentioned earlier, {\bf these fields are static
and they conserve their energy, momentum and angular momentum}.
The entire local force is given in the theory
by the divergence of the Maxwell stress tensor which is a nonliner object,
namely, a bilinear combination of the field components and their derivatives
and all its terms are mutually compenseted in the static case.

As we saw, the Coulomb force gets an admissible interpretation as an
integral characteristic of the system describing some realizable integral
tendences to minimization of the integral interaction energy $U$ of the two
fields at the expense of the kinetic energy of the two particles. Surely,
$\omega$ and $\Omega$ carry some local physical information but in a quite
indirect manner: except spherical symmetry (which, of course, is not specific
only for electric fields) any of these two local objects can not clarifiy the
physical nature of the local changes in the space when charged particles are
around. In other words, from local point of view, we could not identify
$\omega$ as electric field. Any topologically nontrivial region of the kind
"$\mathbb{R}^3$ minus a point" generates such fields, so, the electric nature
of the field can be proved only by means of additional procedures concerning
the integral structure of the system.

The topological interpretation of $\omega$ and $\Omega$ also
suggests that the description is rather integral than local: although
$\omega$ and $\Omega$ are local objects, in fact they are just specially chosen
representatives of integral characteristics of the physical system considered:
they specify the topology of the space where the two fields are defined. For
another example, the Newton gravitation force law looks the same except the
different interpretation of the corresponding topological numbers as masses.
Following the same argument, the Newton gravitation force law is of integral
nature and shows similar tendences except that the masses are always
positive numbers, so, the corresponding interaction energy should be
always negative, which does not allow repulsion. But this
integral difference says too little about the local nature of the two
physically different field structures. \vskip 0.3cm \noindent {\bf Remark}.
 As for the relativistic formulation, $\omega$ and $\Omega$
can also be correctly defined and used, just the topology of the space where
the two fields are defined very slightly changes:
$\Sigma\rightarrow\Sigma\times\mathbb{R}$.
\vskip 0.3cm

The above consideration makes us think that, from theoretical point of view,
the Maxwell stress tensor field is the right object in terms of which local
force fields must be defined, namely, through computing its divergence. If the
field is static and free then this divergence is zero and represents physically
admissible quasistatic local changes, i.e. corresponding equations of motion,
and any additional conditions must be consistent with this zero divergence. In
our static case we have $\mathbf{d}\omega=0,\ \mathbf{d}*\omega=0$, so this
divergence is zero:
$$
\nabla_i M^i_k dx^k=\Big[(*\omega)^{i}(\mathbf{d}*\omega)_{ik}
+\frac12\,\omega^{ij}(\mathbf{d}\omega)_{ijk}\Big]dx^k=
\mathrm{rot}\,\mathbf{E}\times\mathbf{E}+\mathbf{E}\,\mathrm{div}\mathbf{E}=0 ,
$$
where
$\frac12\,\omega^{ij}(\mathbf{d}\omega)_{ijk}dx^k=
\mathbf{E}\,\mathrm{div}\mathbf{E}$,
$(*\omega)^{i}(\mathbf{d}*\omega)_{ik}dx^k=
(\mathrm{rot}\,\mathbf{E})\times\mathbf{E}$,
 $(x^1,x^2,x^3)=(x,y,z)$, and vectors and covectors are identified through the
euclidean metric. Hence, we can not gain energy from the field $\mathbf{E}$ in
local way. Therefore, the Coulomb force has not a local nature.

The situation seriously changes when we are going to consider independent and
self-consistent time-dependent and time-stable spatially finite physical
objects of electromagnetic nature, namely, we have no such topologically
motivated suggestions to choose adequate mathematical objects been able to
represent appropriately the corresponding physical stresses. Hence, the
mathematical model must be created on the basis of assumptions of quite
different nature, for example: requirements for definite and appropriately
defined integrability properties representing the object's time stability;
experimentally proved and traditionally assumed straight-line propagation of the
energy-density; orthogonality of the electric and magnetic components of the
field suggesting absence of local interaction energy between the electric and
magnetic components or their new versions; some notion for internal
energy redistribution during time-evolution, etc. In our view, in such cases,
the eigen and other algebraic properties of the corresponding
stress-energy-momentum tensor field should play a basic role.

In view of this in the next two chapters we consider in a more detail from
formal point of view the general Maxwell stress tensor as a starting physically
meaningful theoretical object .

\section{A non-physical view on Maxwell stress tensor}
The mathematical identities have
always attracted the attention of theorists, in particular, those identities which
involve the derivatives of the objects of interest (differential identities). A well
known such example is the Bianchi identity satisfied by any connection components: this
identity is a second order system of (in general, nonlinear) partial differential
equations. The gauge interpretation of classical Maxwell electrodynamics, as well as the
Yang-Mills theory, substantially make use of this identity. Such identities are of
particular importance when on the two sides of "=" stay correctly (i.e. in a coordinate
free way) defined expressions.

The physical reason to consider a couple of vector fields on $\mathbb{R}^3$ (or
a couple of spatial-directed vector fields on
$[\mathbb{R}^3\times(time-coordinate)]$
as mathematical representation of a free time dependent physical field comes
from the observation that, being free, it propagates translationaly along
straight lines, so the identification properties of the field should be searched
inside the 2-dimensional distribution that is orthogonal to these straight
lines.

We begin with the well known differential relation satisfied by the square of
every vector field $V$ on the euclidean space $\mathbb{R}^3$. Our attention is
directed to the square of $V$ just because of the experimentally suggested
assumption that $V^2$ mesures the energy-density of each of the electric and
magnetic components.

Let $\mathbb{R}^3$ be related to the standard coordinates
$(x^i=x,y,z), i=1,2,3$; we denote by
$"\times"$ the vector product, and make use of the $\nabla$-operator: \[
\frac12\nabla(V^2)=V\times
\mathrm{rot}\,V+(V.\nabla)V =V\times \mathrm{rot}\,V + \nabla_V V.
\]
 Clearly, on the two
sides of this relation stay well defined quantities, i.e. quantities defined in a
coordinate free way. The first term on the right hand side of this identity accounts for
the rotational component of the change of $V$, and the second term accounts mainly for
the translational component of the change of $V$. Making use of component notation we
write down the last term on the right side as follows (summation over the repeated
indices): \[ (\nabla_V V)^j=V^i\nabla_i V^j=\nabla_i(V^iV^j)-V^j\nabla_iV^i=
\nabla_i(V^iV^j)-V^j\mathrm{div}\,V . \] Substituting into the first identity, and making
some elementary transformations we obtain
\[
\nabla_i\left(V^iV^j-\frac12
\delta^{ij}V^2\right)= \big [(\mathrm{rot}\,V)\times V+V\mathrm{div}\,V\big ]^j,
\]
where
$\delta^{ij}=1$ for $i=j$, and $\delta^{ij}=0$ for $i\neq j$ are the euclidean
metric components. If now $W$ is another
vector field it must satisfy the same above identity:
\[ \nabla_i\left(W^iW^j-\frac12
\delta^{ij}W^2\right)= \big [(\mathrm{rot}\,W)\times W+W\mathrm{div}\,W\big ]^j.
\]
Summing up these two identities we obtain the new identity \setlength\arraycolsep{8pt}
\begin{eqnarray}
\lefteqn{ \nabla_iM^{ij}\equiv \nabla_i\left(V^iV^j+W^iW^j-
\delta^{ij}\frac{V^2+W^2}{2}\right)={} } \nonumber\\ & & {}=\big [(\mathrm{rot}\,V)\times
V+ V\mathrm{div}\,V+(\mathrm{rot}\,W)\times W+W\mathrm{div}\,W\big ]^j.
\end{eqnarray}

We note the invariance of $M^{ij}$ with respect to the transformations
$(V,W)\rightarrow
(V\mathrm{cos}\alpha-W\mathrm{sin}\alpha,V\mathrm{sin}\alpha+W\mathrm{cos}\alpha)$
where $\alpha(x,y,z)$ is an arbitrary real function. With respect to the
slightly more general transformation $(V,W)\rightarrow (V\,a-W\,b,V\,b+W\,a)$
where $(a,b)$ are real nonzero functions, we obtain
$M(V,W)\rightarrow(a^2+b^2)M(V,W)$. Hence, the transformations
$(V,W)\rightarrow (V\,a-W\,b,V\,b+W\,a)$ do not change the eigen
directions structure of $M^{ij}$.

The expression inside the round brackets on the left of (2.3), denoted by
$M^{ij}$, looks formally the same as the introduced by Maxwell tensor from
physical considerations concerned with the electromagnetic stress energy
properties of continuous media in presence of external electromagnetic field.
This allows to call formally any such tensor {\bf Maxwell stress tensor}
generated by the two vector fields $(V,W)$. The term "stress" in this general
mathematical setting is not topologically motivated as in the Coulomb case,
but could be justified in the following way. Every vector field on
$\mathbb{R}^3$ generates corresponding flow by means of the trajectories
started from some domain $U_o\subset\mathbb{R}^3$:  at the moment $t>0$ the
domain $U_o$ is diffeomorphically transformed to a new domain
$U_t\subset\mathbb{R}^3$. Having two vector fields on $\mathbb{R}^3$ we obtain
two {\it compatible} flows, so, the points of any domain
$U_o\subset\mathbb{R}^3$ are forced to accordingly move to new positions.

We emphasize the following moments: {\bf first}, the identity (2.3) is purely
mathematical; {\bf second}, on the two sides of (2.3) stay well
defined coordinate free quantities; {\bf third}, there is no $V\leftrightarrow
W$ interaction stress: the full stress is a sum of the $V$-stress and
$W$-stress.

Physically, we say that the corresponding physical medium that occupies the spatial
region $U_o$ and is parametrized by the points of the mathematical subregion
$U_o\subset\mathbb{R}^3$, is subject to {\it compatible} and {\it admissible}
physical "stresses" generated by physical interactions mathematically described
by the couple of vector fields $(V,W)$, and these physical stresses are
quantitatively described by the corresponding physical interpretation of the
tensor $M^{ij}(V,W)$.

We note that the stress tensor $M^{ij}$ in (2.3) is subject to the divergence
operator, and if we interpret the components of $M^{ij}$ as physical stresses,
then the left hand side of (2.3) acquires in general the physical interpretation
of force density. Of course, in the static situation as it is given by relation
(2.3), no energy-momentum propagation is possible, so at every point the forces
mutually compensate: $\nabla_{i}M^{ij}=0$. If propagation is allowed then the
force field is NOT zero: $\nabla_{i}M^{ij}\neq 0$, and we may identify the
right hand side of (2.3) as a {\bf real time-change} of appropriately defined
momentum density $\mathbf{P}$. So, assuming some expression for this momentum
density $\mathbf{P}$ we are ready to write down corresponding field equation of motion of
Newton type through equalizing the spatially directed force densities $\nabla_{i}M^{ij}$
with the momentum density changes along the time coordinate, i.e. equalizing
$\nabla_iM^{ij}$ with the $ct$-derivative of $\mathbf{P}$, where $c=const$ is the
translational propagation velocity of the momentum density flow of the physical system
$(V,W)$. In order to find how to choose $\mathbf{P}$ we have to turn to the intrinsic
physical properties of the field, so, it seems natural to turn to the eigen
properties of $M^{ij}$, since, clearly, namely $M^{ij}$ is assumed to carry the
physical properties of the field.

\section{Notes on the eigen properties of Maxwell stress tensor} We
consider $M^{ij}(\mathbf{E},\mathbf{B})$ at some point $p\in\mathbb{R}^3$ and
assume that in general the vector fields $\mathbf{E}$ and $\mathbf{B}$ are
linearly inependent, so $\mathbf{E}\times\mathbf{B}\neq 0$. Let the coordinate
system be chosen such that the coordinate plane $(x,y)$ to coincide with the
plane defined by $\mathbf{E}(p),\mathbf{B}(p)$. In this coordinate system
$\mathbf{E}=(E_1,E_2,0)$ and $\mathbf{B}=(B_1,B_2,0)$, so, identifying the
contravariant and covariant indices through the Euclidean metric $\delta^{ij}$
(so that $M^{ij}=M^i_j=M_{ij}$), we obtain the following nonzero components of
the stress tensor: $$ M^1_1=(E^1)^2+(B^1)^2-\frac12(\mathbf{E}^2+\mathbf{B}^2);
\ \ M^1_2=M^2_1=E^1\,E_2+B_1\,B^2; $$ $$
M^2_2=(E^2)^2+(B^2)^2-\frac12(\mathbf{E}^2+\mathbf{B}^2); \ \
M^3_3=-\frac12(\mathbf{E}^2+\mathbf{B}^2). $$ Since $M^1_1=-M^2_2$, the trace of $M$ is
$Tr(M)=-\frac12(\mathbf{E}^2+\mathbf{B}^2)$. The eigen value equation acquires the simple
form $\big[(M^1_1)^2-(\lambda)^2\big]+(M^1_2)^2\big](M^3_3-\lambda)=0$. The corresponding
eigen values are
\begin{equation}
\lambda_1=-\frac12(\mathbf{E}^2+\mathbf{B}^2);\ \
\lambda_{2,3}=\pm\sqrt{(M^1_1)^2+(M^1_2)^2}= \pm\frac12\sqrt{(I_1)^2+(I_2)^2} ,
\end{equation}
where
$I_1=\mathbf{B}^2-\mathbf{E}^2,\, I_2=2\mathbf{E}.\mathbf{B}$. The corresponding to
$\lambda_1$ eigen vector $Z_1$ must satisfy the equation
$\mathbf{E}(\mathbf{E}.Z_1)+\mathbf{B}(\mathbf{B}.Z_1)=0$, and since
$(\mathbf{E},\mathbf{B})$ are linearly independent, the two coefficients
$(\mathbf{E}.Z_1)$ and $(\mathbf{B}.Z_1)$ must be equal to
zero, therefore, $Z_1\neq 0$ must be orthogonal to $\mathbf{E}$ and
$\mathbf{B}$, i.e. $Z_1$ must be colinear to $\mathbf{E}\times\mathbf{B}$:

The other two eigen vectors $Z_{2,3}$
satisfy correspondingly the equations
$$ \mathbf{E}(\mathbf{E}.Z_{2,3})+
\mathbf{B}(\mathbf{B}.Z_{2,3})=\Big[\pm\frac12\sqrt{(I_1)^2+(I_2)^2}+
\frac12(\mathbf{E}^2+\mathbf{B}^2)\Big]Z_{2,3}.         \ \ \ \ \ \   (*)
$$
Taking into account the easily verified relation
\begin{equation}
\frac14\Big[(I_1)^2+(I_2)^2\Big]=
\left(\frac{\mathbf{E}^2+\mathbf{B}^2}{2}\right)^2-
|\mathbf{E}\times\mathbf{B}|^2 \ , \ \ \text{so}\ \ \ \,
\frac{\mathbf{E}^2+\mathbf{B}^2}{2}- |\mathbf{E}\times\mathbf{B}|\geq 0 \ ,
\end{equation}
we conclude that the coefficient before $Z_{2,3}$ on the right is always different
from zero, therefore, the eigen vectors $Z_{2,3}(p)$ lie in the plane defined
by $(\mathbf{E}(p),\mathbf{B}(p)), \ p\in \mathbb{R}^3$. In particular,
the above mentioned transformation properties of the
Maxwell stress tensor $M(V,W)\rightarrow (a^2+b^2)M(V,W)$ show that the
corresponding eigen directions do not change under the transformation
$(V,W)\rightarrow (V\,a-W\,b,V\,b+W\,a)$.

The above consideration suggests that {\it the intrinsically determined
potential dynamical abilities of propagation of the field are: translational
along $(\mathbf{E}\times\mathbf{B})$, and rotational inside the plane defined
by} $(\mathbf{E},\mathbf{B})$.

It is natural to ask now under what conditions the very $\mathbf{E}$ and
$\mathbf{B}$ may be eigen vectors of $M(\mathbf{E},\mathbf{B})$? Assuming
$\lambda_2=\frac12\sqrt{(I_1)^2+(I_2)^2}$ and $Z_2=\mathbf{E}$
in the above relation (*) and having in view that
$\mathbf{E}\times\mathbf{B}\neq 0$ we obtain that
$\mathbf{E}(\mathbf{E}^2)+\mathbf{B}(\mathbf{E}.\mathbf{B})$ must be
proportional to $\mathbf{E}$, so, $\mathbf{E}.\mathbf{B}=0$,
i.e. $I_2=0$. Moreover, substituting now $I_2=0$ in that same  relation we obtain
$$
\mathbf{E}^2=\frac12(\mathbf{B}^2-\mathbf{E}^2)+
\frac12(\mathbf{E}^2+\mathbf{B}^2)=\mathbf{B}^2, \ \ \text{i.e.}, \ \ I_1=0.
$$
The case
"-" sign before the square root, i.e. $\lambda_3=-\frac12\sqrt{(I_1)^2+(I_2)^2}$,
leads to analogical conclusions just the role of $\mathbf{E}$ and $\mathbf{B}$
is exchanged.

\vskip 0.2cm
{\bf Corollary}. $\mathbf{E}$ and $\mathbf{B}$ may be eigen
vectors of $M(\mathbf{E},\mathbf{B})$ only if
 $I_1=I_2=0$.
\vskip 0.2cm

The above notices suggest to consider in a more detail the case
$\lambda_2=-\lambda_3=0$ for the vacuum case. We shall show, making use of the
Lorentz transformation in 3-dimensional form that, if
these two relations do not hold then under $\mathbf{E}\times\mathbf{B}\neq 0$
the translational velocity of propagation
is less then the speed of light in vacuum $c$. Recall first the
transformation laws of the electric and magnetic vectors under Lorentz
transformation defined by the 3-velocity vector $\mathbf{v}$ and corresponding
parameter $\beta=v/c, v=|\mathbf{v}|$. If $\gamma$ denotes the factor
$1/\sqrt{1-\beta^2}$ then we have $$
\mathbf{E'}=\gamma\,\mathbf{E}+\frac{1-\gamma}{v^2}\mathbf{v}(\mathbf{E}.
\mathbf{v})+\frac{\gamma}{c}\mathbf{v}\times\mathbf{B} , $$ $$
\mathbf{B'}=\gamma\,\mathbf{B}+\frac{1-\gamma}{v^2}\mathbf{v}(\mathbf{B}.
\mathbf{v})-\frac{\gamma}{c}\mathbf{v}\times\mathbf{E} . $$

Assume first that $I_2=2\mathbf{E}.\mathbf{B}=0$, i.e. $\mathbf{E}$ and $\mathbf{B}$ are
orthogonal, so, in general, in some coordinate system we shall have
$\mathbf{E}\times\mathbf{B}\neq 0$ .

If $I_1>0$, i.e. $|\mathbf{E}|<|\mathbf{B}|$, we shall show that the
conditions $\mathbf{E'}=0, \mathbf{v}.\mathbf{B}=0, \infty>\gamma>0$ are
compatible. In fact, these assumptions lead to
$\gamma\,\mathbf{v}.\mathbf{E}+(1-\gamma)(\mathbf{E}.\mathbf{v})=0$, i.e.
$\mathbf{E}.\mathbf{v}=0$. Thus,
 $c|\mathbf{E}|=v|\mathbf{B}||\mathrm{sin}(\mathbf{v},\mathbf{B})|$, and since
$\mathbf{v}.\mathbf{B}=0$ then $|\mathrm{sin}(\mathbf{v},\mathbf{B})|=1$.
It follows that the speed
$v=c\frac{|\mathbf{E}|}{|\mathbf{B}|}<c$ is allowed.

If $I_1<0$, i.e. $|\mathbf{E}|>|\mathbf{B}|$, then the choice $\mathbf{B'}=0$ and
$\mathbf{v}.\mathbf{E}=0$ analogically lead to the conclusion that the speed
$v=c\frac{|\mathbf{B}|}{|\mathbf{E}|}<c$ is allowed.

Assume now that $I_2=2\mathbf{E}.\mathbf{B}\neq 0$. We are looking for a reference frame
$K'$ such that $\mathbf{E'}\times\mathbf{B'}=0$, while in the reference frame $K$ we have
$\mathbf{E}\times\mathbf{B}\neq 0$. We choose the relative velocity $\mathbf{v}$ such
that $\mathbf{v}.\mathbf{E}= \mathbf{v}.\mathbf{B}=0$. Under these conditions the
equation $\mathbf{E'}\times\mathbf{B'}=0$ reduces to $$
\mathbf{E}\times\mathbf{B}+\frac{\mathbf{v}}{c}(\mathbf{E}^2+\mathbf{B}^2)=0 ,\ \
\text{so}, \ \ \frac{v}{c}=|\mathbf{E}\times\mathbf{B}|/(\mathbf{E}^2+\mathbf{B}^2). $$
Now, from the above mentioned inequality
$\mathbf{E}^2+\mathbf{B}^2-2|\mathbf{E}\times\mathbf{B}|\geq 0$ it follows that
$\frac{v}{c}<1$.

Physically, these considerations show that under nonzero $I_1$ and $I_2$ the
translational velocity of propagation of the field, and of the field energy
density of course, will be less than $c$. Hence, the only realistic choice for
the vacuum case (where this velocity is assumed by definition to be equal to
$c$), is $I_1=I_2=0$, which is equivalent to
$\mathbf{E}^2+\mathbf{B}^2=2|\mathbf{E}\times\mathbf{B}|$. Hence, assuming
$|Tr(M)|$ to be the energy density of the field, the names "electromagnetic
energy flux" for the quantity $c\mathbf{E}\times\mathbf{B}$, and "momentum" for
the quantity $\frac1c\mathbf{E}\times\mathbf{B}$, seem well justified without
turning to any field equations.

These considerations show also that if $I_1=0$, i.e.
$|\mathbf{E}|^2=|\mathbf{B}|^2$ during propagation, then the electric and
magnetic components of the field carry always the same energy density, so, a
local mutual energy exchange between $\mathbf{E}$ and $\mathbf{B}$ is not
forbidden in general, but, if it takes place, it must be {\it simultanious} and
in {\it equal quantities}. Hence, under zero invariants $I_1=0$ and
$I_2=2\mathbf{E}.\mathbf{B}=0$, internal energy redistribution among possible
subsystems of the field is allowed but such an exchange should occur {\it
without available interaction energy} because the full energy density is always
equal to the sum of the energy carried by the electric and magnetic components
of the field. However, the required time stability and propagation
with velocity "c" of the field suggest/imply also available internal momentum
exchange since under these conditions the energy density is always equal to the
momentum magnitude $|\mathbf{E}\times\mathbf{B}|$, and $\mathbf{E}$ and
$\mathbf{B}$ can not carry momentum separately. Moreover, besides
$(\mathbf{E},\mathbf{B})$, another subsystem of the field has to be constructed
out of $(\mathbf{E},\mathbf{B})$ such, that both these two subsystems to carry
always the same quantity of energy-momentum, so the exchange also must be in
equal quantities.

After these preliminary considerations we procede to write down
dynamical equations for the field through specializing how the internal local
momentum exchange is realized keeping always in mind that the free field energy
density propagates translationaly with the speed of light, so the relations $$
I_1=I_2=0, \\ \ \ \text{i.e.} \ \ \
\mathbf{E}^2+\mathbf{B}^2=2|\mathbf{E}\times\mathbf{B}|
$$
must {\bf always} hold.

\section{Nonlinear equations for the electromagnetic field}

We are going to consider time dependent fields, and
begin with noting once again that the assumption that the energy
density of the field coincides with
$|Tr(M)|=\frac12[\mathbf{E}^2+\mathbf{B}^2]$ presupposes that {\it there is NO
interaction energy} between the electric and magnetic components of the field:
the full stress tensor (and the energy density, in particular) is a sum of the
stress tensors determined separately by $\mathbf{E}$ and $\mathbf{B}$. Of
course, this does NOT mean that there is no energy exchange between the
electric and magnetic components, but if such an exchange takes place, it must
occur simultaniously and in equal quantities..

Now, following the above stated idea that the field momentum density could be
responsible for such an internal energy-momentum exchange, we have to find at
least two appropriate subsystems of the field which subsystems are NOT the
electric $\mathbf{E}$ and magnetic $\mathbf{B}$ ones, but are constructed out
of them. Note that the assumption that the field momentum is given by
$\frac1c\mathbf{E}\times\mathbf{B}$, i.e. it is a bilinear function of the
electric and magnetic components, and that the local energy is always equal
to $|\mathbf{E}\times\mathbf{B}|$,
suggests that {\it the electromagnetic
momentum of the field is of interaction nature}. The point now is to get some
clarification how such a local momentum exchange (and the corresponding energy
exchange) takes place and to find appropriate mathematical representatives of
the corresponding partners realizing such special kind of energy exchange, since
{\it neither} $\mathbf{E}$ {\it nor} $\mathbf{B}$ {\it are able to carry
momentum separately} (although each of them may carry energy independently of
the other).

In view of the above we shall assume that the field keeps its identity through
adopting some special and appropriate dynamical behavior according to its
intrinsic capabilities. Hence, the corresponding dynamical/field equations must
be consistent with the intrinsic stress-energy-momentum nature of the
field. So, our basic assumption is that the Maxwell stress tensor
$M(\mathbf{E},\mathbf{B})$ should play the basic role, and its zero-divergence
in the static case should suggest how to determine the allowed dynamics.

Recalling that any member of the family
$$
(\mathcal{E},\mathcal{B})=(\mathbf{E}\,\mathrm{cos}\,\alpha-
\mathbf{B}\,\mathrm{sin}\,\alpha; \ \mathbf{E}\,\mathrm{sin}\,\alpha+
\mathbf{B}\,\mathrm{cos}\,\alpha), \ \ \ \alpha=\alpha(x,y,z,t),
$$
generate the same Maxwell stress tensor, the most natural assumption should
read like this: {\bf the field}
$(\mathbf{E},\mathbf{B})$ {\bf is looking for an energy-momentum exchanging
partner inside the $\alpha(x,y,z,t)$-familly of
$(\mathbf{E},\mathbf{B})$-couples, and any such couple identifies itself
through appropriate (local) interaction, defining in this way corresponding
dynamical behavior}.

Replacing $(V,W)$ in (2.3) with $(\mathbf{E},\mathbf{B})$ we obtain
\setlength\arraycolsep{8pt} \begin{eqnarray} \lefteqn{ \nabla_iM^{ij}\equiv
\nabla_i\left(\mathbf{E}^i\mathbf{E}^j+\mathbf{B}^i\mathbf{B}^j-
\delta^{ij}\frac{\mathbf{E}^2+\mathbf{B}^2}{2}\right)={} } \nonumber\\ & & {}= \big
[(\mathrm{rot}\,\mathbf{E})\times \mathbf{E}+ \mathbf{E}\mathrm{div}\,\mathbf{E}+
(\mathrm{rot}\,\mathbf{B})\times \mathbf{B}+ \mathbf{B}\mathrm{div}\,\mathbf{B}\big ]^j.
\end{eqnarray}

As we mentioned, in the static case, i.e. when the vector fields
$(\mathbf{E},\mathbf{B})$ do not depend on the time coordinate $\xi=ct$, NO propagation
of field momentum density $\mathbf{P}$ should take place, so, at every point, where
$(\mathbf{E},\mathbf{B})\neq 0$, the  stress generated forces must mutually compensate,
i.e. the divergence $\nabla_iM^{ij}$ should be equal to zero: $\nabla_iM^{ij}=0$. In this
static case Maxwell vacuum equations
\[
\mathrm{rot}\,\mathbf{E}+\frac{\partial\mathbf{B}}{\partial \xi}=0,\quad
\mathrm{rot}\,\mathbf{B}-\frac{\partial\mathbf{E}}{\partial \xi}=0,\quad
\mathrm{div}\,\mathbf{E}=0,\quad \mathrm{div}\,\mathbf{B}=0 \ \ \ \ \ \ \ (*)
\]
give:
$\mathrm{rot}\mathbf{E}=\mathrm{rot}\mathbf{B}=0;\,
\mathrm{div}\mathbf{E}=\mathrm{div}\mathbf{B}=0$, so, all static solutions to Maxwell
equations determine a sufficient, but NOT necessary, condition that brings to zero the
right hand side of (2.6) through forcing each of the four vectors there
to get zero values.

In the non-static case, i.e. when $\frac{\partial\mathbf{E}}{\partial t}\neq 0;
\,\frac{\partial\mathbf{B}}{\partial t}\neq 0$, time change and propagation of field
momentum density should take place, so, a full mutual compensation of the generated by
the Maxwell stresses at every spatial point local forces may NOT be possible, which means
$\nabla_iM^{ij}\neq 0$ in general. These local forces generate time-dependent momentum
propagation at the spatial points. Therefore, if we
want to describe this physical process of field energy-momentum density time change and
spatial propagation we have to introduce explicitly the dependence
$\mathbf{P}(\mathbf{E},\mathbf{B})$. If we follow the classical (nonrelativistic) way of
consideration and denote by $\mathfrak{F}$ the vector field with components
$\mathfrak{F}^j=\nabla_iM^{ij}$, we can write down the {\it force flow} across some
finite 2-surface $S$ in the usual (and widely spread in almost all textbooks) way as
$\int_{S}\mathfrak{F}.\mathbf{ds}$ (from modern point of view we should write
$i_{\mathfrak{F}}(dx\wedge dy\wedge dz)$ instead of $\mathfrak{F}.\mathbf{ds}$ under the
integral, where $i_{\mathfrak{F}}$ denotes the inner product between the vector field
$\mathfrak{F}$ and the volume form $dx\wedge dy\wedge dz$, i.e. to make use of the
Poincare isomorphism between vector fields and 2-forms on $\mathbb{R}^3$). This flow
generates changes of the momentum density flow across $S$ which should be equal to
$\frac{d}{dt}\int_{S}\mathbf{P}(\mathbf{E},\mathbf{B}).\mathbf{ds}$. We obtain \[
\frac{d}{dt}\int_{S}\mathbf{P}(\mathbf{E},\mathbf{B}).\mathbf{ds}=
\int_{S}\mathfrak{F}.\mathbf{ds} \ . \] \vskip 0.3cm The explicit expression for
$\mathbf{P}(\mathbf{E},\mathbf{B})$, paying due respect to J.Poynting [13], and
to J.J.Thomson, H.Poincare, M. Abraham [15], and in view of the huge, a century
and a half available experence, has to be introduced by the following \vskip
0.4cm \noindent {\bf Assumption}: {\it The entire field momentum density is
given by $\mathbf{P}:=\frac1c\mathbf{E}\times\mathbf{B}$} .
 \vskip 0.4cm
According to the {\bf Assumption} and the above interpretation of the relation
$\nabla_iM^{ij}\neq 0$, and in view of the arbitrariness of the 2-surface $S$ we come to
the vector differential equation
\[ \frac{\partial}{\partial
\xi}\left(\mathbf{E}\times\mathbf{B}\right)=\mathfrak{F}, \ \ \ \xi\equiv ct,   \ \ \ \ \
\ \ (**)
\]
which according to relation (2.6) is equivalent to
\begin{equation}
\left(\mathrm{rot}\,\mathbf{E}+\frac{\partial\mathbf{B}}{\partial \xi}\right)\times
\mathbf{E}+ \mathbf{E}\mathrm{div}\,\mathbf{E}+    
\left(\mathrm{rot}\,\mathbf{B}-\frac{\partial\mathbf{E}}{\partial \xi}\right)\times
\mathbf{B}+ \mathbf{B}\mathrm{div}\,\mathbf{B}=0.
\end{equation}
This last equation (2.7) we write down in the following equivalent way:
\begin{equation}
\left(\mathrm{rot}\,\mathbf{E}+\frac{\partial\mathbf{B}}{\partial \xi}\right)\times
\mathbf{E}+\mathbf{B}\mathrm{div}\,\mathbf{B}=     
-\left[\left(\mathrm{rot}\,\mathbf{B}-\frac{\partial\mathbf{E}}{\partial
\xi}\right)\times\mathbf{B}+\mathbf{E}\mathrm{div}\,\mathbf{E}\right].
\end{equation}
The
above relation (**) and the corresponding differential relation (2.7)/(2.8) we
consider as mathematical adequate in momentum-change terms of the
electric-magnetic and magnetic-electric induction phenomena in the charge free
case. We recall that these induction phenomena are described in what we call
"Faraday-Maxwell theory" by the following well known integral and differential
equations
\[
\frac{d}{d\xi}\int_{S}\mathbf{B}.\mathbf{ds}=-
\int_{S}\mathrm{rot}\mathbf{E}.\mathbf{ds}\ \ \ \rightarrow \ \ \
\frac{\partial\mathbf{B}}{\partial \xi}=-\mathrm{rot}\mathbf{E}, \ \ \ \text{(the Faraday
induction law)},
\]
\[
\frac{d}{d\xi}\int_{S}\mathbf{E}.\mathbf{ds}=
\int_{S}\mathrm{rot}\mathbf{B}.\mathbf{ds}\ \ \ \rightarrow \ \ \
\frac{\partial\mathbf{E}}{\partial \xi}=\mathrm{rot}\mathbf{B},  \ \ \ \text{(the Maxwell
displacement current law)}.
\]
We stress once again that these last Faraday-Maxwell
relations have NO {\it direct} energy-momentum change-propagation (i.e. force flow)
nature, so they could not be experimentally verified in a {\it direct} way. Our feeling
is that, in fact, they are stronger than needed. So, on the corresponding solutions of
these equations we'll be able to write down {\it formally adequate} energy-momentum
change expressions, but the correspondence of these expressions with the
experiment will crucially depend on the nature of these solutions. As we
already mentioned, the nature of the free solutions (with no boundary
conditions) to Maxwell vacuum equations with spatially finite and smooth enough
initial conditions requires strong time-instability (the Poisson theorem for
the D'Alembert wave equation). And time-stability of time-dependent vacuum
solutions usually requires spatial infinity (plane waves), which is physically
senseless. Making calculations with spatially finite parts of these spatially
infinite solutions may be practically acceptable, but from theoretical
viewpoint assuming these equations for {\it basic} ones seems not acceptable
since the relation "time stable physical object - exact free solution" is
strongly violated.

Before to go further we write down the right hand side bracket expression of
(2.8) in the following two equivalent ways:
\begin{equation}
\left[\left(\mathrm{rot}\,\mathbf{B}+\frac{\partial\mathbf{(-E)}}{\partial
\xi}\right)\times \mathbf{B}+ \mathbf{(-E)}\mathrm{div}\,\mathbf{(-E)}\right];\,
\left[\left(\mathrm{rot}\,\mathbf{(-B)}+\frac{\partial\mathbf{E}}
{\partial\xi}\right)\times\mathbf{(-B)}+\mathbf{E}\mathrm{div}\,\mathbf{E}\right].
\end{equation}
These last two expressions (2.9) can be considered as obtained from the left
hand side of (2.8) under the substitutions
$(\mathbf{E},\mathbf{B})\rightarrow(\mathbf{B},\mathbf{-E})$ and
$(\mathbf{E},\mathbf{B})\rightarrow(\mathbf{-B},\mathbf{E})$ respectively.
Hence, the field $(\mathbf{E},\mathbf{B})$ chooses as a partner-field
one of the fields $(\mathbf{-B},\mathbf{E})$, or $(\mathbf{B},\mathbf{-E})$.

We may resume this in the following way:
\vskip 0.3cm
{\bf An adequate mathematical representation of a time dependent free
electromagnetic field requires a collection of two fields} :
$\Big[(\mathbf{E},\mathbf{B});(\mathbf{-B},\mathbf{E})\Big]$, or
$\Big[(\mathbf{E},\mathbf{B});(\mathbf{B},\mathbf{-E})\Big]$. We could also say
that {\bf a real free field consists of two interacting subsystems described by
two partner-fields inside the $\alpha(x,y,z,t)$-family
$$
(\mathcal{E},\mathcal{B})=(\mathbf{E}\,\mathrm{cos}\,\alpha-
\mathbf{B}\,\mathrm{sin}\,\alpha; \ \mathbf{E}\,\mathrm{sin}\,\alpha+
\mathbf{B}\,\mathrm{cos}\,\alpha)
$$
giving the same Maxwell stress-energy
tensor. Each partner-field has electric and magnetic components, and each
partner-field is determined by the other through $(\pm\frac{\pi}{2})$ -
rotation-like transformation.  Both partner-fields carry  the same
energy-momentum and minimize the relation $I_1^2+I_2^2\geqslant 0$}. This view and
relation (2.7/2.8) suggest, in turn, that {\bf the intrinsic dynamics of free
real time-dependent electromagnetic fields could be considered as establishing
and maintaining local energy-momentum exchange partnership between two fields
called above partner-fields}, and, since $(\mathbf{E},\mathbf{B})$ and
$(\mathbf{-B},\mathbf{E})$ carry {\bf always} the same stress-energy-momentum,
{\bf the allowed inter-exchange is necessarily simultaneous and in equal
quantities}, so, {\bf each partner-field conserves its energy-momentum}.

\vskip 0.3cm
In order to find how much is the locally exchanged energy-momentum
we are going to interpret the equation
(2.8) in accordance with the view on equations of motion as stated in Sec.2.2.
Our object of interest $\Phi$, representing the integrity of a real time
dependent electromagnetic field, is the couple
$\Big[(\mathbf{E},\mathbf{B});(\mathbf{-B},\mathbf{E})\Big]$ (the other case
$\Big[(\mathbf{E},\mathbf{B});(\mathbf{B},\mathbf{-E})\Big]$ is considered
analogically). In view of the above considerations our equations should
directly describe admissible energy-momentum exchange between these recognized
two subsystems, i.e. from formal point of view, between the two
partner-fields. Hence, we have to define the corresponding change-objects
$D(\mathbf{E},\mathbf{B})$ and $D(\mathbf{-B},\mathbf{E})$ for each
partner-field, and their self-"projections" and their mutual "projections".

The change object $D(\mathbf{E},\mathbf{B})$ for the first partner-field
$(\mathbf{E},\mathbf{B})$ we naturally define as $$ D(\mathbf{E},\mathbf{B}):=
\left(\mathrm{rot}\mathbf{E}+ \frac{\partial\mathbf{B}}{\partial \xi}; \,
\mathrm{div}\mathbf{B}\right). $$ The corresponding "projection" of
$D(\mathbf{E},\mathbf{B})$ on $(\mathbf{E},\mathbf{B})$ $$
\mathfrak{P}\left[D(\mathbf{E},\mathbf{B}); (\mathbf{E},\mathbf{B})\right]=
\mathfrak{P}\left[\left(\mathrm{rot}\mathbf{E}+ \frac{\partial\mathbf{B}}{\partial \xi};
\, \mathrm{div}\mathbf{B}\right); (\mathbf{E},\mathbf{B})\right] $$ is suggested by the
left hand side of (8) and we define it by : $$
\mathfrak{P}\left[\left(\mathrm{rot}\mathbf{E}+ \frac{\partial\mathbf{B}}{\partial \xi};
\, \mathrm{div}\mathbf{B}\right); (\mathbf{E},\mathbf{B})\right]:=
\left(\mathrm{rot}\,\mathbf{E}+\frac{\partial\mathbf{B}}{\partial \xi}\right)\times
\mathbf{E}+\mathbf{B}\mathrm{div}\,\mathbf{B}. $$ For the second partner-field
$(-\mathbf{B},\mathbf{E})$, following the same procedure we obtain: $$
\mathfrak{P}\left[D(\mathbf{-B},\mathbf{E}); (\mathbf{-B},\mathbf{E})\right]=
\mathfrak{P}\left[\left(\mathrm{rot}\mathbf{(-B)}+ \frac{\partial\mathbf{E}}{\partial
\xi}; \, \mathrm{div}\mathbf{E}\right); (\mathbf{-B},\mathbf{E})\right]= $$ $$
=\left(\mathrm{rot}\,\mathbf{(-B)}+\frac{\partial\mathbf{E}}{\partial \xi}\right)\times
\mathbf{(-B)}+\mathbf{E}\mathrm{div}\,\mathbf{E}=
\left(\mathrm{rot}\,\mathbf{B}-\frac{\partial\mathbf{E}}{\partial \xi}\right)\times
\mathbf{B}+\mathbf{E}\mathrm{div}\,\mathbf{E}. $$ Hence, relation (2.7) looks
like $$ \mathfrak{P}\left[D(\mathbf{E},\mathbf{B});
(\mathbf{E},\mathbf{B})\right]+ \mathfrak{P}\left[D(\mathbf{-B},\mathbf{E});
(\mathbf{-B},\mathbf{E})\right]=0 . $$

The accepted two-component view on a real time dependent electromagnetic field allows in
principle admissible energy-momentum exchange with the outside world through any of the
two partner-fields. Hence, the above calculations suggest to interpret the two
sides of (2.8) as momentum quantities that each partner-field
$(\mathbf{E},\mathbf{B})$, or $(\mathbf{-B},\mathbf{E})$, is potentially able
to give to some other physical object withot destroying itself,
 and these quantities are expressed in
terms of $\mathbf{E},\mathbf{B}$ and their derivatives only. In the case of
free field, since no energy-momentum is lost by the field, there are two
possibilities: first, there is NO energy-momentum exchange between the two
partner-fields, second, each of the partner-fields changes its
energy-momentum at the expense of the other through simultanious and in equal
quantities exchanges. Such kind of mutual exchange is in correspondence with
the mathematical representatives of the two subsystems: the partner-fields
$(\mathbf{E},\mathbf{B})$ and $(\mathbf{-B},\mathbf{E})$ being members of the
above mentioned $\alpha(x,y,z,t)$-family, obviously carry the
same energy $\frac14(\mathbf{E^2}+\mathbf{B^2})$ and momentum
$\frac{1}{2c}(\mathbf{E}\times\mathbf{B})$. If we denote by $\Delta_{11}$ and
by $\Delta_{22}$ the allowed energy-momentum changes of the two
component-fields, by $\Delta_{12}$ the energy-momentum that the first
partner-field receives from the second partner-field, and by $\Delta_{21}$
the energy-momentum that the second partner-field receives from the first
partner-field, then according to the energy-momentum local conservation law
we may write the following equations:
\[
\Delta_{11}=\Delta_{12}+\Delta_{21}; \
\ \Delta_{22}=-\left(\Delta_{21}+\Delta_{12}\right),
\]
which is in accordance with the equation (2.8): $\Delta_{11}+\Delta_{22}=0$.

We determine now how the mutual momentum exchange between the two partner-fields
$\mathbf{P}_{(\mathbf{E},\mathbf{B})}\rightleftarrows
\mathbf{P}_{(\mathbf{-B},\mathbf{E})}$, or,
$\mathbf{P}_{(\mathbf{E},\mathbf{B})}\rightleftarrows
\mathbf{P}_{(\mathbf{B},\mathbf{-E})}$ is performed, i.e. the explicit expressions for
$\Delta_{12}$ and $\Delta_{21}$, keeping in mind that
$|\mathbf{P}_{(\mathbf{E},\mathbf{B})}|= |\mathbf{P}_{(\mathbf{-B},\mathbf{E})}|=
|\mathbf{P}_{(\mathbf{B},\mathbf{-E})}|=
|\frac12\mathbf{P}_{[(\mathbf{E},\mathbf{B});(\mathbf{-B},\mathbf{E})]}|$. The formal
expressions are easy to obtain. In fact, in the case
$\mathbf{P}_{(\mathbf{E},\mathbf{B})}\rightarrow \mathbf{P}_{(\mathbf{-B},\mathbf{E})}$,
i.e. the quantity $\Delta_{21}$, we have to "project" the change object for the second
partner-field given by
$$
D(\mathbf{-B},\mathbf{E}):= \left(\mathrm{rot}\mathbf{(-B)}+
\frac{\partial\mathbf{E}}{\partial \xi}; \, \mathrm{div}\mathbf{E}\right)
$$
on the first partner-field $(\mathbf{E},\mathbf{B})$. We obtain:
\begin{equation}
\Delta_{21}=
\left(\mathrm{rot}\,(\mathbf{-B})+\frac{\partial\mathbf{E}}{\partial \xi}\right)\times
\mathbf{E}+\mathbf{B}\mathrm{div}\,\mathbf{E}=
-\left(\mathrm{rot}\,\mathbf{B}-\frac{\partial\mathbf{E}}{\partial \xi}\right)\times
\mathbf{E}+\mathbf{B}\mathrm{div}\,\mathbf{E} \ .
\end{equation}
In the reverse case
$\mathbf{P}_{(\mathbf{-B},\mathbf{E})}\rightarrow \mathbf{P}_{(\mathbf{E},\mathbf{B})}$,
i.e. the quantity $\Delta_{12}$, we have to project the change-object for the first
partner-field $(\mathbf{E},\mathbf{B})$ given by
$$
D(\mathbf{E},\mathbf{B}):= \left(\mathrm{rot}\mathbf{E}+
\frac{\partial\mathbf{B}}{\partial \xi}; \, \mathrm{div}\mathbf{B}\right)
$$
on the second partner-field $(\mathbf{-B},\mathbf{E})$. We obtain
\begin{equation}
\Delta_{12}= \left(\mathrm{rot}\,\mathbf{E}+\frac{\partial\mathbf{B}}{\partial
\xi}\right)\times (\mathbf{-B})+\mathbf{E}\mathrm{div}\,\mathbf{B}=
-\left(\mathrm{rot}\,\mathbf{E}+\frac{\partial\mathbf{B}}{\partial
\xi}\right)\times\mathbf{B}+\mathbf{E}\mathrm{div}\,\mathbf{B}.
\end{equation}
So, the
internal local momentum balance is governed by the equations
\vskip 0.2cm
\begin{equation}
\left(\mathrm{rot}\,\mathbf{E}+\frac{\partial\mathbf{B}}{\partial
\xi}\right)\times \mathbf{E}+\mathbf{B}\mathrm{div}\,\mathbf{B}= \\
-\left(\mathrm{rot}\,\mathbf{E}+\frac{\partial\mathbf{B}}{\partial \xi}\right)\times
\mathbf{B}+\mathbf{E}\mathrm{div}\,\mathbf{B}-
\left(\mathrm{rot}\,\mathbf{B}-\frac{\partial\mathbf{E}}{\partial \xi}\right)\times
\mathbf{E}+\mathbf{B}\mathrm{div}\,\mathbf{E},
 \end{equation}
\vskip 0.3cm
\begin{equation}
\left(\mathrm{rot}\,\mathbf{B}-\frac{\partial\mathbf{E}}{\partial
\xi}\right)\times \mathbf{B}+\mathbf{E}\mathrm{div}\,\mathbf{E}= \\
\left(\mathrm{rot}\,\mathbf{B}-\frac{\partial\mathbf{E}}{\partial \xi}\right)\times
\mathbf{E}-\mathbf{B}\mathrm{div}\,\mathbf{E}+
\left(\mathrm{rot}\,\mathbf{E}+\frac{\partial\mathbf{B}}{\partial \xi}\right)\times
\mathbf{B}-\mathbf{E}\mathrm{div}\,\mathbf{B}.
\end{equation}
\vskip 0.2cm
These two vector equations (2.12)-(2.13) we consider as natural Newton type
field equations. According to them the intrinsic dynamics of a free
electromagnetic field is described by two couples of vector fields,
$[(\mathbf{E},\mathbf{B}); (\mathbf{-B},\mathbf{E})]$, or
$[(\mathbf{E},\mathbf{B}); (\mathbf{B},\mathbf{-E})]$, and this intrinsic
dynamics could be interpreted as a direct energy-momentum exchange between two
appropriately individualized subsystems mathematically described by these two
partner-fields.

A further natural specilization of the above two vector equations (2.12)-(2.13)
could be made if we recall that this internal energy-momentum exchange realizes
a special kind of dynamical equilibrium between the two partner-fields,
namely, the two partner-fields necessarily carry always the same energy and
momentum : $M^{ij}(\mathbf{E},\mathbf{B})=M^{ij}(\mathcal{E},\mathcal{B})$,
so each partner-field must conserve its momentum :
$\Delta_{11}=\Delta_{22}=0$. In such a dynamical situation each partner-field
loses as much as it gains during any time period, so, equations (2.12)-(2.13)
reduce to
\begin{equation}
\Delta_{11}\equiv\left(\mathrm{rot}\,\mathbf{E}+\frac{\partial\mathbf{B}}{\partial
\xi}\right)\times \mathbf{E}+\mathbf{B}\mathrm{div}\,\mathbf{B}=0,     
\end{equation}
\begin{equation}
\Delta_{22}\equiv
\left(\mathrm{rot}\,\mathbf{B}-\frac{\partial\mathbf{E}}{\partial      
\xi}\right)\times \mathbf{B}+\mathbf{E}\mathrm{div}\,\mathbf{E}=0,
\end{equation}
\begin{equation}
\Delta_{12}+\Delta_{21}\equiv\left(\mathrm{rot}\,\mathbf{E}+\frac{\partial\mathbf{B}}{\partial
\xi}\right)\times \mathbf{B}-\mathbf{E}\mathrm{div}\,\mathbf{B}+        
\left(\mathrm{rot}\,\mathbf{B}-\frac{\partial\mathbf{E}}{\partial \xi}\right)\times
\mathbf{E}-\mathbf{B}\mathrm{div}\,\mathbf{E}=0.
\end{equation}
Equation (2.16) fixes,
namely, that {\it the exchange of energy-momentum density between the two
partner-fields is {\bf simultanious} and in {\bf equal} quantities}, i.e. a
{\bf permanent dynamical equilibrium} between the two partner-fields holds:
$\mathbf{P}_{(\mathbf{E},\mathbf{B})}\rightleftarrows
\mathbf{P}_{(\mathbf{-B},\mathbf{E})}$, or,
$\mathbf{P}_{(\mathbf{E},\mathbf{B})}\rightleftarrows
\mathbf{P}_{(\mathbf{B},\mathbf{-E})}$.

Note that, if equations (2.14) and (2.15) may be considered as
field-equivalents to the zero force field (eqn. (2.14)) and its dual (eqn.
(2.15)), this double-field viewpoint and the corresponding mutual
energy-momentum exchange described by equation (2.16) are essentially new
moments. Equations (2.14)-(2.16) also suggest that the corresponding fields are
able to exchange energy-momentum with other physical systems in three ways. If
such an exchange has been done, then the exchanged energy-momentum
quantities can be given in terms of the characteristics of the other physical
system (or in terms of the characteristics of the both systems) and to be
correspondingly equalized to the left hand sides of equations (2.14)-(2.16) in
accordance with the local energy-momentum conservation law.

\subsection{Some Properties of the nonlinear solutions}
Clearly, all solutions to Maxwell pure field equations (*) are solutions to
our nonlinear equations (2.14)-(2.16), we shall call these solutions linear, and
will not be interested of them just because the notion for stress-energy-momentum
partnership between $(\mathbf{E},\mathbf{B})$ and $(-\mathbf{B},\mathbf{E})$ is
missing. Therefore,  we shall concentrate on those
solutions of (2.14)-(2.16) which satisfy the conditions
\[
\mathrm{rot}\,\mathbf{E}+\frac{\partial\mathbf{B}}{\partial \xi}\neq 0,\quad
\mathrm{rot}\,\mathbf{B}-\frac{\partial\mathbf{E}}{\partial \xi}\neq 0,\quad
\mathrm{div}\,\mathbf{E}\neq 0,\quad \mathrm{div}\,\mathbf{B}\neq 0.
\]
These solutions we call further nonlinear (among them there are no constant
ones as it is in the class of linear ones). We note some of the properties
they have.
\vskip 0.3cm
$1.\ \mathbf{E}.\mathbf{B}=0;$ \vskip 0.3cm $2.\
\left(\mathrm{rot}\,\mathbf{E}+ \frac{\partial\mathbf{B}}{\partial
\xi}\right).\mathbf{B}=0$; \ \ $\left(\mathrm{rot}\,\mathbf{B}-
\frac{\partial\mathbf{E}}{\partial \xi}\right).\mathbf{E}=0$.

\noindent
From these two relations the classical Poynting energy-momentum balance equation
follows.
 \vskip 0.3cm
The above two properties are obvious from equations (2.14) and (2.15).
 \vskip 0.3cm
3. If $(\mathbf{E},\mathbf{B})$ defines a solution then $(\mathbf{E}',\mathbf{B}')=
(a\mathbf{E}-b\mathbf{B};\  b\mathbf{E}+a\mathbf{B})$, where $a,b\in\mathbb{R}$, defines
also a solution. This property is immediately verified through substitution.
 \vskip 0.3cm
4. $\mathbf{E}^2=\mathbf{B}^2$.

\noindent
To prove this, we first multiply equation (2.14) on the
left by $\mathbf{E}$ and equation (2.16) by $\mathbf{B}$ (scalar products). Then
we make use of the above properties 1 and 2, of the vector algebra relation
$X.(Y\times Z)=Z.(X\times Y)$, and of the assumed nonlinear values of the
divergences of $\mathbf{E}$ and $\mathbf{B}$.

Properties 1. and 4. say that all nonlinear solutions to (2.14)-(2.16) are {\it
null fields}, i.e. the two well known relativistic invariants
$I_1=\mathbf{B}^2-\mathbf{E}^2$ and $I_2=2\mathbf{E}.\mathbf{B}$ of the field
are zero, and this property leads to optimisation of the inequality
$I_1^2+I_2^2\geqslant 0$ (recall the eigen properties of Maxwell stress tensor,
Sec.2.3), which in turn guarantees $\alpha(x,y,z,t)$-invariance of $I_1$
and $I_2$.

\vskip 0.3cm 5.\ \ $\mathbf{B}.\left(\mathrm{rot}\,\mathbf{B}-
\frac{\partial\mathbf{E}}{\partial \xi}\right)- \mathbf{E}.
\left(\mathrm{rot}\,\mathbf{E}+ \frac{\partial\mathbf{B}}{\partial \xi}\right)=
\mathbf{B}.\mathrm{rot}\mathbf{B}-\mathbf{E}.\mathrm{rot}\mathbf{E}=0.$

\noindent To prove this property we first multiply (vector product) (2.14) from
the right by $\mathbf{E}$, recall property 1., then multiply (scalar product)
from the left by $\mathbf{E}$, recall again $\mathbf{E}.\mathbf{B}=0$, then
multiply from the right (scalar product) by $\mathbf{B}$ and recall property 4.

Property 5. suggests the following consideration. If $\mathbf{V}$ is an
arbitrary vector field on $\mathbb{R}^3$ then the quantity
$\mathbf{V}.\mathrm{rot}\mathbf{V}$ is known as {\it local helicity} and its
integral over the whole (compact) region occupied by $\mathbf{V}$ is known as
{\it integral helicity}, or just as {\it helicity} of $\mathbf{V}$. Hence,
property 5. says that the electric and magnetic components of a nonlinear
solution generate the same helicities. If we consider (through the euclidean
metric) $\mathbf{E}$ as 1-form on $\mathbb{R}^3$ and denote by $\mathbf{d}$ the
exterior derivative on $\mathbb{R}^3$
, then $\mathbf{E}\wedge\mathbf{d}\mathbf{E}=
\mathbf{E}.\mathrm{rot}\mathbf{E}\,dx\wedge dy\wedge dz$, so, the zero helicity
says that the 1-form $\mathbf{E}$ defines a completely integrable Pfaff system:
$\mathbf{E}\wedge\mathbf{d}\mathbf{E}=0$.
The nonzero helicity says that the 1-form
$\mathbf{E}$ defines non-integrable 1d Pfaff system, so the nonzero helicity defines
corresponding curvature. Therefore the equality between the $\mathbf{E}$-helicity and the
$\mathbf{B}$-helicity suggests to consider the corresponding integral helicities
$\int_{\mathbb{R}^3}\mathbf{E}\wedge\mathbf{d}\mathbf{E}
=\int_{\mathbb{R}^3}\mathbf{B}\wedge\mathbf{d}\mathbf{B}$ (when they take finite nonzero
 values)
as a measure of the spin properties
of the solution. \vskip 0.3cm 6.\ \ Example of nonlinear solution:
\begin{align*} &\mathbf{E}=\left[\phi(x,y,ct+\varepsilon z)
\mathrm{cos}(-\kappa\frac{z}{l_o}+const), \, \phi(x,y,ct+\varepsilon
z)\mathrm{sin}(-\kappa\frac{z}{l_o}+const),\,0\right];\\ &\mathbf{B}=\left[\varepsilon
\phi(x,y,ct+\varepsilon z)\, \mathrm{sin}(-\kappa\frac{z}{l_o}+const),\, -\varepsilon
\phi(x,y,ct+\varepsilon z) \mathrm{cos}(-\kappa\frac{z}{l_o}+const),\,0\right],
\end{align*}
where $\phi(x,y,ct+\varepsilon z)$ is an arbitrary positive function,
$l_o<\infty$ is an arbitrary positive constant with physical dimension of
length, and $\varepsilon$ and $\kappa$ take values $\pm1$ independently.
Modifying the helicity 3-forms to $\kappa
\frac{4l^2_o}{c}\mathbf{E}\wedge\mathbf{d}\mathbf{E}= \kappa
\frac{4l^2_o}{c}\mathbf{B}\wedge\mathbf{d}\mathbf{B}$, then the corresponding
3d integral gives $\kappa TE$, where $\kappa=\pm 1$, $T=4l_o/c$ and
$E=\int{\phi^2}dxdydz$ is the integral energy of the solution.

\section{Discussion} The main idea of this part of the paper
is that carrying out the Newton way for writing down dynamical equations for
particles in mechanics to writing down dynamical equations for continuous field
systems should naturally result to nonlinear partial differential equations
even in non-relativistic theories. Moreover, clarifying the sense of the
information included in these dynamical equations according to the Newton
approach, we come to the conclusion formulated in the Introduction, namely, we
have to mathematically describe those changes of the object considered which
are qualified as {\it admissible} and {\it consistent} with the system's
identification and with the local energy-momentum balance relations.  In the
case of$\,$ "free" systems these relations represent the local energy-momentum
exchange/conservation properties of the system.  The energy-momentum
characteristics are chosen because of their two important properties: they are
physically {\it universal} and {\it conservative}. This means that {\it every}
physical object carries nonzero energy-momentum and, vice versa, {\it every}
quantity of energy-momentum is carried by some physical object. Hence, if a
physical object loses/gains some quantity of energy-momentum then some other
physical object necessarily gains/loses the same quantity of energy-momentum.
If this viewpoint is assumed, then the problem of finding appropriate dynamical
equations for an object reduces mainly to: {\bf first}, getting knowledge of
the potential abilities of the object considered to lose and gain
energy-momentum; {\bf second}, to create adequate mathematical quantities
describing locally these abilities.

The electromagnetic field, considered as a continuous physical object of special kind,
gives a good example in this direction since, thanks to Maxwell's fundamental and
summarizing works, all the information needed is available. The notices of
Poynting [13], and Thomson, Poincare and Abraham [14], showing the importance
of the (deduced from Maxwell equations) vector
$\frac1c\mathbf{E}\times\mathbf{B}$ from local energy-momentum propagation
point of view, has completed the resource of adequate and appropriate
mathematical objects since it appears as natural complement of Maxwell stress
tensor, and allows to write down dynamical field equations having direct local
energy-momentum balance sense. However, looking back in time, we see that this
viewpoint for writing down field equations has been neglected, theorists have
paid more respect and attention to the "linear part" of Maxwell theory,
enjoying, for example, the {\it exact} but not realistic, and even {\it
physically senseless} in many respects, plane wave solutions in the pure field
case.

Therefore, not so long after the appearance of Maxwell equations the photoeffect
experiments showed the nonadequateness of the linear part of Maxwell theory as a
mathematical model of electromagnetic fields producing realistic model-solutions of free
time-dependent fields. Although the almost a century long time development of standard
quantum and relativistic quantum theories that followed, a reasonable model-solutions
describing individual photons, considered as basic, spatially finite and time-stable
objects, these theories have not presented so far. Nobody doubts nowadays that photons
really exist, and this very fact suggests to try first classical field approach in
finding equations admitting 3d-finite and time stable solutions with appropriate
properties.

The historical perspective suggests to follow the 4-potential approach, but
modern knowledge and experience, and even the Maxwell stress tensor
achievements, suggest some different views. In fact, we have all reasons to
consider the microobjects as real as all other physical objects, so, no
point-like charges and infinite field model-solutions should be considered as
adequate. Since the 4-potential approach in Maxwell theory does not allow
spatially finite and time stable pure field solutions with photon-like
structure and behavior its interpretation as a basic concept does not seem to
be appreciable. Also, the 4-potential approach excludes many solutions of the
charge free Maxwell equations. For example, in relativistic terms the well
known field $F=\frac{q}{r^2}dr\wedge d\xi, \ \mathbf{d}F=0$, has global
4-potential, and its Minkowski-dual $*F=q\sin\,\theta\, d\theta\wedge d\varphi,
\ \mathbf{d}*F=0$, has NO global 4-potential. Now, the 2-parameter family of
2-forms $(\mathfrak{F},*\mathfrak{F})=(aF-b*F; bF+a*F), a,b\in\mathbb{R}$,
gives an infinite number of solutions to Maxwell equations
$\mathbf{d}\mathfrak{F}=0, \mathbf{d}*\mathfrak{F}=0$ admitting NO global
4-potential. This suggests the view that the 4-potential can be used as a {\it
working tool} (wherever it causes no controversies) but {\it not as a basic
concept}.

In conclusion, paying due respect to the Newton view on dynamical equations and
to the local energy-momentum conservation law we based our approach on the
Maxwell stress tensor and on the Poynting vector as natural quantities carrying
the physically meaningful energy-momentum characteristics of the
electromagnetic field. The natural description in these terms leads to the
assumption that any real time-dependent electromagnetic field consists of two
interacting subsystems mathematically represented by the two partner-fields:
$[(\mathbf{E},\mathbf{B})],[(\mathbf{-B},\mathbf{E})]$/
$[(\mathbf{E},\mathbf{B})],[(\mathbf{B},\mathbf{-E})]$, or any couple
$[(\mathcal{E},\mathcal{B}),(-\mathcal{B},\mathcal{E})]$ inside the considered
$\alpha(x,y,z,t)$-family of fields.
These two partner-fields carry always the same stress-energy-momentum, and a
dynamical equilibrium between these two subsystems is realized through a
simultanious mutual energy-momentum exchange in equal quantities. The equations
obtained represent formally this dynamical equilibrium, i.e. they show that
{\bf partner-fields identify/recognize each other through appropriate local
energy-momentum exchange partnership minimizing the quantity}
$\frac12\sqrt{I_1^2+I_2^2}=\frac12(\mathbf{E}^2+\mathbf{B}^2)-
|\mathbf{E}\times\mathbf{B}|\geqslant 0$. Accordinly, all nonlinear solutions
have zero invariants $I_1=I_2=0$, and can not be constant. Among these
zero-invariant nonlinear solutions there are time-stable and spatially finite
ones with helical spatial structure, having photon-like properties and
behavior. An analog of the Planck relation $E=h\nu$ holds for these solutions,
where the constant $h$ appears as an integral helicity of such a solution.

\chapter{Relativistic Considerations}
\section{The Extended Electrodynamics approach}
The generalization of Classical Electrodynamics (CED) known as Extended
Electrodynamics (EED) [11], starts with the conviction that
CED surely carries inside the potential ability to
be extended in such a way, that spatially finite and time-stable solutions of
photon-like nature to be incorporated, and it exploits mainly two ideas:
the idea for a direct local energy-momentum exchange sense of the
new dynamical equations, and the well known dual symmetry (mentioned above) of
the vacuum CED-equations and local conservation laws in the frame of
relativistic formalism. Let's consider first the elementary physical approach.

Maxwell vacuum equations
\[
\mathrm{rot}\,\mathbf{E}+\frac{\partial\mathbf{B}}{\partial \xi}=0,\quad
\mathrm{rot}\,\mathbf{B}-\frac{\partial\mathbf{E}}{\partial \xi}=0,\quad
\mathrm{div}\,\mathbf{E}=0,\quad \mathrm{div}\,\mathbf{B}=0 \ \ \ \ \ \ \ (*)
\]
clearly suggest that a free electromagnetic field has two vector components:
electric $\mathbf{E}$ and magnetic $\mathbf{B}$. On one hand,
from physical viewpoint, these equations imply also interaction, i.e.
energy-momentum exchange between the electric and magnetic components of the
electromagnetic field, and on the other hand, the energy density expression
$\frac12(\mathbf{E}^2+\mathbf{B}^2)$ in the theory does not contain interaction
energy term: the full energy density is the sum of the electric and magnetic
energy densities. Now, since any of these two components can NOT carry momentum
separately (the field momentum is given by
$\frac1c\mathbf{E}\times\mathbf{B}$), then $\mathbf{E}$ and $\mathbf{B}$ can
exchange only energy and NO momentum. But the field propagates translationally
along null straight-line directions with the speed of light "$c$", so it
necessarily carries momentum being numerically equal (in energy units) to the
energy-density, the so called {\it electromagnetic energy flux}. Hence, {\it
the field energy is of entirely dynamical nature and any internal energy
exchange between subsystems necessarily implies corresponding momentum
exchange}. Therefore, from energy-momentum exchange point of view, we {\it
should be interested in finding such subsystems of the field, which are able to
carry and exchange simultaniously both energy and momentum}.

In order to come to appropriate mathematical representatives in relativistic
terms of such subsystems
we recall that under null character of the local energy-momentum, i.e. when
$T_{\mu\nu}T^{\mu\nu}=0$, the
translational propagation requires zero invariants:
$I_1=\mathbf{B}^2-\mathbf{E}^2=0$ and $I_2=2\mathbf{E}.\mathbf{B}=0$,
 which in terms of the relativistic 2-form
formalism is equivalent respectively
to $I_1=\frac12F_{\alpha\beta}F^{\alpha\beta}=0$ and
$I_2=\frac12F_{\alpha\beta}(*F)^{\alpha\beta}=0$, where $*$ is defined by the
Minkowski (pseudo)metric:
$\alpha\wedge *\beta=
-\eta(\alpha,\beta)\sqrt{|det(\eta_{\mu\nu}|}dx^1\wedge dx^2\wedge dx^3\wedge
dx^4$. Now, recall the well known identity, being in force for any two 2-forms
$F$ and $G$ in Minkowski space-time $(M,\eta)$: $$
\frac12F_{\alpha\beta}G^{\alpha\beta}\delta_{\mu}^{\nu}=
F_{\mu\sigma}G^{\nu\sigma}-(*G)_{\mu\sigma}(*F)^{\nu\sigma}.
$$
Under $G=F$ and $\frac12F_{\alpha\beta}F^{\alpha\beta}=0$, it follows
$F_{\mu\sigma}F^{\nu\sigma}=(*F)_{\mu\sigma}(*F)^{\nu\sigma}$.
In view of the canonical stress-energy-momentum tensor of the field
(we omit the coefficient $\frac{1}{4\pi}$)
$$
T_{\mu}^{\nu}=-\frac12\Big[F_{\mu\sigma}F^{\nu\sigma}+
(*F)_{\mu\sigma}(*F)^{\nu\sigma}\Big]=
$$
$$
=\frac14F_{\alpha\beta}F^{\alpha\beta}\delta_{\mu}^{\nu}-
F_{\mu\sigma}F^{\nu\sigma}=
\frac14(*F)_{\alpha\beta}(*F)^{\alpha\beta}\delta_{\mu}^{\nu}-
(*F)_{\mu\sigma}(*F)^{\nu\sigma},
$$
satisfying the Rainich condition $T_{\mu\sigma}T^{\nu\sigma}=
\frac14T_{\alpha\beta}T^{\alpha\beta}\delta_{\mu}^{\nu}=
\frac14(I_1^2+I_2^2)\delta_{\mu}^{\nu}$(= 0 in our case),
this physically means that $F(\mathbf{E},\mathbf{B})$ and
$*F(-\mathbf{B},\mathbf{E})$ can carry both energy and momentum, moreover,
which is very important, they carry {\it always
the same stress-energy-momentum}. Formally, this is partially hidden in the
obvious invariance of $T_{\mu}^{\nu}$ with respect to $F\rightarrow *F$. Now,
since there is no other physical object participating in the energy-momentum
exchange, we come to the conclusion that for adequate mathematical
representatives of the two subsystems we are looking for, namely $F$ and $*F$
can be chosen. Moreover, since they necessarily carry always the same
stress-energy-momentum, they may exchange locally energy-momentum {\bf only
simultaniously and in equal quantities}. Therefore, in view of
$\nabla_\nu T_\mu^\nu=F^{\alpha\beta}(\mathbf{d}F)_{\alpha\beta\mu}+
(*F)^{\alpha\beta}(\mathbf{d}*F)_{\alpha\beta\mu}=0, \ \alpha<\beta $,
the most natural dynamical equations should read
$$
F^{\alpha\beta}(\mathbf{d}F)_{\alpha\beta\mu}=0, \ \
(*F)^{\alpha\beta}(\mathbf{d}*F)_{\alpha\beta\mu}=0, \ \
(*F)^{\alpha\beta}(\mathbf{d}F)_{\alpha\beta\mu}+
F^{\alpha\beta}(\mathbf{d}*F)_{\alpha\beta\mu}=0,\ \ \alpha<\beta,
$$
where the first two equations require that $F$ and $*F$ conserve the
energy-momentum they carry, and the third equation establishes the local
dynamical equilibrium between $F$ and $*F$:
$(*F)^{\alpha\beta}(\mathbf{d}F)_{\alpha\beta\mu}$ and
$F^{\alpha\beta}(\mathbf{d}*F)_{\alpha\beta\mu}$ denote respectively the
allowed from the consevation laws energy-momentum gains and losses of $*F$ and
$F$, which gains and losses are forbidden by the old equations
$\mathbf{d}F=0, \ \mathbf{d}*F=0$.

Extended Electrodynamics gives the following mathematical picture of
this field dynamics. Recall that if $(F,*F)$ is a CED vacuum solution, i.e.
$\mathbf{d}F=0, \ \mathbf{d}*F=0$, then the combinations
$\mathcal{F}=a\,F-b\,*F, \ \mathcal{F}^*=b\,F+a*F$, where $(a,b)$ are two
arbitrary real numbers, also give a CED vacuum solution and, since on Minkowski
space the reduced to 2-forms Hodge star $*$ satisfies the relation
$*^2=-id_{\Lambda^2(M)}$, we obtain $\mathcal{F}^*=*\mathcal{F}$. The two
corresponding energy tensors are related by $T(\mathcal{F},\mathcal{F}^*)=
(a^2+b^2)\,T(F,*F)$. Recall the real representation of complex numbers
$z=aI+bJ$ where $I$ is the unit matrix in $\mathbb{R}^2$ and $J$ is the
standard complex structure matrix in $\mathbb{R}^2$ with columns
$(0,-1);(1,0)$. So, we obtain an action of the linear group of
matrices $\alpha=aI+bJ$ on the CED vacuum solutions. This is a commutative
group $G$ and its Lie algebra $\mathcal{G}$ just adds the zero $(2\times 2)$
matrix to $G$, and $(I,J)$ define a natural basis of $\mathcal{G}$. So, having
a CED vacuum solution, we have in fact a 2-parameter family of vacuum
solutions. Hence, we can define a $\mathcal{G}$-valued 2-form $\Omega$ on $M$
by $\Omega=F\otimes\,I+*F\otimes\,J$, and the equation $\mathbf{d}\Omega=0$ is
equivalent to $\mathbf{d}F=0, \ \mathbf{d}*F=0$.

Consider the new basis $(I',J')$ of $\mathcal{G}$ given by
\[
I'=(aI+bJ),\quad J'=(-bI+aJ).
\]
Accordingly, the
"new" solution $\Omega'$, i.e. the old solution in the new basis of
$\mathcal{G}$, will be
\[
\Omega'=F\otimes I'+*F\otimes J'=F\otimes (aI+bJ)+*F\otimes (-bI+aJ)=
(a\,F-b\,*F)\otimes I+(b\,F+a\,*F)\otimes J.
\]
In view of this we may  consider this transformation as {\it nonessential},
i.e. we may consider $(F,*F)$ and
$(\mathcal{F},\mathcal{F}^*)$ as two different representations in corresponding
 bases of $\mathcal{G}$  of the same solution.

Such an interpretation is approporiate and useful if the field shows
some {\it invariant properties} with respect to this class of
transformations. For example, if the Lorentz invariants
$$
I_1=\frac12\,F_{\mu\nu}F^{\mu\nu}=(\mathbf{B}^2-\mathbf{E}^2),
\quad I_2=\frac12\,F_{\mu\nu}(*F)^{\mu\nu}=2\mathbf{E}.\mathbf{B},
$$
are zero: $I_1=I_2=0$,
then all the above transformations keep unchanged these zero-values of $I_1$ and
$I_2$.  In fact, under such a transformation
$(F,*F)\rightarrow(\mathcal{F},\mathcal{F}^*)$
the two Lorentz invariants transform to $(I_1',I_2')$ in the following way:
\[
I_1'=(a^2-b^2)\,I_1+2ab\,I_2,\quad I_2'=-2ab\,I_1+(a^2-b^2)\,I_2,
\]
and the determinant of this transformation is $(a^2+b^2)^2\neq 0$. So, a null
field stays a null field under these transformations. Moreover, NO
non-null field can be transformed to a null field by means of these
transformations, and, conversely, NO null field can be transformed to a
non-null field in this way. Hence, the Lorentz invariance and the dual
invariance of $I_1$ and $I_2$ hold simultanoiusly only in the null-field case.
This observation distinguishes once again the null-field case.

In order to come to the new equations we recall that
every bilinear map $\varphi:\mathcal{G}\times\mathcal{G}\rightarrow W$,
where $W$ is some linear space with basis $\{e_i\}, i=1,2,\dots$, defines
corresponding product in the $\mathcal{G}$-valued differential forms by means
of the relation \[ \varphi(\Omega_1^i\otimes e_i,\Omega_2^j\otimes e_j)=
\Omega_1^i\wedge\Omega_2^j\otimes\varphi(e_i,e_j).
\]
Now, let $\varphi=\vee$, where "$\vee$" is the symmetrized tensor product in
$\mathcal{G}$. We consider the expression $\vee(\Omega,*\mathbf{d}\Omega)$.
$$
\vee(\Omega,*\mathbf{d}\Omega)=(F\wedge *\mathbf{d}F)\otimes I\vee I+
(*F\wedge *\mathbf{d}*F)\otimes J\vee J+
(F\wedge *\mathbf{d}*F\otimes +*F\wedge *\mathbf{d}F)\otimes I\vee J .
$$
The vacuum EED equations are $\vee(\Omega,*\mathbf{d}\Omega)=0$, or
equavalently,
$$
F\wedge *\mathbf{d}F=0,\ \ (*F)\wedge *\mathbf{d}*F=0,\ \
F\wedge *\mathbf{d}*F+(*F)\wedge *\mathbf{d}F=0.
$$
In terms of the codifferential $\delta=*\mathbf{d}*$ these equations look like
$$
\delta *F\wedge F=0, \ \ \delta F\wedge *F=0, \ \ \delta F\wedge F-\delta
*F\wedge *F=0.
$$
In components we obtain correspondingly
$$
 F^{\alpha\beta}(\mathbf{d}F)_{\alpha\beta\mu}\equiv
(*F)_{\mu\nu}(\delta *F)^\nu=0,  \ \ \
(*F)^{\alpha\beta}(\mathbf{d}*F)_{\alpha\beta\mu}\equiv F_{\mu\nu}(\delta F)^\nu=0, \ \
\alpha<\beta; $$ $$ (*F)^{\alpha\beta}(\mathbf{d}F)_{\alpha\beta\mu}+
F^{\alpha\beta}(\mathbf{d}*F)_{\alpha\beta\mu}\equiv (\delta *F)^\nu F_{\nu\mu}+(\delta
F)^\nu (*F)_{\nu\mu}=0, \ \ \alpha<\beta.
$$
It is easy to see that these equations are equivalent to the equations
$\Delta_{11}=\Delta_{22}=0,\ \Delta_{12}+\Delta_{21}=0$ as given
in terms of $(\mathbf{E},\mathbf{B})$ in Sec.2.4. Moreover, all
nonlinear solutions to these EED vacuum equations, i.e. those satisfying
$\mathbf{d}F\neq 0,  \mathbf{d}*F\neq 0$, have zero invariants: $I_1=I_2=0$
(for the case EED in presence of media see [11]).

As for the energy-momentum tensor $T_{\mu\nu}$ of the vacuum solutions,
considered as a symmetric 2-form on $M$, it is defined in terms of $\Omega$
as follows:
\[
T(X,Y)=\frac12 *g\big[i(X)\Omega,*i(Y)\Omega\big]=
-\frac12\,X^\mu
Y^\nu\Big[F_{\mu\sigma}F_{\nu}\,^{\sigma}+
(*F)_{\mu\sigma}(*F)_{\nu}\,^{\sigma}\Big]=
X^\mu Y^\nu T_{\mu\nu},
\]
where
$(X,Y)$ are two arbitrary vector fields on $M$, $g$ is the metric in
$\mathcal{G}$ defined by $g(\alpha,\beta)=\frac12 tr(\alpha.\beta^*)$, and
$\beta^*$ is the transposed to $\beta$. Note that $g(I,J)=
g(aI+bJ,-bI-aJ)=0$, which
elliminates the corresponding coefficient in $T(X,Y)$, which reads
$F_{\mu\sigma}(*F)^{\nu\sigma}+
(*F)_{\mu\sigma}F^{\nu\sigma}=\frac12F_{\alpha\beta}(*F)^{\alpha\beta}\delta_\mu^\nu$,
so, in a $g$-NONorthogonal basis of $\mathcal{G}$ this coefficient will appear.
Now, since $\frac12F_{\alpha\beta}(*F)^{\alpha\beta}=
2(\mathbf{E}.\mathbf{B})$, choosing $g$-orthogonal basis in $\mathcal{G}$
corresponds to mutual orthogonality of $(\mathbf{E},\mathbf{B})$ in this
context, and choosing $g$-nonorthogonal basis of $\mathcal{G}$ will
formally result in some interaction between $\mathbf{E}$ and $\mathbf{B}$.

Finally, recall the generalization of Lie derivative $\mathcal{L}_{K}$ with
respect to the $k$-vector $K$, acting in the exterior
algebra of differential forms according to the formula
$\mathcal{L}_{K}=i(K)\mathbf{d}-(-1)^{k}\mathbf{d}i(K)$ [12]. Then, in view of
the relations $F_{\mu\nu}F^{\mu\nu}=(*F)_{\mu\nu}F^{\mu\nu}=0$, the above
equations acquire the form $$ \mathcal{L}_{\bar{F}}F=0, \ \
\mathcal{L}_{\bar{*F}}(*F)=0, \ \
\mathcal{L}_{\bar{F}}(*F)+\mathcal{L}_{(\bar{*F})}F=0, $$ where $\bar{F}$ and
$\bar{*F}$ are the $\eta$-corresponding 2-vectors. In terms of $\Omega$ and
$\bar{\Omega}=\bar{F}\otimes e_1+\bar{*F}\otimes e_2$ these three equations can
be represented as one relation as follows:
$$
\mathcal{L}^{\vee}_{\bar{\Omega}}\Omega=\mathcal{L}_{\bar{F}}F\otimes e_1\vee
e_1+ \mathcal{L}_{\bar{*F}}*F\otimes e_2\vee e_2+
(\mathcal{L}_{\bar{F}}*F+\mathcal{L}_{\bar{*F}}F)\otimes e_1\vee e_2=0.
$$

The above consideration is based on the assumption that the
$\mathcal{G}$-valued 2-form $\Omega(\mathbf{F},*\mathbf{F})$ represents
mathematically the {\bf wholeness+structural integrity} of the electromagnetic
field through the equations $\mathcal{L}^{\vee}_{\bar{\Omega}}\Omega=0$, and
direct physical motivation for such an assumption was not given. In order to
motivate looking for photon-like solutions of these equations we now present
our notion for photon-like object(s), and further we show how this notion
leads to corresponding mathematics by means of which we could separate the
desired subclass of solutions.

\section{The Notion of Photon-like Object(s)}

\subsection{Introduction} At the very dawn of the 20th century Planck [16]
proposed and a little bit later Einstein [17] appropriately
used the well known and widely used through the whole last century simple
formula $E=h\nu$, $h=const>0$.  This formula marked the beginning of a new era
and became a real symbol of the physical science during the following years.
According to the Einstein's interpretation it gives the full energy $E$ of {\it
really existing} light quanta of frequency $\nu=const$, and in this way a new
understanding of the nature of the electromagnetic field was introduced:
{\it the field has structure}, which contradicts the description given by
Maxwell vacuum equations.  After De Broglie's [18] suggestion for the
particle-wave nature of the electron obeying the same energy-frequency
relation, one could read Planck's formula in the following way:  {\it there are
physical objects in Nature the very existence of which is strongly connected to
some periodic (with time period $T=1/\nu$) process of intrinsic for the object
nature and such that the Lorentz invariant product $ET$ is equal to $h$}. Such
a reading should suggest that these objects do NOT admit point-like
approximation since the relativity principle for free point particles requires
straight-line uniform motion, hence, no periodicity should be allowed.

Although the great (from pragmatic point of view) achievements of the developed
theoretical approach, known as {\it quantum theory}, the great challenge to
build an adequate description of individual representatives of these objects,
especially of light quanta called by Lewis {\it photons} [19], is still to be
appropriately met since  the efforts made in this direction, we have to admit,
still have not brought satisfactory results. Recall that Einstein in his late
years recognizes [20] that "the whole fifty years of conscious brooding have
not brought me nearer to the answer to the question "what are light quanta",
and now, half a century later, theoretical physics still needs progress to
present a satisfactory answer to the question "what is a photon". We consider
the corresponding theoretically directed efforts as necessary and even {\it
urgent} in view of the growing amount of definite experimental skills in
manipulation with individual photons, in particular, in connection with the
experimental advancement in the "quantum computer" project.  The dominating
modern theoretical view on microobjects is based on the notions and concepts of
quantum field theory (QFT) where the structure of the photon (as well as of any
other microobject) is accounted for mainly through the so called {\it
structural function}, and highly expensive and delicate collision experiments
are planned and carried out namely in the frame of these concepts and methods
(see the 'PHOTON' Conferences Proceedings, some recent review papers: [21-24]).
Going not in details we just note a special feature of this QFT approach: if
the study of a microobject leads to conclusion that it has structure, i.e. it
is not point-like, then the corresponding constituents of this structure are
considered as point-like, so the point-likeness stays in the theory just in a
lower level.

According to our view on PhLO we follow here an approach based on the
assumption that the description of the available (most probably NOT arbitrary)
spatial structure of photon-like objects can be made by {\it continuous
finite/localized} functions of the three space variables.  The difficulties met
in this approach consist mainly, in our view, in finding adequate enough
mathematical objects and solving appropriate PDE.  The lack of sufficiently
reliable corresponding information made us look into the problem from as
general as possible point of view on the basis of those properties of
photon-like objects which may be considered as most undoubtedly trustful, and
in some sense, {\it identifying}. The analysis made suggested that such a
property seems to be {\it the available and
intrinsically compatible translational-rotational dynamical structure}, so we
shall focus on this property in order to see what useful for our purpose
suggestions could be deduced and what appropriate structures could be
constructed. All these suggestions and structures should be the building
material for a step-by-step creation of a {\it self-consistent} system. From
physical point of view this should mean that the corresponding properties may
combine to express a dynamical harmony in the inter-existence of appropriately
defined subsystems of a finite and time stable larger physical system.

\subsection{The notion of photon-like object} We begin with recalling our view
that any notion of a physical object must unify two kinds of properties of the
object considered: {\it identifying} and {\it kinematical}. The identifying
properties being represented by quantities and relations, stay unchanged
throughout the existence, i.e. throughout the time-evolution, of the object,
they represent all the intrinsic structure and relations.  The kinematical
properties describe those changes, called {\it admissible}, which do NOT lead
to destruction of the object, i.e. to the destruction of any of the identifying
properties. Correspondingly, physics introduces two kinds of quantities and
relations, identifying and kinematical. From theoretical point of view the more
important quantities used turn out to be the {\it dynamical} quantities which,
as a rule, are functions of the identifying and kinematical ones, and the joint
relations they satisfy represent the necessary interelations between them in
order this object to survive under external influence. This view suggests to
introduce the following notion of Photon-like object(s) (we shall use the
abbreviation "PhLO" for "Photon-like object(s)"):

\begin{center} {\bf PhLO are real massless time-stable physical objects with an
intrinsically compatible translational-rotational dynamical structure}.
\end{center}

We give now some explanatory comments, beginning with the term {\bf real}. {\bf
First} we emphasize that this term means that we consider PhLO as {\it really}
existing {\it physical} objects, not as appropriate and helpful but imaginary
(theoretical) entities.  Accordingly, PhLO {\bf necessarily carry
energy-momentum}, otherwise, they could hardly be detected by physical means.
{\bf Second}, PhLO can undoubtedly be {\it created} and {\it destroyed}, so, no
point-like and infinite models are reasonable: point-like objects are assumed
to have no structure, so they can not be destroyed since there is no available
structure to be destroyed; creation of infinite physical objects (e.g. plane
waves) requires infinite quantity of energy to be transformed from one kind to
another during finite time-periods, which seems also unreasonable. Accordingly,
PhLO are {\it spatially finite} and have to be modeled like such ones, which is
the only possibility to be consistent with their "created-destroyed" nature. It
seems hardly reasonable to believe that PhLO can not be created and destroyed,
and that spatially infinite and indestructible physical objects may exist at
all. {\bf Third}, "spatially finite" implies that PhLO may carry only {\it
finite values} of physical (conservative or non-conservative) quantities.  In
particular, the most universal physical quantity seems to be the
energy-momentum, so the model must allow finite integral values of
energy-momentum to be carried by the corresponding solutions. {\bf Fourth},
"spatially finite" means also that PhLO {\it propagate}, i.e.  they do not
"move" like classical particles along trajectories, therefore, partial
differential equations should be used to describe their evolution in time.

The term "{\bf massless}" characterizes physically the way of propagation in
terms of appropriate dynamical quantities: the {\it integral}
4-momentum $p$ of a PhLO should satisfy the relation $p_\mu p^\mu=0$, meaning
that its integral energy-momentum vector {\it must be isotropic}, i.e. to
have zero module with respect to Minkowski (pseudo)metric in
$\mathbb{R}^4$. If the object considered has spatial and time-stable structure,
so that the translational velocity of every point where the corresponding field
functions are different from zero must be equal to $c$, we have in fact null
direction in the space-time {\it intrinsically determined} by a PhLO. Such a
direction is formally defined by a null vector field $\bar{\zeta},\bar{\zeta}^2=0$. The
integral trajectories of this vector field are isotropic (or null) {\it straight
lines} as is traditionally assumed in physics, except in presence of gravity.
It follows that with every PhLO a null straight line direction is {\it
necessarily} associated, so, canonical coordinates
$(x^1,x^2,x^3,x^4)=(x,y,z,\xi=ct)$ on $\mathbb{R}^4$ may be chosen such that in
the corresponding coordinate frame $\bar{\zeta}$ to have only two non-zero
components of magnitude $1$: $\bar{\zeta}^\mu=(0,0,-\varepsilon, 1)$, where
$\varepsilon=\pm 1$ accounts for the two directions along the coordinate $z$
(further such a coordinate system will be called $\bar{\zeta}$-adapted and will
be of main usage). It seems important to emphasize that our PhLO propagates
{\it as a whole} along the $\bar{\zeta}$-direction, so the corresponding
energy-momentum tensor field $T_{\mu\nu}(x,y,z,\xi)$ of the model must satisfy
the corresponding {\it local isotropy (null) condition}, namely,
$T_{\mu\nu}T^{\mu\nu}=0$ (summation over the repeated indices is throughout
used).

The term "{\bf translational-rotational}" means that besides translational
component along $\bar{\zeta}$, the propagation necessarily demonstrates some
rotational (in the general sense of this concept) component in such a way that
{\it both components are compatible and exist simultaneously}, and this is an
{\it intrinsic} property. It seems reasonable to expect that such kind of
dynamical behavior should require some distinguished spatial shapes. Moreover,
if the Planck relation $E=h\nu$ must be respected throughout the evolution, the
rotational component of propagation should have {\it time-periodical} nature
with time period $T=\nu^{-1}=h/E=const$, and one of the two possible, {\it
left} or {\it right}, orientations. It seems reasonable also to expect spatial
periodicity of PhLO, which somehow to be related to the time periodicity.

The term "{\bf dynamical structure}" means that the propagation is supposed to
be necessarily accompanied by an {\it internal energy-momentum redistribution},
which may be considered in the model as energy-momentum exchange between (or
among) some appropriately defined subsystems.  It could also mean that PhLO
live in a dynamical harmony with the outside world, i.e.  {\it any outside
directed energy-momentum flow should be accompanied by a parallel inside
directed energy-momentum flow}.

Finally, note that if the time periodicity and the spatial periodicity should
be consistent with each other somehow, the simplest integral feature of such
compatability would seem like this: the spatial size $\lambda$ along the
translational component of propagation is equal to $cT$: $\lambda=cT$, where
$\lambda$ is some finite positive characteristic constant of the corresponding
solution. This would mean that every individual PhLO determines its own
length/time scale.

It is important to note now the following. We
don't know what mathematical objects are appropriate for describing PhLO, so,
our first task is to come to such mathematical objects having in view what we
mean under PhLO. The next Section is devoted namely to find mathematical
structures that are adequate enough to the above introduced notion for PhLO and
carring rich enough flexability to meet all requirements for a field theory of
spatially finite and time-stable physical objects with dynamical structure. Our
hope is that the ideas and concepts connected with the Frobenius integrability
theory seem to represent the most adequate part of mathematics for this
purpose.

\section{Curvature of Distributions and Physical Interaction}

\subsection{The general idea for geometrization of local physical interaction} We
begin with a short motivation for this choice of mathematics directed to the
readers already acquanted with Frobenius integrability theory, and right after
this we shall carefully introduce the necessary mathematics.

Any physical system with a dynamical structure is characterized by some internal
energy-momentum redistributions, i.e. internal energy-momentum fluxes, during
evolution. Any time-stable compatible system of energy-momentum fluxes (as well
as fluxes of other interesting for the case physical quantities subject to
change during evolution, but we limit ourselves just to energy-momentum fluxes
here) can be considered mathematically as generated by a compatible system of
vector fields. A {\it physically isolated} and {\it interelated time-stable}
system of energy-momentum fluxes can be considered to correspond directly or
indirectly to a completely integrable distribution $\Delta$ of vector fields
(or differential system [25]) according to the principle: {\it some local
objects can generate integral object}. Every nonintegrable distribution on a
manifold defines its own curvature form (given further in the section). Let
$\Delta_1$ and $\Delta_2$ be two nonintegrable distributions on the same
manifold with corresponding curvature forms $\Omega_1$ and $\Omega_2$, each of
them carries couples of vector fields inside their distributions outside
$\Delta_1$ and $\Delta_2$ correspondingly, i.e. $\Omega_1(Y_1,Y_2)\neq 0$ is
out of $\Delta_1$ and $\Omega_2(Z_1,Z_2)\neq 0$ is out of $\Delta_2$, where
$(Y_1,Y_2)$ live in $\Delta_1$ and $(Z_1,Z_2)$ live in $\Delta_2$. Let now
$\Delta_1$ and $\Delta_2$ characterize two locally interacting physical systems,
or two locally interacting subsystems of a larger physical system. It seems
reasonable to assume as a workong tool the following geometrization of the
concept of local physical interaction: {\it two nonintegrable distributions
$\Delta_1$ and $\Delta_2$ on a manifold will be said to interact
infinitesimally (or locally) if some of the nonzero values of the corresponding
two curvature forms $\Omega_1$/$\Omega_2$ live respectively in
$\Delta_2$/$\Delta_1$}.

The above geometric concept of {\it infinitesimal interaction} is motivated by
the fact that, in general, an integrable distribution $\Delta$ may contain
various {\it nonintegrable} subdistributions $\Delta_1, \Delta_2, \dots$ which
subdistributions may be associated physically with interacting subsytems of a
larger time stable physical system. Any physical interaction between 2
subsystems is necessarily accompanied with available energy-momentum exchange
between them, this could be understood mathematically as nonintegrability of
each of the two subdistributions of $\Delta$ and could be naturally measured
directly or indirectly by the corresponding curvatures. For example, if
$\Delta$ is an integrable 3-dimensional distribution represented by the vector
fields $(X_1,X_2,X_3)$ then we may have, in general, three non-integrable, i.e.
geometrically interacting, 2-dimensional subdistributions $(X_1,X_2),
(X_1,X_3), (X_2,X_3)$. Finally, some interaction with the outside world can be
described by curvatures of distributions (and their subdistributions) in which
elements of $\Delta$ and vector fields outside $\Delta$ are involved (such
processes will not be considered in this paper).

The above considerations launch the general idea to {\it consider the
concept of Frobenius curvature as a natural and universal mathematical
tool for describing local physical interaction between/among the relatively
stable subsystems of the physical world}. In other words, the {\bf Frobenius
curvature appears as appropriate mathematical tool describing formally the
possible ability two continuous systems to recognize each other as physically
interacting partners}.

Two formal aspects of the obove idea exist. The first applies directly
the {\bf Frobenius integrability machinary} [25], while the second one (been
developed recently) is known as {\bf nonlinear connections} [26].
We consider now briefly the first one.

\subsection{Frobenius integrability, curvature and local physical interaction}
A $p$-dimensional distribution $\Delta_{p}$ on a $n$-dimensional manifold $M^n$
is defined by associating to each point $x\in M^n$ a $p$-dimensional subspace
of the tangent space at this point: $\Delta^{p}_{x}\subset T_xM^n, x\in M^n,
1\leq p<n.$ Let the system of vector fields $\left\{X_1,X_2,\dots, X_p\right\}$
represent this distribution, so $\left\{X_1(x),X_2(x),\dots, X_p(x)\right\}$,
$x\in M^n$, $1\leq p<n$, satisfy $X_1(x)\wedge X_2(x)\wedge \dots ,\wedge\,
X_p(x)\neq 0, \,x\in M^n$, and represent a basis of $\Delta^{p}_{x}$.
According to the Frobenius integrability theorem
(further all manifolds are assumed smooth and finite dimensional and all
objects defined on $M^n$ are also assumed smooth) $\Delta_p$ is completely
integrable, i.e. through every point $x\in M^n$ passes a $p$-dimensional
submanifold $N^p$ such that all elements of $\Delta_p$ are tangent to $N^p$,
iff all Lie brackets $\left[X_i,X_j\right], \ i,j=1,2,\dots, p$, are
representable linearly through the very $X_i, i=1,2,\dots, p:
\left[X_i,X_j\right]=C^k_{ij}X_k$, where $C^k_{ij}$ are functions. Clearly, an
easy way to find out if a distribution is completely integrable is to check if
the exterior products \begin{equation} [X_i,X_j]\wedge X_1(x)\wedge
X_2(x)\wedge \dots ,\wedge\, X_p(x), \,x\in M^n;\ \ \ i,j=1,2,\dots,p
\end{equation}
are identically zero. If this is not the case (which means
that at least one such Lie bracket "sticks out" of the distribution $\Delta_p$)
then the corresponding coefficients, which are multilinear combinations of the
components of the vector fields and their derivatives, represent the
corresponding curvatures. We note finally that if two subdistributions contain
at least one common vector field it seems naturally to expect interaction.

In the dual formulation of Frobenius theorem in terms of differential 1-forms
(i.e. Pfaff forms), having the distribution $\Delta_p$ , we
look for $(n-p)$-Pfaff forms $(\alpha^1, \alpha^2, \dots, \alpha^{n-p}$), i.e. a
$(n-p)$-codistribution $\Delta^*_p$ , such that $ \langle\alpha^m,X_j\rangle=0,\
\ \text{and} \ \ \alpha^1(x)\wedge\alpha^2(x)\wedge\dots
\wedge\alpha^{n-p}(x)\neq 0, $ $ m=1,2,\dots,n-p, \ \ j=1,2,\dots,p , x\in M^n.
$ Then the integrability of the distribution $\Delta_p$ is equivalent to the
requirements \begin{equation}
\mathbf{d}\alpha^m\wedge\alpha^1\wedge\alpha^2\wedge\dots\wedge\alpha^{n-p} =0,\ \ \ m=1,2,\dots,
(n-p),
\end{equation}
where $\mathbf{d}$ is the exterior derivative.

Since the idea of curvature associated with, for example, an arbitrary
2-dimensional
distribution $(X,Y)$ is to find out if the Lie bracket $[X,Y](p)$ has components
along vectors outside the 2-plane defined by $(X_p,Y_p)$, in our case
 we have to evaluate
the quantities $\langle\alpha^m,[X,Y]\rangle$, where all linearly independent
1-forms $\alpha^m$ annihilate
$(X,Y):\langle\alpha^m,X\rangle=\langle\alpha^m,Y\rangle=0$. In view of the
formula
$$ \mathbf{d}\alpha^m(X,Y)=X(\langle\alpha^m,Y\rangle)-
Y(\langle\alpha^m,X\rangle) -\langle\alpha^m,[X,Y]\rangle=
-\langle\alpha^m,[X,Y]\rangle
$$
we may introduce explicitly the curvature 2-form for the distribution
$\Delta(X)=(X_1,\dots,X_p)$. In
fact, if $\Delta(Y)=(Y_1,\dots,Y_{n-p})$ define a distribution which is
complimentary (in the sense of direct sum) to $\Delta(X)$ and
$\langle\alpha^m,X_i\rangle=0$,
$\langle\alpha^m,Y_n\rangle=\delta^m_n$, i.e. $(Y_1,\dots,Y_{n-p})$ and
$(\alpha^1, \dots, \alpha^{n-p})$ are dual bases, then the
corresponding curvature 2-form $\Omega_{\Delta(X)}$ should be defined by
\begin{equation}
\Omega_{\Delta(X)}=-\mathbf{d}\alpha^m\otimes Y_m, \ \ \text{since} \ \
\Omega_{\Delta(X)}(X_i,X_j)=-\mathbf{d}\alpha^m(X_i,X_j) Y_m=
\langle\alpha^m,[X_i,X_j]\rangle Y_m ,
\end{equation}
where it is meant here that
$\Omega_{\Delta(X)}$ is restricted to the distribution $(X_1,\dots,X_p)$.
Hence, if we call the distribution $(X_1,\dots,X_p)$ {\it horizontal} and the
complimentary distribution $(Y_1,\dots,Y_{n-p})$ {\it vertical}, then the
corresponding curvature 2-form acquires the status of {\it vertical bundle
valued 2-form}. We see that the curvature 2-form distinguishes those couples of
vector fields inside $\Delta(X)$ the Lie brackets of which define outside
$\Delta(X)$ directed flows, and so, do not allowing to find integral manifold
of $\Delta(X)$. Clearly, the supposition here for dimensional complementarity
of the two distributions $\Delta(X)$ and $\Delta(Y)$ is not essential for the
idea of geometrical interaction, i.e. the distribution $\Delta(Y)\neq\Delta(X)$
may be any other distribution on the same manifold with dimension smaller than
$(n-p)$, so that $m=1,2,\dots,q<(n-p)$ in general, the important moment is that
the two distributions (or subdistributions) can "communicate" {\it
differentially} through their curvature 2-forms.

Hence, from physical point of view, if the quantities
$\Omega_{\Delta(X)}(X_i,X_j)$ are meant to be used for building the components
of the energy-momentum locally transferred from the system $\Delta(X)$ to the
system $\Delta(Y)$, then, naturally, we have to make use of the quantities
$\Omega_{\Delta(Y)}(Y_m,Y_n)$ to build the components of the energy-momentum
transferred from $\Delta(Y)$ to $\Delta(X)$.

It deserves to note that this formalism allows a dynamical
equilibrium between the two systems $\Delta(Y)$ and
$\Delta(X)$ to be described: each system to gain from the other as much
energy-momentum as it loses, and this to take place at every space-time point.
Therefore, if $W_{(X,Y)}$ denotes the energy-momentum transferred locally from
$\Delta(X)$ to $\Delta(Y)$, $W_{(Y,X)}$ denotes the energy-momentum transferred
locally from $\Delta(Y)$ to $\Delta(X)$, and $\delta W_{(X)}$ and $\delta
W_{(Y)}$ denote respectively the local energy-momentum changes of the two
systems $\Delta(X)$ and $\Delta(Y)$, then according to the local
energy-momentum conservation law we can write \[ \delta
W_{(X)}=W_{(Y,X)}+W_{(X,Y)}, \ \ \delta W_{(Y)}=-(W_{(X,Y)}+W_{(Y,X)})=-\delta
W_{(X)} , \] i.e. $\Delta(X)$ and $\Delta(Y)$ are {\it phisically compatible},
or {\it able to interact}. For the case of dynamical equilibrium we have
$W_{(X,Y)}=-W_{(Y,X)}$, so in such a case we obtain \begin{equation} \delta
W_{(X)}=0,\ \ \ \delta W_{(Y)}=0,\ \ \ W_{(Y,X)}+W_{(X,Y)}=0. \end{equation} As
for how to build explicitly the corresponding representatives of the
energy-momentum fluxes, probably, universal procedure can not be offered. The
most simple procedure seems to be to "project" the curvature values
$\Omega_{\Delta(X)}(X_i,X_j)$ and $\Omega_{\Delta(Y)}(Y_m,Y_n)$ on the
corresponding co-distribution volume forms, i.e. to consider the corresponding
inner products $i(\Omega(X_i,X_j))(\alpha^1\wedge\alpha^2\wedge\dots
\wedge\alpha^{n-p})$. For every special case, however, appropriate quantities
constructed out of the members of the introduced distributions and
co-distributions must be worked out.

\section{PhLO Dynamical Structure in Terms of Frobenius \newline Curvature}
We consider the Minkowski space-time $M=(\mathbb{R}^4,\eta)$ with
signature $sign(\eta)=(-,-,-,+)$ related to the standard global
coordinates $(x^1,x^2,x^3,x^4)=(x,y,z,\xi=ct)$,the natural volume
form $\omega_o=\sqrt{|\eta|}dx^1\wedge dx^2\wedge dx^3\wedge dx^4=
dx\wedge dy\wedge dz\wedge d\xi$, and the
Hodge star $*$ defined by $\alpha\wedge *\beta=-\eta(\alpha,\beta)\omega_o$.

In view of our concept of PhLO we introduce the null vector field
$\bar{\zeta},\ \bar{\zeta}^2=0$, which in the $\bar{\zeta}$-adapted coordinates
(throughout used further) is assumed to look as follows: \begin{equation}
\bar{\zeta}=-\varepsilon\frac{\partial}{\partial z} +          
\frac{\partial}{\partial \xi}, \ \ \varepsilon=\pm 1.
\end{equation}
Let's denote the corresponding to $\bar{\zeta}$ completely integrable
3-dimensional Pfaff system by $\Delta^*(\bar{\zeta})$. Thus,
$\Delta^*(\bar{\zeta})$ can be generated by any three linearly independent
1-forms $(\alpha_1,\alpha_2,\alpha_3)$ which annihilate $\bar{\zeta}$, i.e. $$
\alpha_1(\bar{\zeta})=\alpha_2(\bar{\zeta})=\alpha_3(\bar{\zeta})=0; \ \
\alpha_1\wedge \alpha_2\wedge \alpha_3\neq 0.
$$
Instead of $(\alpha_1,\alpha_2,\alpha_3)$ we introduce the notation
$(A, A^*, \zeta)$ and define $\zeta$ to be the $\eta$-corresponding 1-form to
$\bar{\zeta}$:
\begin{equation}
\zeta=\varepsilon dz+d\xi, \ \ \text{so},
\ \ \langle\zeta,\bar{\zeta}\rangle=0 ,                 
\end{equation}
where $\langle , \rangle$ is the coupling between forms and vectors.

Now, since $\zeta$ is closed, it defines 1-dimensional completely
integrable Pfaff system, so, we have the corresponding completely integrable
distribution $(\bar{A},\bar{A^{*}},\bar{\zeta}):
\langle\zeta,\bar{A}\rangle=\langle\zeta,\bar{A^{*}}\rangle=0$.
We shall restrict our further study to
PhLO of electromagnetic nature according to the following
\vskip 0.2cm
{\bf{Definition}}: We shall call a PhLO {\it electromagnetic} if the following
conditions hold:

1. the vector fields $(\bar{A},\bar{A^{*}})$ have no components along
$\bar{\zeta}$,

2. $(\bar{A},\bar{A^{*}})$ are $\eta$-corresponding to $(A, A^*)$
respectively .

3. $\langle A,\bar{A^{*}}\rangle=0,\ \
\langle A,\bar{A}\rangle =\langle A^{*},\bar{A^{*}}\rangle$ .
\vskip 0.2cm
\noindent
{\bf Remark}. These relations formalize knowledge from Classical
electrodynamics (CED). In fact, our vector fields $(\bar{A},\bar{A^{*}})$ are
meant to represent what we call in CED {\it electric and magnetic
components} of a free time-dependent electromagnetic field, where, as we have
mentioned several times, the translational propagation of
the field energy-momentum along a fixed null direction with the velocity "$c$"
is possible only if the two invariants $I_1=\mathbf{B}^2-\mathbf{E}^2$ and
$I_2=2\mathbf{E}.\mathbf{B}$ are zero, because only in such a case the
electromagnetic energy-momentum tensor $T_{\mu\nu}$ satisfies
$T_{\mu\nu}T^{\mu\nu}=0$ and has {\it unique} null eigen direction. So it seems
naturally to consider this property as {\it intrinsic} for the field and to
choose it as a starting point. Moreover, in such a case the relation
$(I_1)^2+(I_2)^2=0$ is equivalent to
$\mathbf{E}^2+\mathbf{B}^2=2|\mathbf{E}\times\mathbf{B}|$ and this relation
shows that this is the only case when the field momentum can
not be made equal to zero by means of frame change. Together with the fact
that the spatial direction of translational energy-momentum propagation is
determined by $\mathbf{E}\times\mathbf{B}$, this motivates to introduce the
vector field $\bar{\zeta}$ in this form and to assume the properties 1-3 in the
above definition. \vskip 0.3cm
 From the above conditions it follows that in the
$\bar{\zeta}$-adapted coordinate system we have
\[
A=u\,dx + p\,dy, \ \
A^*=-\varepsilon\,p\,dx + \varepsilon\,u\,dy; \ \
\bar{A}=-u\,\frac{\partial}{\partial x} -          
p\,\frac{\partial}{\partial y}, \ \
\bar{A^*}=\varepsilon\,p\,\frac{\partial}{\partial x} -            
\varepsilon\,u\,\frac{\partial}{\partial y},
\]
where $\varepsilon=\pm 1$, and $(u,p)$ are two smooth functions on $M$.

The completely integrable 3-dimensional Pfaff system $(A, A^*, \zeta)$
contains three 2-dimensional subsystems: $(A,A^{*}), (A,\zeta)$ and
$(A^*,\zeta)$. We have the following \vskip 0.3cm {\bf Proposition 1}. The
following relations hold:
\[
\mathbf{d}A\wedge A\wedge A^*=0;\ \
\mathbf{d}A^*\wedge A^*\wedge A=0;\ \                              
\]
\[ \mathbf{d}A\wedge A\wedge \zeta= \varepsilon\big[u(p_\xi-\varepsilon
p_z)- p(u_\xi-\varepsilon u_z)\big]\omega_o; \] \[ \mathbf{d}A^*\wedge
A^*\wedge \zeta=                                  
\varepsilon\big[u(p_\xi-\varepsilon p_z)- p(u_\xi-\varepsilon
u_z)\big]\omega_o.
\]
{\bf Proof.} Immediately verified.
\vskip 0.3cm
\noindent
These relations say that the 2-dimensional Pfaff system
$(A,A^*)$ is completely integrable for any choice of the two functions
$(u,p)$, while the two 2-dimensional Pfaff systems $(A,\zeta)$ and
$(A^*,\zeta)$ are NOT completely integrable in general, and the same
curvature factor $$ \mathbf{R}=u(p_\xi-\varepsilon
p_z)-p(u_\xi-\varepsilon u_z) $$ determines their nonintegrability.

Correspondingly, the 3-dimensional completely
integrable distribution (or differential system) $\Delta(\bar{\zeta})$
contains three 2-dimensional subsystems:
$(\bar{A},\bar{A^*})$, $(\bar{A},\bar{\zeta})$ and $(\bar{A^*},\bar{\zeta})$.
We have the
\vskip 0.3cm
{\bf Proposition 2}. The following relations hold (recall that $[X,Y]$ denotes the Lie
bracket):
\begin{equation}
[\bar{A},\bar{A^*}]\wedge\bar{A}\wedge\bar{A^*}=0 ,
\end{equation}
\begin{equation}
[\bar{A},\bar{\zeta}]=(u_\xi-\varepsilon u_z)\frac{\partial}{\partial x}+
(p_\xi-\varepsilon p_z)\frac{\partial}{\partial y} ,            
\end{equation}
\begin{equation}
[\bar{A^*},\bar{\zeta}]=
-\varepsilon(p_\xi-\varepsilon p_z)\frac{\partial}{\partial x}+
\varepsilon(u_\xi-\varepsilon u_z)\frac{\partial}{\partial y}.            
\end{equation}

{\bf Proof.} Immediately verified.
\vskip 0.3cm

From these last relations (3.7-3.9)
it follows that the distribution $(\bar{A},\bar{A^*})$
is completely integrable, and it can be easily shown that the
two distributions $(\bar{A},\bar{\zeta})$
and
$(\bar{A^*},\bar{\zeta})$ would be completely integrable only if the same
curvature factor
\begin{equation}
\mathbf{R}=u(p_\xi-\varepsilon p_z)-p(u_\xi-\varepsilon u_z)     
\end{equation}
is zero (the elementary proof is omitted).

As it should be, the two projections
$$
\langle A,[\bar{A^*},\bar{\zeta}]\rangle=
-\langle A^*,[\bar{A},\bar{\zeta}]\rangle=
\varepsilon u(p_\xi-\varepsilon p_z)-\varepsilon p(u_\xi-\varepsilon u_z)=
-\varepsilon\,\mathbf{R}
$$
are nonzero and give (up to a sign) the same factor $\mathbf{R}$. The same curvature factor
appears, of course, as coefficient in the exterior products
$
[\bar{A^*},\bar{\zeta}]\wedge \bar{A^*}\wedge\bar{\zeta}$ and
$[\bar{A},\bar{\zeta}]\wedge \bar{A}\wedge\bar{\zeta}$.
In fact, we obtain
\[
[\bar{A^*},\bar{\zeta}]\wedge \bar{A^*}\wedge\bar{\zeta}=
-[\bar{A},\bar{\zeta}]\wedge \bar{A}\wedge\bar{\zeta}=
-\varepsilon\mathbf{R}\,\frac{\partial}{\partial x}
\wedge\frac{\partial}{\partial y}\wedge
\frac{\partial}{\partial z}+
\mathbf{R}\,\frac{\partial}{\partial x}\wedge\frac{\partial}{\partial y}\wedge
\frac{\partial}{\partial \xi} .
\]
On the other hand, for the other two projections we obtain
\begin{equation}
\langle
A,[\bar{A},\bar{\zeta}]\rangle=                          
\langle
A^*,[\bar{A^*},\bar{\zeta}]\rangle= \frac12\big[(u^2+p^2)_\xi-\varepsilon
(u^2+p^2)_z\big].
\end{equation}
Clearly, the last relation (3.11) may be put in
terms of the Lie derivative $L_{\bar{\zeta}}$ as
$$
\frac 12L_{\bar{\zeta}}(u^2+p^2)=
-\frac12L_{\bar{\zeta}}\langle A,\bar{A}\rangle=
-\langle A,L_{\bar{\zeta}}\bar{A}\rangle=
-\langle A^*,L_{\bar{\zeta}}\bar{A^*}\rangle.
$$

{\bf Remark}. Further in the paper we shall denote $\sqrt{u^2+p^2}\equiv
\phi$.
 \vskip 0.3cm {\bf Proposition 3.} There is a function $\psi(u,p)$ such, that

$$L_{\bar{\zeta}}\psi=\frac
{u(p_\xi-\varepsilon p_z)-p(u_\xi-\varepsilon u_z)}{\phi^2}=
\frac{\mathbf{R}}{\phi^2} .
$$
\vskip 0.4cm
{\bf Proof}. It is immediately verified that $\psi=\arctan\frac pu$
is such one.
\vskip 0.4cm
We note that the function $\psi$ has a natural
interpretation of {\it phase} because of the easily verified now relations
$u=\phi\cos\psi$, $p=\phi \sin\psi $,
and $\phi$ acquires the status of {\it amplitude}, i.e. energy density. Since
the transformation $(u,p)\rightarrow (\phi,\psi)$ is non-degenerate this allows
to work with the two functions $(\phi,\psi)$ instead of $(u,p)$.

From {\bf Prop.3} we have
\begin{equation}
\mathbf{R}=\phi^2L_{\bar{\zeta}}\psi= \
\phi^2(\psi_\xi-\varepsilon\psi_z) \ \ \
\rightarrow \ \ L_{\bar{\zeta}}\psi=
\frac{\mathbf{R}}{T(\partial_{\xi},\partial_{\xi})}=
\frac{*(\mathbf{d}A\wedge A\wedge A^*)}{T(\partial_{\xi},\partial_{\xi})},
\end{equation}
where $T(\partial_{\xi},\partial_{\xi})$ is the coordinate-free definition of
the energy density.

 This last formula (3.12) shows something very important: at
any $\phi\neq 0$ the curvature $\mathbf{R}$ will NOT be zero only if
$L_{\bar{\zeta}}\psi\neq 0$, which admits in principle availability of
rotation. In fact, lack of rotation would mean that $\phi$ and $\psi$ are
running waves along $\bar{\zeta}$. The relation $L_{\bar{\zeta}}\psi\neq 0$
means, however, that rotational properties are possible in general, and some of
these properties are carried by the phase $\psi$. It follows that in such a
case the translational component of propagation along $\bar{\zeta}$ (which is
supposed to be available) must be determined essentially, and most probably
entirely, by $\phi$.  In particular, we could expect the relation
$L_{\bar{\zeta}}\phi=0$ to hold, and if this happens, then the rotational
component of propagation will be represented entirely by the phase $\psi$, and,
more specially, by the curvature factor $\mathbf{R}\neq 0$, so, the objects we
are going to describe may have compatible translational-rotational dynamical
structure. Finally, (3.12) may be considered as a definition for the phase
function $\psi$.

We are going now to represent some relations, analogical to the energy-momentum
relations in classical electrodynamics, determined by some 2-form $F$, in
terms of the Frobenius curvatures given above.

The two nonintegrable Pfaff systems $(A,\zeta)$ and
$(A^*,\zeta)$ carry two volume 2-forms:
$$
G=A\wedge\zeta \ \
\text{and} \ \ G^*=A^*\wedge\zeta ,
$$
and the two corresponding distributions define the 2-vectors
$$
\bar{G}=\bar{A}\wedge\bar{\zeta}, \ \ \text{and} \ \ \
\bar{G^*}=\bar{A^*}\wedge\bar{\zeta} .
$$
Making use now of the Hodge $*$-operator,
 we can verify the relation: $G^*=*G$.
Now $G$ and $\bar{G}^*$ define the (1,1)-tensor, called
stress-energy-momentum tensor $T_{\mu}^{\nu}$, according to the rule
$$
T_{\mu}^{\nu}=-\frac12\big[G_{\mu\sigma}\bar{G}^{\nu\sigma}+
(G^*)_{\mu\sigma}(\bar{G}^*)^{\nu\sigma}\big] ,
$$
and the divergence of this tensor field can be represented in the form
$$
\nabla_\nu T_{\mu}^{\nu}=\big[i(\bar{G})\mathbf{d}G\big]_{\mu}+
\big[i(\bar{G^*})\mathbf{d}G^*\big]_\mu,
$$
where $\bar{G}$ and $\bar{G^*}$ coincide with the metric-corresponding
contravarint tensor fields, and $i(\bar{G})=i(\bar{\zeta})\circ i(\bar{A})$,
$i(\bar{G^*})=i(\bar{\zeta})\circ i(\bar{A^*})$,
$i(X)$ is the standard insertion operator in the exterior
algebra of differential forms on $\mathbb{R}^4$ defined by the vector field
$X$.
So, we shall need the quantities
$$
i(\bar{G})\mathbf{d}G, \ \ i(\bar{G^*})\mathbf{d}G^*,
\ \
i(\bar{G^*})\mathbf{d}G, \ \ i(\bar{G})\mathbf{d}G^* .
$$
Having in view the
explicit expressions for $A,A^*,\zeta,\bar{A},\bar{A^*}$ and $\bar{\zeta}$ we
obtain
\begin{equation}
i(\bar{G})\mathbf{d}G=i(\bar{G}^*)\mathbf{d}G^*=
\frac12 L_{\bar{\zeta}}\left(\phi^2\right).
\,\zeta \ ,
\end{equation}
also, we obtain
\[
i(\bar{G^*})\mathbf{d}G=-i(\bar{G})\mathbf{d}G^* =
\]
\begin{equation}
=\Big[u(p_\xi-\varepsilon p_z)-p(u_\xi-\varepsilon u_z)\Big]dz+
\varepsilon\Big[u(p_\xi-\varepsilon p_z)-p(u_\xi-\varepsilon u_z)\Big]d\xi
=\varepsilon\mathbf{R}\,\zeta .
\end{equation}
If $F$ and $H$ are correspondingly 2 and 3 forms on $M$ we have the relation
$$
*(F\wedge *H)=i(\bar{F})H=F^{\mu\nu}H_{\mu\nu\sigma}dx^\sigma, \mu<\nu.
$$
Therefore, since $G^*=*G$,
\begin{equation}
i(\bar{G^*})\mathbf{d}G=-*(\delta *G\wedge *G), \ \
-i(\bar{G})\mathbf{d}G^*=-*(\delta G\wedge G), \ \ \text{so}, \ \
\delta *G\wedge *G=\delta G\wedge G=\varepsilon\mathbf{R}*\zeta .
\end{equation}

In the following formulae we must keep in mind the relations
$\mathbf{d}\zeta=0, \langle A,\bar{A^*}\rangle=\langle A^*,\bar{A}\rangle=
\langle\zeta,\bar{A^*}\rangle=\langle\zeta,\bar{A}\rangle=0$, and $(\bar{A})^2=
(\bar{A}^*)^2=\langle A,\bar{A}\rangle=\langle A^*,\bar{A^*}\rangle=
-(u^2+p^2)=-\Phi^2=-|A|^2=-|A^*|^2=-|\bar{A}|^2=-|\bar{A^*}|^2$.

In view of these formulae and the required duality in the definition of the
curvature form (3.3), the two distributions $(\bar{A},\bar{\zeta})$ and
$(\bar{A^*},\bar{\zeta})$ determine the following two curvature forms $\Omega$
and $\Omega^*$:
$$
\Omega=-\mathbf{d}\frac{-A^*}{|A^*|}\otimes\frac{\bar{A^*}}{|\bar{A^*}|}=
\mathbf{d}\frac{A^*}{|A^*|}\otimes\frac{\bar{A^*}}{|\bar{A^*}|}
,\ \ \
\Omega^*=-\mathbf{d}\frac{-A}{|A|}\otimes\frac{\bar{A}}{|\bar{A}|}=
\mathbf{d}\frac{A}{|A|}\otimes\frac{\bar{A}}{|\bar{A}|}.
$$
Denoting $Z_{\Omega}\equiv\Omega(\bar{A},\bar{\zeta})$,
$Z^*_{\Omega}\equiv\Omega(\bar{A^*},\bar{\zeta})$,
$Z_{\Omega^*}\equiv\Omega^*(\bar{A},\bar{\zeta})$ and
$Z^*_{\Omega^*}\equiv\Omega^*(\bar{A^*},\bar{\zeta})$
we obtain
\begin{equation}
Z_{\Omega}=-\frac{\varepsilon\mathbf{R}}{\phi^2}\bar{A^*},\ \
Z^*_{\Omega}=-\frac{\bar{A^*}}{2\phi^2}L_{\bar{\zeta}}(\phi^2), \ \
Z_{\Omega^*}=-\frac{\bar{A}}{2\phi^2}L_{\bar{\zeta}}(\phi^2), \ \
Z^*_{\Omega^*}=\frac{\varepsilon\mathbf{R}}{\phi^2}\bar{A}.
\end{equation}
The following relations express the connection between the curvatures and the
energy-momentum characteristics.
\begin{align}
i(Z_{\Omega})(A\wedge\zeta)=0, \ \
i(Z_{\Omega})(A^*\wedge\zeta)=\varepsilon\mathbf{R}.\zeta=
-i(\bar{G})\mathbf{d}G^*=i(\bar{G^*})\mathbf{d}G, \\
i(Z_{\Omega^*})(A^*\wedge\zeta)=0, \ \
i(Z^*_{\Omega^*})(A\wedge\zeta)=-\varepsilon\mathbf{R}.\zeta=
i(\bar{G})\mathbf{d}G^*=-i(\bar{G^*})\mathbf{d}G, \\
i(Z^*_{\Omega})(A\wedge\zeta)=0, \ \
i(Z^*_{\Omega})(A^*\wedge\zeta)=\frac12L_{\bar{\zeta}}(\phi^2).\zeta=
i(\bar{G})\mathbf{d}G=i(\bar{G^*})\mathbf{d}G^{*}, \\
i(Z^*_{\Omega^*})(A^*\wedge\zeta)=0, \ \
i(Z_{\Omega^*})(A\wedge\zeta)=\frac12L_{\bar{\zeta}}(\phi^2).\zeta=
i(\bar{G})\mathbf{d}G=i(\bar{G^*})\mathbf{d}G^{*}.
\end{align}

It follows from these relations that in case of dynamical equilibrium we shall
have
\[
L_{\bar{\zeta}}(\phi^2)=0,\ \
i(\bar{G})\mathbf{d}G=0,\ \ i(\bar{G^*})\mathbf{d}G^*=0, \ \
i(\bar{G^*})\mathbf{d}G+i(\bar{G})\mathbf{d}G^*=0.
\]

Resuming, we can say that Frobenius integrability viewpoint suggests to
make use of one completely integrable 3-dimensional distribution (resp.
Pfaff system) consisting of one isotropic and two space-like vector fields
(resp. 1-forms), such that the corresponding 2-dimensional spatial
subdistribution $(\bar{A},\bar{A^*})$ (resp. Pfaff system $(A,A^*)$)
defines a completely integrable system, and the rest two 2-dimensional
subdistributions $(\bar{A},\bar{\zeta})$ and $(\bar{A^*},\bar{\zeta})$
(resp. Pfaff systems $(A,\zeta )$ and $(A^*,\zeta )$) are NON-integrable
in general and give the same curvature. This curvature may be used to
build quantities, physically interpreted as energy-momentum
internal exchanges between the corresponding
two subsystems $(\bar{A},\bar{\zeta})$ and $(\bar{A^*},\bar{\zeta})$
(resp.$(A,\zeta)$ and $(A^*,\zeta))$. Moreover, rotational component of
propagation will be available only if the curvature $\mathbf{R}$ is nonzero,
i.e. only if an internal energy-momentum exchange takes place. We see that
all physically important characteristics and relations, describing the
translational and rotational components of propagation, can be expressed in
terms of the corresponding Frobenius curvature. We'll see that this
holds also for some integral characteristics of PhLO.

\section{PhLO Dynamical Structure in Terms of Non-linear \newline Connections}
\subsection{Projections and algebraic curvatures}
The projections are linear maps $P$ in a linear space $W^n$ (under linear space
we mean here {\it module over a ring}, or {\it vector space over a field})
sending all
elements of $W^n$ to some subspace $P(W^n)\subset W^n$, such that $P\circ P=P$.
Let $(e_1,\dots,e_p,\dots,e_n)$ and
$(\varepsilon^1,\dots,\varepsilon^p,\dots,\varepsilon^n)$ be two dual bases:
$<\varepsilon^\mu,e_\nu>=\delta^\mu_\nu, \ \mu,\nu=1,\dots,n$, and let $N_i^a, \
i=1,\dots,p\,; \ a=p+1,\dots,n$ be the corresponding to $P$ $[p\times(n-p)]$
matrix of rank $(n-p)$. We define another couple of dual bases:
$$
k_\mu=(e_i+N_i^ae_a,\,e_a)\,; \ \
\omega^\nu=(\varepsilon^i,\,\varepsilon^b-N^b_j\varepsilon^j),\ \
\ j=1,\dots,p\,; \ a,b=p+1,\dots,n.
$$
Now the identity map $id_{W^n}=\omega^\nu\otimes k_\nu$ acquires the form
\begin{equation}
id_{W^n}=\omega^\nu\otimes k_\nu=
\omega^i\otimes k_i+\omega^b\otimes k_b=
\varepsilon^i\otimes(e_i+N_i^ae_a)+
(\varepsilon^b-N^b_j\varepsilon^j)\otimes e_b.
\end{equation}
We obtain two projections: $P_V=(\varepsilon^b-N^b_j\varepsilon^j)\otimes e_b$
and $P_H=\varepsilon^i\otimes(e_i+N_i^ae_a)$ such that $KerP_V=ImP_H$ and
$KerP_H=ImP_V$, also, $P_H=id_{W^n}-P_V$. Hence, $W^n=KerP_V\oplus ImP_V=
KerP_H\oplus ImP_H$. Usually $P_V$ is called vertical projection, and $P_H$ is
called horizontal projection.

Let now $\phi$ and $\psi$ be two arbitrary linear maps in a module
$\mathfrak{M}$, $\mathfrak{B}:\mathfrak{M}\times
\mathfrak{M}\rightarrow\mathfrak{M}$ be a binar map satisfying
$\mathfrak{B}(\mathbf{x+z},\mathbf{y})=\mathfrak{B}(\mathbf{x},\mathbf{y})+
\mathfrak{B}(\mathbf{z},\mathbf{y})$ and
$\mathfrak{B}(\mathbf{x},\mathbf{y+z})=\mathfrak{B}(\mathbf{x},\mathbf{y})+
\mathfrak{B}(\mathbf{x},\mathbf{z})$, and
$(\mathbf{x},\mathbf{y},\mathbf{z})$ be three arbitrary elements of
$\mathfrak{M}$. We consider the expression
\[
\mathcal{A}(\mathfrak{B};\phi,\psi)(\mathbf{x},\mathbf{y})\equiv
\frac12\Big[\mathfrak{B}(\phi(\mathbf{x}),\psi(\mathbf{y}))+
\mathfrak{B}(\psi(\mathbf{x}),\phi(\mathbf{y}))+
\phi\circ\psi(\mathfrak{B}(\mathbf{x},\mathbf{y}))+
\psi\circ\phi(\mathfrak{B}(\mathbf{x},\mathbf{y}))
\]
\[
-\phi(\mathfrak{B}(\mathbf{x},\psi(\mathbf{y})))-
\phi(\mathfrak{B}(\psi(\mathbf{x}),\mathbf{y}))-
\psi(\mathfrak{B}(\mathbf{x},\phi(\mathbf{y})))-
\psi(\mathfrak{B}(\phi(\mathbf{x}),\mathbf{y}))\Big]\ \ .
\]
Assuming $\phi=\psi $ are projections
in $\mathfrak{M}$
denoted by $P$, this expression becomes
\[\mathcal{A}(\mathfrak{B};P)(\mathbf{x},\mathbf{y})\equiv
P(\mathfrak{B}(\mathbf{x},\mathbf{y}))+\mathfrak{B}(P(\mathbf{x}),P(\mathbf{y}))-
P(\mathfrak{B}(\mathbf{x},P(\mathbf{y})))-P(\mathfrak{B}(P(\mathbf{x}),\mathbf{y})) \ .
\]
Denoting the identity map of $\mathfrak{M}$ by $id$ and adding and subtracting
$P\Big[\mathfrak{B}\big(P(\mathbf{x}),P(\mathbf{y})\big)\Big]$, after some
elementary transformations we obtain
\[
\mathcal{A}(\mathfrak{B};P)(\mathbf{x},\mathbf{y})\equiv
P\Big[\mathfrak{B}\big[(id-P)(\mathbf{x}),(id-P)(\mathbf{y})\big]\Big]+
(id-P)\Big[\mathfrak{B}\big[P(\mathbf{x}),P(\mathbf{y}\big]\Big] .
\]
Recalling that $P$ and $(id-P)$ project on two subspaces of $\mathfrak{M}$,
the direct sum of which generates $\mathfrak{M}$,
and naming $P$ as {\it vertical} projection denoted by $V$, then
$(id-P)$, denoted by $H$, gets naturally the name {\it horizontal}
projection. So the above expression gets the final form of
\begin{equation}
\mathcal{A}(\mathfrak{B};P)(\mathbf{x},\mathbf{y})\equiv
V\Big[\mathfrak{B}\big[H(\mathbf{x}),H(\mathbf{y})\big]\Big]
+H\Big[\mathfrak{B}\big[V(\mathbf{x}),V(\mathbf{y})\big]\Big]=
\mathcal{R}_{P}(\mathfrak{B};\mathbf{x},\mathbf{y})+
\bar{\mathcal{R}}_{P}(\mathfrak{B};\mathbf{x},\mathbf{y}).
\end{equation}
Hence, the first term on the right,
$\mathcal{R}_{P}(\mathfrak{B};\mathbf{x},\mathbf{y})$, which may be called {\it
$\mathfrak{B}$-algebraic curvature of $P$}, measures the vertical component of
the $\mathfrak{B}$-image of the horizontal projections of
$(\mathbf{x},\mathbf{y})$, and then the second term
$\bar{\mathcal{R}}_{P}(\mathfrak{B};\mathbf{x},\mathbf{y})$, acquiring the name
of  $\mathfrak{B}$-{\it algebraic cocurvature of $P$}, measures the horizontal
component of the $\mathfrak{B}$-image of the vertical projections of
$(\mathbf{x},\mathbf{y})$.

We carry now this pure algebraic construction to the tangent bundle of a smooth
manifold $M^n$, where the above binar map $\mathfrak{B}$ will be interpreted as
the Lie bracket of vector fields, and the linear maps will be just linear
endomorphisms of the tangent/cotangent bundles of $M^n$. Under these
assumptions the quantity $\mathcal{A}(\Phi,\Psi)$ is called Nijenhuis bracket
of the two linear endomorphisms $\Phi$ and $\Psi$, and is usually denoted by
$[\Phi,\Psi]$. It has two important for us properties: the first one is that
$[\Phi,\Psi]$ is linear with respect to the smooth functions on the manifold,
so, the Nijenhuis bracket allows, starting with two $(1,1)$-tensors on $M^n$,
to construct through differentiations a 2-form that is valued in the tangent
bundle of $M^n$; the second property is that if $\Phi=\Psi$ then $[\Phi,\Phi]$
is not necessarily zero.

\vskip 0.3cm \noindent
 \subsection{Nonlinear connections}

\noindent
Let now $(x^1,\dots,x^n)$ be any local coordinate system on our real manifold
$M^n$. We have the corresponding local frames $\{dx^1,\dots,dx^n\}$ and
$\{\partial_{x^1},\dots,\partial_{x^n}\}$. Let for each $x\in M$ we are given a
projection $P_x$ of the same constant rank $(n-p)$, i.e. $p$ does not depend on
$x$, in every tangent space $T_x(M)$. The space $Ker(P_x)\subset T_x(M)$ is
usually called $P$-{\it horizontal}, and the space $Im(P_x)\subset T_x(M)$ then
is called $P$-{\it vertical}. Thus, we have two distributions on $M$  the
direct sum of which gives the tangent bundle: $T(M)=Ker(P)\oplus Im(P)$. The
above algebraic construction shows that each of these two distributions can be
endowed with corresponding 2-form, valued in the other distribution, and
depending on the same binar operation in $TM^n$. As we mentioned, the choice
$\mathfrak{B}$= {\it Lie bracket} leads to tensor field. Therefore, assuming
this choice, we say that $P$ defines a nonlinear connection on $M$. Denoting by
$\mathcal{R}$ the so defined curvature 2-form of $P$ and by
$\tilde{\mathcal{R}}$ the corresponding cocurvature 2-form of $P$, by $V_{P}$
 and $H_{P}$ the corresponding vertical and horizontal prjections, we can write
\begin{equation}
[P,P](X,Y)=\mathcal{R}(X,Y)+\bar{\mathcal{R}}(X,Y),
\end{equation}
where
$$
\mathcal{R}(X,Y)=V_{P}\big([H_{P}X,H_{P}Y]\big), \ \ \
\tilde{\mathcal{R}}(X,Y)=H_{P}\big([V_{P}X,V_{P}Y]\big),
$$
$(X,Y)$ are any two
vector fields and the Lie bracket is denoted by $[ , ]$. Recalling the contents
of the preceding section, it can be shown that $\mathcal{R}(X,Y)\neq 0$
measures the nonintegrability of the corresponding horizontal distribution, and
$\mathcal{\tilde{R}}(X,Y)\neq 0$ measures the nonintegrability of the
corresponding vertical distribution.

If the vertical distribution is given before-hand and is completely integrable,
i.e. $\mathcal{\tilde{R}}=0$,
then $\mathcal{R}(X,Y)$ is called {\it curvature} of the
nonlinear connection $P$ if there exist at least one couple of horizontal vector
fields $(X,Y)$ such that $\mathcal{R}(X,Y)\neq 0$. \vskip 0.5 cm
\subsection{Photon-like nonlinear connections}
We assume now that our manifold is $\mathbb{R}^4$ endowed with standard
coordinates $(x^1,x^2,x^3,x^4=x,y,z,\xi=ct)$, and make some preliminary
considerations in order to make the choice of our projection
$P:T\mathbb{R}^4\rightarrow T\mathbb{R}^4$ consistent with the introduced
concept of PhLO. The intrinsically defined straight-line translational
component of propagation of the PhLO will be assumed to be parallel to the
coordinate plane $(z,\xi)$. Also, $\frac{\partial}{\partial x}$ and
$\frac{\partial}{\partial y}$ will be vertical coordinate fields, so every
vertical vector field $Y$ can be represented by  $Y=u\,\frac{\partial}{\partial
x}+p\,\frac{\partial}{\partial y}$, where $(u,p)$ are two functions on $\mathbb{R}^4$.
It is easy to check that any two such linearly independent vertical vector
fields $Y_1$ and $Y_2$ define an integrable distribution, hence, the
corresponding curvature will be zero. It seems very natural to choose $Y_1$ and
$Y_2$ to coincide correspondingly with the vertical projections
$P(\frac{\partial}{\partial z})$ and $P(\frac{\partial}{\partial \xi})$.
Moreover, let's restrict ourselves to PhLO of electromagnetic nature and denote
further the verical projection by $V$. Then, since this vertical structure is
meant to be smoothly straight-line translated along the plane $(z,\xi)$ with
the velocity of light, a natural suggestion comes to mind these two projections
$Y_1=V(\frac{\partial}{\partial z})$ and $Y_2=V(\frac{\partial}{\partial \xi})$
to be physically interpreted as representatives of the electric and magnetic
components. Now we know from classical electrodynamics that the situation
described corresponds to zero invariants of the electromagnetic field,
therefore, we may assume that $Y_1$ and $Y_2$ are ortogonal to each other and
with the same modules with respect to the euclidean metric in the 2-dimensional
space spent by $\frac{\partial}{\partial x}$ and $\frac{\partial}{\partial y}$.
It follows that the essential components of $Y_1$ and $Y_2$ should be
expressible only with two independent functions $(u,p)$. The conclusion is that
our projection should depend only on $(u,p)$. Finally, we note that these
assumptions lead to the horizontal nature of $dz$ and $d\xi$.

Note that if the translational component of propagation is along the vector
field $\bar{\zeta}$ then we can define two new distributions :
$(Y_1,\bar{\zeta})$ and $(Y_2,\bar{\zeta})$, which do not seem to be integrable
in general even if $\bar{\zeta}$ has constant components as it will be in our
case. Since these two distributions are nontrivially intersected (they have a
common member $\bar{\zeta}$), it is natural to consider them as geometrical
images of two interacting physical subsystems of our PhLO. Hence,
we must introduce two projections with the same image space but with different
kernal spaces, and the components of both projections must depend only on the
two functions $(u,p)$.

Let now $(u,p)$ be two smooth functions on $\mathbb{R}^4$ and
$\varepsilon=\pm 1$ . We introduce two projections $V$ and $\tilde{V}$ in
$T\mathbb{R}^4$ as follows:
\begin{equation}
V=dx\otimes\frac{\partial}{\partial
x}+dy\otimes\frac{\partial}{\partial y}
-\varepsilon\,u\,dz\otimes\frac{\partial}{\partial x}-
u\,d\xi\otimes\frac{\partial}{\partial x}-
\varepsilon\,p\,dz\otimes\frac{\partial}{\partial y}-
p\,d\xi\otimes\frac{\partial}{\partial y},
\end{equation}
\begin{equation}
\tilde{V}=
dx\otimes\frac{\partial}{\partial
x}+dy\otimes\frac{\partial}{\partial y} +p\,dz\otimes\frac{\partial}{\partial
x} +\varepsilon p\,d\xi\otimes\frac{\partial}{\partial x} -
u\,dz\otimes\frac{\partial}{\partial y}- \varepsilon
u\,d\xi\otimes\frac{\partial}{\partial y}.
\end{equation}
So, in both cases we consider $(\frac{\partial}{\partial
x},\frac{\partial}{\partial y})$ as vertical vector fields, and $(dz,d\xi)$ as
horizontal 1-forms. By corresponding transpositions we can determine
projections $V^*$ and $\tilde{V}^*$ in the cotangent bundle $T^*\mathbb{R}^4$.
\[ V^*=dx\otimes\frac{\partial}{\partial x}+ dy\otimes\frac{\partial}{\partial
y} -\varepsilon\,u\,dx\otimes\frac{\partial}{\partial z}-
u\,dx\otimes\frac{\partial}{\partial \xi}-
\varepsilon\,p\,dy\otimes\frac{\partial}{\partial z}-
p\,dy\otimes\frac{\partial}{\partial \xi}, \] \[ \tilde{V}^*=
dx\otimes\frac{\partial}{\partial x}+ dy\otimes\frac{\partial}{\partial y}+
p\,dx\otimes\frac{\partial}{\partial z} + \varepsilon
p\,dx\otimes\frac{\partial}{\partial \xi} -
u\,dy\otimes\frac{\partial}{\partial z}- \varepsilon
u\,dy\otimes\frac{\partial}{\partial \xi}. \] The corresponding horizontal
projections, denoted by $(H,\tilde{H};H^*\tilde{H}^*)$ look as follows: \[
H=dz\otimes\frac{\partial}{\partial z}+d\xi\otimes\frac{\partial}{\partial \xi}
+\varepsilon\,u\,dz\otimes\frac{\partial}{\partial x}+
u\,d\xi\otimes\frac{\partial}{\partial x}+
\varepsilon\,p\,dz\otimes\frac{\partial}{\partial y}+
p\,d\xi\otimes\frac{\partial}{\partial y}, \] \[
\tilde{H}=dz\otimes\frac{\partial}{\partial
z}+d\xi\otimes\frac{\partial}{\partial \xi}-
p\,dz\otimes\frac{\partial}{\partial x} - \varepsilon
p\,d\xi\otimes\frac{\partial}{\partial x}+ u\,dz\otimes\frac{\partial}{\partial
y}+ \varepsilon u\,d\xi\otimes\frac{\partial}{\partial y}, \] \[
H^*=dz\otimes\frac{\partial}{\partial z}+ d\xi\otimes\frac{\partial}{\partial
\xi} +\varepsilon\,u\,dx\otimes\frac{\partial}{\partial z}+
u\,dx\otimes\frac{\partial}{\partial \xi}+ \varepsilon
p\,dy\otimes\frac{\partial}{\partial z}+ p\,dy\otimes\frac{\partial}{\partial
\xi}, \] \[ \tilde{H}^*=dz\otimes\frac{\partial}{\partial z}+
d\xi\otimes\frac{\partial}{\partial \xi}- p\,dx\otimes\frac{\partial}{\partial
z} - \varepsilon p\,dx\otimes\frac{\partial}{\partial \xi}+
u\,dy\otimes\frac{\partial}{\partial z}+ \varepsilon
u\,dy\otimes\frac{\partial}{\partial \xi}. \]

The corresponding matrices look like:
\[ V=
\begin{Vmatrix}1 & 0 & -\varepsilon\,u & -u \\
0 & 1 & -\varepsilon\,p  & -p \\
0 & 0 & 0 & 0 \\
0 & 0 & 0 & 0 \end{Vmatrix} ,
\ \ H=
\begin{Vmatrix}0 & 0 & \varepsilon\,u & u \\
0 & 0 & \varepsilon\,p & p \\
0 & 0 & 1 & 0 \\
0 & 0 & 0 & 1 \end{Vmatrix} ,
\]

\[ V^*=
\begin{Vmatrix}1 & 0 & 0 & 0\\
0 & 1 & 0 & 0 \\
-\varepsilon\,u & -\varepsilon\,p & 0 & 0 \\
-u & -p & 0 & 0
\end{Vmatrix} ,
\ \ H^*=
\begin{Vmatrix}0 & 0 & 0 & 0\\
0 & 0 & 0 & 0 \\
\varepsilon\,u & \varepsilon\,p & 1 & 0 \\
u & p & 0 & 1 \end{Vmatrix} ,
\]

\vskip 0.4cm
\[
\tilde{V}= \begin{Vmatrix}1 & 0 & p & \varepsilon\,p \\
0 & 1 & -u & -\varepsilon\,u \\
0 & 0 & 0 & 0 \\
0 & 0 & 0 & 0 \end{Vmatrix} ,
\ \
\tilde{H}= \begin{Vmatrix}
0 & 0 & -p & -\varepsilon\,p \\
0 & 0 & u & \varepsilon\,u \\
0 & 0 & 1 & 0 \\
0 & 0 & 0 & 1 \end{Vmatrix} ,
\]
\vskip 0.4cm
\[
\tilde{V}^*= \begin{Vmatrix}1 & 0 & 0 & 0 \\
0 & 1 & 0 & 0\\
p & -u & 0 & 0 \\
\varepsilon\,p & -\varepsilon\,u & 0 & 0 \end{Vmatrix} ,
\ \
\tilde{H}^*= \begin{Vmatrix}
0 & 0 & 0 & 0 \\
0 & 0 & 0 & 0 \\
-p & u & 1 & 0 \\
-\varepsilon\,p & \varepsilon\,u & 0 & 1 \end{Vmatrix} .
\]
The
projections of the coordinate bases are:
\[
\left(\frac{\partial}{\partial x},
\frac{\partial}{\partial y}, \frac{\partial}{\partial z},
\frac{\partial}{\partial \xi}\right).V=
\left(\frac{\partial}{\partial x}, \frac{\partial}{\partial y},
-\varepsilon u\frac{\partial}{\partial x}
-\varepsilon p\frac{\partial}{\partial y},
-u\frac{\partial}{\partial x}
-p\frac{\partial}{\partial y}\right);
\]
\vskip 0.3cm
\[
\left(\frac{\partial}{\partial x},\frac{\partial}{\partial y},
\frac{\partial}{\partial z},\frac{\partial}{\partial \xi}\right).H=
\left(0,0,\varepsilon u\frac{\partial}{\partial x}
+\varepsilon p\frac{\partial}{\partial y}+\frac{\partial}{\partial z},
u\frac{\partial}{\partial x}
+p\frac{\partial}{\partial y}+
\frac{\partial}{\partial \xi}\right);
\]
\[
\left(dx,dy,dz,d\xi\right).V^*=
\left(dx-\varepsilon udz-ud\xi, dy-\varepsilon pdz-pd\xi,0,0\right)
\]
\[
\left(dx,dy,dz,d\xi\right).H^*=
\left(\varepsilon udz+ud\xi,\varepsilon pdz+pd\xi,dz,d\xi\right)
\]
\[
\left(\frac{\partial}{\partial x},
\frac{\partial}{\partial y}, \frac{\partial}{\partial z},
\frac{\partial}{\partial \xi}\right).\tilde{V}=
\left(\frac{\partial}{\partial x}, \frac{\partial}{\partial y},
p\frac{\partial}{\partial x}
-u\frac{\partial}{\partial y},
\varepsilon\,p\frac{\partial}{\partial x}
-\varepsilon\,u\frac{\partial}{\partial y}\right);
\]
\[
\left(\frac{\partial}{\partial x},\frac{\partial}{\partial y},
\frac{\partial}{\partial z},\frac{\partial}{\partial \xi}\right).\tilde{H}=
\left(0,0,-p\frac{\partial}{\partial x}
+u\frac{\partial}{\partial y}+\frac{\partial}{\partial z},
-\varepsilon\,p\frac{\partial}{\partial x}
+\varepsilon\,u\frac{\partial}{\partial y}+
\frac{\partial}{\partial \xi}\right);
\]
\[
\left(dx,dy,dz,d\xi\right).\tilde{V}^*=
\left(dx+p\,dz+\varepsilon\,pd\xi, dy-u\,dz-\varepsilon\,ud\xi,0,0\right)
\]
\[
\left(dz,d\xi,dx,dy\right).\tilde{H}^*=
\left(-p\,dz-\varepsilon\,p\,d\xi,u\,dz+\varepsilon\,u\,d\xi,dz,d\xi\right) .
\]
We compute now the two curvature 2-forms $\mathcal{R}$ and
$\tilde{\mathcal{R}}$. The components $\mathcal{R}^\sigma_{\mu\nu}$ of
$\mathcal{R}$ in coordinate basis are given by
$V^\sigma_\rho\Big(\big[H\frac{\partial}{\partial
x^{\mu}},H\frac{\partial}{\partial x^{\nu}}\big]^{\rho}\Big)$, and the only
nonzero components are just
$$
\mathcal{R}^{x}_{z\xi}=\mathcal{R}^{1}_{34}=
-\varepsilon(u_{\xi}-\varepsilon\,u_z),\ \ \
\mathcal{R}^{y}_{z\xi}=\mathcal{R}^{2}_{34}=
-\varepsilon(p_{\xi}-\varepsilon\,p_z)  .
$$
For the nonzero components of $\tilde{\mathcal{R}}$ we obtain
$$
\tilde{\mathcal{R}}^{x}_{z\xi}=\tilde{\mathcal{R}}^{1}_{34}=
(p_{\xi}-\varepsilon\,p_z),\ \ \
\tilde{\mathcal{R}}^{y}_{z\xi}=\tilde{\mathcal{R}}^{2}_{34}=
-(u_{\xi}-\varepsilon\,u_z)  .
$$

The corresponding two curvature forms are:
\begin{equation}
\mathcal{R}=-\varepsilon(u_\xi-\varepsilon u_z)dz\wedge d\xi\otimes
\frac{\partial}{\partial x}-
\varepsilon(p_\xi-\varepsilon p_z)dz\wedge d\xi\otimes
\frac{\partial}{\partial y}
\end{equation}
\begin{equation}
\mathcal{\tilde{R}}=(p_\xi-\varepsilon p_z)dz\wedge d\xi\otimes
\frac{\partial}{\partial x}-
(u_\xi-\varepsilon u_z)dz\wedge d\xi\otimes
\frac{\partial}{\partial y} .
\end{equation}
We obtain (in our coordinate system): $-\frac12tr\left(V\circ H^*\right)=
-\frac12tr\left(\tilde{V}\circ\tilde{H}^*\right)=u^2+p^2$, and
$$
V\left(\left[H\left(\frac{\partial}{\partial z}\right),
H\left(\frac{\partial}{\partial \xi}\right)\right]\right)=
\left[H\left(\frac{\partial}{\partial z}\right),
H\left(\frac{\partial}{\partial \xi}\right)\right]=
-\varepsilon(u_{\xi}-\varepsilon u_z)\frac{\partial}{\partial x}
-\varepsilon(p_{\xi}-\varepsilon p_z)\frac{\partial}{\partial y}\equiv Z_1,
$$
$$
\tilde{V}\left(\left[\tilde{H}\left(\frac{\partial}{\partial z}\right),
\tilde{H}\left(\frac{\partial}{\partial \xi}\right)\right]\right)=
\left[\tilde{H}\left(\frac{\partial}{\partial z}\right),
\tilde{H}\left(\frac{\partial}{\partial \xi}\right)\right]=
(p_{\xi}-\varepsilon p_z)\frac{\partial}{\partial x}
-(u_{\xi}-\varepsilon u_z)\frac{\partial}{\partial y}\equiv Z_2,
$$
where $Z_1$ and $Z_2$ coincide with the values of the two curvature forms
$\mathcal{R}$ and $\tilde{\mathcal{R}}$ on the coordinate vector fields
$\frac{\partial}{\partial z}$ and $\frac{\partial}{\partial \xi}$ respectively:
$$
Z_1=\mathcal{R}\left(\frac{\partial}{\partial z},
\frac{\partial}{\partial \xi}\right),\ \ \
Z_2=
\tilde{\mathcal{R}}\left(\frac{\partial}{\partial z},
\frac{\partial}{\partial \xi}\right) .
$$
We evaluate now the vertical 2-form
$V^*(dx)\wedge V^*(dy)$ on the bivector $Z_1\wedge Z_2$ and obtain
$\varepsilon\,\mathcal{K}^2$, where
$$
\mathcal{K}^2=(u_{\xi}-\varepsilon u_z)^2+(p_{\xi}-\varepsilon p_z)^2.
$$
An important parameter, having dimension
of length (the coordinates are assumed to have dimension of length)
and denoted by $l_o$, turns out to be the square root of the
quantity
$$ \frac{-\frac12tr\left(V\circ H^*\right)}{\mathcal{K}^2}=
\frac{u^2+p^2}{(u_{\xi}-\varepsilon
u_z)^2+(p_{\xi}-\varepsilon p_z)^2}.
$$
Clearly, if $l_o$ is finite constant it could be interpreted as some parameter
of extension of the PhLO described, so it could be used as identification
parameter in the dynamical equations and in lagrangians, but only if
$(u_{\xi}-\varepsilon u_z)\neq 0$ and $(p_{\xi}-\varepsilon p_z)\neq 0$. This
goes along with our concept of PhLO which does not admit spatially infinite
extensions.
Finally we'd like to note that the right-hand side of the above relation does
not depend on which projection $V$ or $\tilde{V}$ is used, i.e.
$[\tilde{V}^*(dx)\wedge\tilde{V}^*(dy)](Z_1\wedge Z_2)=
\varepsilon\,\mathcal{K}^2$ too, so
\begin{equation}
l_o^2=\frac{-\frac12tr\left(\tilde{V}\circ\tilde{H}^*\right)}{\mathcal{K}^2}=
\frac{-\frac12tr\Big(V\circ H^*\Big)}{\mathcal{K}^2}=
\frac{u^2+p^2}
{(u_{\xi}-\varepsilon u_z)^2+(p_{\xi}-\varepsilon p_z)^2}.
\end{equation}

The parameter $l_o$ has the following symmetry. Denote by
$V_o=dx\otimes\frac{\partial}{\partial x}+
dy\otimes\frac{\partial}{\partial y}$, then $V=V_o+V_1$ and
$\tilde {V}=V_o+\tilde{V}_1$, where, in our coordinates, $V_1$ and $\tilde
{V}_1$ can be seen above how they look like. We form now $W=aV_1-b\tilde{V}_1$
and $\tilde{W}=bV_1+a\tilde{V}_1$, where $(a,b)$ are two arbitrary real
numbers. The components of the corresponding linear maps $P_W=V_o+W$ and
$P_{\tilde{W}}=V_o+\tilde{W}$ can be obtained through the substitutions:
$u\rightarrow(au+\varepsilon bp); \ p\rightarrow (\varepsilon bp-ap)$,
and, obviously, $P_W$ and $P_{\tilde{W}}$ are projections. Now,
$-\frac12tr(V\circ H^*)$ transforms to $(a^2+b^2)(u^2+p^2)$ and $\mathcal{K}^2$
transforms to $(a^2+b^2)[(u_{\xi}-\varepsilon u_z)^2+(p_{\xi}-\varepsilon
p_z)^2]$, so, $l_o(V,\tilde{V})=l_o(W,\tilde{W})$.
This corresponds in some sense to the dual
symmetry of classical vacuum electrodynamics. We note also that the squared
modules of the two curvature forms $|\mathcal{R}|^2$ and
$|\mathcal{\tilde{R}}|^2$ are equal to $(u_{\xi}-\varepsilon
u_z)^2+(p_{\xi}-\varepsilon p_z)^2$ in our coordinates, therefore, the nonzero
values of $|\mathcal{R}|^2$ and $|\mathcal{\tilde{R}}|^2$, as well as the
finite value of $l_o$ guarantee that the two functions $u$ and $p$ are NOT
plane waves. Finally, the phase function may be defined by the relations
$$
L_{\bar{\zeta}}\psi=\frac{<A,Z_2>}{-\frac12tr\left(\tilde{V}\circ\tilde{H}^*\right)}
=-\frac{<A^*,Z_1>}{-\frac12tr\left(V\circ H^*\right)}.
$$

 \vskip 0.5cm
\subsection{Electromagnetic PhLO in terms of non-linear connections}

Recall that the relativistic formulation of classical
electrodynamics in vacuum ($\rho=0$) is based on the following assumptions. The
configuration space is the Minkowski space-time $M=(\mathbb{R}^4,\eta)$ where
$\eta$ is the pseudometric with $sign(\eta)=(-,-,-,+)$ with the corresponding
volume 4-form $\omega_o=dx\wedge dy\wedge dz\wedge d\xi $ and
Hodge star $*$ defined by $\alpha\wedge *\beta=-\eta(\alpha,\beta)\omega_o$. The
electromagnetic filed is describe by two closed 2-forms $(F,*F):\mathbf{d}F=0,
\ \mathbf{d}*F=0$. The physical characteristics of the field are deduced from
the following stress-energy-momentum tensor field \begin{equation}
T_{\mu}{^\nu}(F,*F)=-\frac12\big[F_{\mu\sigma}F^{\nu\sigma}+
(*F)_{\mu\sigma}(*F)^{\nu\sigma}\big].
\end{equation}
In the non-vacuum case the allowed energy-momentum exchange with other physical
systems is given in general by the divergence
\begin{equation}
\nabla_\nu\,T_{\mu}^{\nu}
=\frac12 \Big[F^{\alpha\beta}(\mathbf{d}F)_{\alpha\beta\mu}
+(*F)^{\alpha\beta}(\mathbf{d}*F)_{\alpha\beta\mu}\Big] =
F_{\mu\nu}(\delta F)^\nu + (*F)_{\mu\nu}(\delta *F)^\nu,
\end{equation}
where $\delta=*\mathbf{d}*$ is the coderivative.
If the field is free: $\mathbf{d}F=0, \mathbf{d}*F=0$, this divergence is
obviously equal to zero on the vacuum solutions since its both terms are zero.
Therefore, energy-momentum exchange between the two partner-fields $F$ and
$*F$, which should be expressed by the terms
$(*F)^{\alpha\beta}(\mathbf{d}F)_{\alpha\beta\mu}$ and
$F^{\alpha\beta}(\mathbf{d}*F)_{\alpha\beta\mu}$ is NOT allowed on the
solutions of $\mathbf{d}F=0, \mathbf{d}*F=0$. This shows that the
widely used 4-potential approach (even if two 4-potentials $A,A^*$ are
introduced so that $\mathbf{d}A=F, \ \mathbf{d}A^*=*F$ locally)
to these equations excludes any possibility to individualize two
energy-momentum exchanging time-stable subsystems of the field that are
mathematically represented by $F$ and $*F$.

On the contrary, as we have mentioned several times, our concept of PhLO does
NOT exclude such two physically interacting subsystems of the field to really
exist, and therefore, to be mathematically individualized. The intrinsically
connected two projections $V$ and $\tilde{V}$ and the corresponding two
curvature forms give the mathematical realization of this idea: $V$ and
$\tilde{V}$ individualize the two subsystems, and the corresponding two
curvature 2-forms $\mathcal{R}$ and $\mathcal{\tilde{R}}$ represent the
instruments by means of which the available mutual local energy-momentum
exchange between these two subsystems could be described. We should not forget
that, as we have already emphasized several times, the energy-momentum tensor
for a PhLO must satisfy the additional local isotropy (null) condition
$T_{\mu\nu}(F,*F)T^{\mu\nu}(F,*F)=0$.

So, we have to construct appropriate quantities and relations
having direct physical sense in terms of the introduced and considered two
projections $V$ and $\tilde{V}$. The above well established in electrodynamics
relations say that we need two 2-forms to begin with.

Recall that our coordinate 1-forms $dx$ nd $dy$ have the following vertical and
horizontal projections:
\[
 V^*(dx)=dx-\varepsilon u\,dz-u\,d\xi, \ \
H^*(dx)=\varepsilon u\,dz+u\,d\xi\ , \]
\[
 V^*(dy)=dy-\varepsilon
p\,dz-p\,d\xi, \ \ H^*(dy)=\varepsilon p\,dz+p\,d\xi .
\]
We form now the 2-forms
$V^*(dx)\wedge H^*(dx)$ and $V^*(dy)\wedge H^*(dy)$:
$$
V^*(dx)\wedge
H^*(dx)=\varepsilon\,u\,dx\wedge dz+u\,dx\wedge d\xi,
$$
$$ V^*(dy)\wedge
H^*(dy)=\varepsilon\,p\,dy\wedge dz+p\,dx\wedge d\xi .
$$
Summing up these last
two relations and denoting the sum by $F$ we obtain
\begin{equation}
F=\varepsilon\,u\,dx\wedge dz+u\,dx\wedge d\xi+
\varepsilon\,p\,dy\wedge dz+p\,dy\wedge d\xi .
\end{equation}
Doing the same steps with $\tilde{V}^*$ and $\tilde{H}^*$ we obtain
\begin{equation}
\tilde{F}=-p\,dx\wedge dz-\varepsilon\,p\,dx\wedge d\xi+
u\,dy\wedge dz+\varepsilon u\,dy\wedge d\xi .
\end{equation}
Noting that our definition of the Hodge star requires
$(*F)_{\mu\nu}=-\frac12\,\varepsilon_{\mu\nu}\,^{\sigma\rho}F_{\sigma\rho}$,
it is now easy to verify that $\tilde{F}=*F$. Moreover, introducing the
notations
$$
A=u\,dx+p\,dy, \ \ A^*=-\varepsilon\,p\,dx+\varepsilon\,u\,dy ,\ \
\zeta=\varepsilon\,dz+d\xi ,
$$
we can represent $F$ and $\tilde{F}$ in the form
$$
F=A\wedge \zeta , \ \ \tilde{F}=*F=A^*\wedge \zeta .
$$
From these last relations
we see that $F$ and $*F$ are isotropic: $F\wedge F=0, F\wedge *F=0$, i.e. the
field $(F,*F)$ has zero invariants:
$F_{\mu\nu}F^{\mu\nu}=F_{\mu\nu}(*F)^{\mu\nu}=0$. The following relations are
now easy to verify:
\begin{equation}
V^*(F)=H^*(F)=V^*(*F)=H^*(*F)=\tilde{V}^*(F)=\tilde{H}^*(F)=
\tilde{V}^*(*F)=\tilde{H}^*(*F)=0,
\end{equation}
i.e. $F$ and $*F$ have zero vertical and horizontal projections with respect
to $V$ and $\tilde{V}$. Since, obviously, $\zeta $ is horizontal with respect
to $V$ and $\tilde{V}$ it is interesting to note that $A$ is vertical with
respect to $\tilde{V}$ and $A^*$ is vertical with respect to $V$:
$\tilde{V}^*(A)=A$, $V(A^*)=A^*$. In fact, for example, $$
\tilde{V}^*(A)=\tilde{V}^*(u\,dx+p\,dy)=u\tilde{V}^*(dx)+p\tilde{V}^*(dy)=
$$
$$
u[dx+p\,dz+\varepsilon p\,d\xi]+p[dy-u\,dz-\varepsilon u\,d\xi]=
u\,dx+p\,dy.
$$

We are going to establish now that there is real energy-momentum exchange
between the $F$-component and the $*F$-component of the field. To come to this
we compute the quantities $i(Z_1)F, i(Z_2)*F,\ \ i(Z_1)*F,\ \
i(Z_2)F$. We obtain: \[ i(Z_1)F=i(Z_2)*F=\langle A,Z_1\rangle\zeta=\langle
A^*,Z_2\rangle\zeta=
\frac12\big[(u^2+p^2)_{\xi}-\varepsilon\,(u^2+p^2)_{z}\big]\zeta=
\]
\begin{equation}
=\frac12F^{\sigma\rho}(\mathbf{d}F)_{\sigma\rho\mu}dx^{\mu}=
\frac12(*F)^{\sigma\rho}(\mathbf{d}*F)_{\sigma\rho\mu}dx^{\mu}
=\frac12\nabla_\nu\,T_{\mu}^{\nu}(F,*F) ,
\end{equation}
\[
i(Z_1)*F=-i(Z_2)F=\langle A^*,Z_1\rangle\zeta =-\langle A,Z_2\rangle\zeta=
\big[u(p_{\xi}-\varepsilon\,p_{z})-p(u_{\xi}-\varepsilon\,u_{z})\big]\zeta=
\]
\begin{equation}
=-\frac12F^{\sigma\rho}(\mathbf{d}*F)_{\sigma\rho\mu}dx^{\mu}=
\frac12(*F)^{\sigma\rho}(\mathbf{d}F)_{\sigma\rho\mu}dx^{\mu} .
\end{equation}
If our field is free then $\nabla_\nu\,T_{\mu}^{\nu}(F,\tilde{F})=0$.
Moreover, in view of the divergence of the stress-energy-momentum tensor given
above, these last relations show that some real energy-momentum exchange between
$F$ and $*F$ takes place: the magnitude of the energy-momentum, transferred from
$F$ to $*F$ and given by
$i(Z_1)*F=\frac12(*F)^{\sigma\rho}(\mathbf{d}F)_{\sigma\rho\mu}dx^{\mu}$,
is equal to that, transferred from $*F$ to $F$, which is
given by
$-i(Z_2)F=-\frac12F^{\sigma\rho}(\mathbf{d}*F)_{\sigma\rho\mu}dx^{\mu}$.
On the other hand, as it is well known, the $*$-invariance
of the stress-energy-momentum tensor in case of zero invariants leads to
$F_{\mu\sigma}F^{\nu\sigma}=(*F)_{\mu\sigma}(*F)^{\nu\sigma}$, so, $F$ and $*F$
carry equal and conserved quantities of stress-energy-momentum.

We interpret physically this as follows.
The electromagnetic PhLO exist through a special internal dynamical
equilibrium between the two subsystems of the field, represented by $V$ and
$\tilde{V}$, namely, both subsystems carry the same
stress-energy-momentum and the mutual energy-momentum exchange between them is
always in equal quantities. This individualization does NOT mean that any of
the two subsystems can exist separately, independently on the other.
Moreover, NO spatial "part" of PhLO should be considered to represent a
real physical object.

\section{Electromagnetic PhLO in terms of electromagnetic strain}
The concept of {\it strain}
is introduced in studying elastic materials subject to external forces of
different nature: mechanical, electromagnetic, etc. In nonrelativistic
continuum physics the local representatives of the external forces in this
context are usually characterized in terms {\it stresses}. Since the force means
energy-momentum transfer leading to corresponding mutual energy-momentum
change of the interacting objects, then according to the energy-momentum
conservation law the material must react somehow to the external
interference in accordance with its structure and reaction abilities. The
classical strain describes mainly the abilities of the material to bear
force-action from outside through deformation, i.e. through changing its
shape, or, configuration. The term {\it elastic} now means that any two
allowed configurations can be deformed to each other without appearence of
holes and breakings, in particular, if the material considered has
deformed from configuration $C_1$ to configuration $C_2$ it is able to
return smoothly to its configuration $C_1$.

The general geometrical description [27] starts with the assumption
that an elastic material is a continuum $\mathbb{B}\subset\mathbb{R}^3$ which
can {\it smoothly deform} inside the space $\mathbb{R}^3$, so, it can be
endowed with differentiable structure, i.e. having an elastic material is
formally equivalent to have a smooth real 3-dimensional submanifold
$\mathbb{B}\subset\mathbb{R}^3$. The deformations are considered as smooth maps
(mostly embeddings) $\varphi: \mathbb{B}\rightarrow\mathbb{R}^3$. The spaces
$\mathbb{B}$ and $\mathbb{R}^3$ are endowed with riemannian metrics
$\mathbf{G}$ and $g$ respectively (and corresponding riemannian co-metrics
$\mathbf{G}^{-1}$ and $g^{-1}$), and induced isomorphisms
$\tilde{\mathbf{G}}$ and $\tilde{g}$
between the corresponding tangent and cotangent spaces .
This allows to define linear map inside every tangent space of
$\mathbb{B}$ in the following way: a tangent vector $V\in T_x\mathbb{B},\,
x\in\mathbb{B},$ is sent through the differential $d\varphi$ of $\varphi$ to
$(d\varphi)_x(V)\in T_{\varphi(x)}\mathbb{R}^3$, then by means of the
isomorphism $\tilde{g}$ we determine the corresponding 1-form (i.e. we "lower
the index"), this 1-form is sent to the dual space $T^*_x\mathbb{B}$ of
$T_x\mathbb{B}$ by means of the dual linear map $(d\varphi)^*:
T^*_{\varphi(x)}\mathbb{\mathbb{R}}^3\rightarrow T^*_x\mathbb{B}$, and finally,
we determine the corresponding tangent vector by means of the isomorphism
$\tilde{\mathbf{G}}^{-1}$ (i.e. we "raise the index" correspondingly). The so
obtained linear map
$$
\mathbf{C}_x :=
\big[\tilde{\mathbf{G}}^{-1}\circ
(d\varphi)^*\circ\tilde{g}\circ(d\varphi)\big]_x:
T_x\mathbb{B}\rightarrow T_x\mathbb{B}
$$
(which is denoted in [27] by $(\mathbf{F^TF})_x$), extended to the
whole $\mathbb{B}$, is called {\it Caushy-Green deformation tensor field}. Now,
the combination
$$
\mathbf{E}_x:=\frac12\big[(\tilde{\mathbf{G}}\circ\mathbf{C}-\mathbf{G})\big]_x=
\frac12\big[(d\varphi)^*\circ\tilde{g}\circ(d\varphi)-\mathbf{G}\big]_x:
T_x\mathbb{B}\times T_x\mathbb{B}\rightarrow \mathbb{R}
$$
is called {\it Lagrangian strain tensor field}.
 Note that if we denote by
$\varphi^*g$ the induced on $\mathbb{B}$ metric from the metric $g$ (usually
euclidean) on $\mathbb{R}^3$ then $\mathbf{E}=\frac12(\varphi^*g-\mathbf{G})$.

We could look at the problem also as
follows. The mathematical counterparts of the allowed (including
reversible) deformations are the diffeomorphisms $\varphi$ of a riemannian
manifold $(M,g)$, and every $\varphi(M)$ represents a possible configuration of
the material considered. But some diffeomorphisms do not lead to
deformation (i.e. to shape changes), so, a criterion must be introduced to
separate those diffeomorphisms which should be considered as essential. For
such a criterion is chosen the distance change: {\it if the distance between
any two fixed points does not change during the action of the external force
field, then we say that there is no deformation}. Now, every essential
diffeomorphism $\varphi$ must transform the metric $g$ to some new metric
$\varphi^*g$, such that $g\neq\varphi^*g$. The naturally arising tensor field
$e=(\varphi^*g-g)\neq 0$ appears as a measure of the physical abilities of the
material to withstand external force actions.

Since the external force is assumed to act locally and the material
considered gets the corresponding to the external force field final
configuration in a smooth way, i.e. passing smoothly through a family of
allowed configurations, we may introduce a localization of the above scheme, such
that the isometry doffeomorphisms to be eliminated. This is done by means
of introducing 1-parameter group $\varphi_t, t\in [a,b]\subset\mathbb{R}$
of local diffeomorphisms, so, $\varphi_a(M)$ and $\varphi_b(M)$ denote
correspondingly the initial and final configurations. Now $\varphi_t$
generates a family of metrics $\varphi_t^*\,g$, and a corresponding family
of tensors $e_t$. According to the local analysis every local 1-parameter
group of diffeomorphisms is generated by a vector field on $M$. Let the
vector field $X$ generate $\varphi_t$. Then the quantity
$$
\frac12\,L_{X}g:=\frac12\,\lim_{t\rightarrow 0}\frac{\varphi_t^*\,g-g}{t} \ ,
$$
i.e. one half of the {\it Lie derivative} of $g$ along $X$, is called
(infinitesimal) {\it strain tensor}, or {\it deformation tensor}. \vskip
0.2cm {\bf Remark}. Further in the paper we shall work with $L_{X}\,g$,
i.e. the factor $1/2$ will be omitted. \vskip 0.2cm

In our further study we shall call $L_X\,g$, where $g=\eta$ is the
Minkowski (pseudo)metric, just {\it strain tensor}.
Clearly, the term "material" is not appropriate for
PhLO because no static situations are admissible,
{\bf our objects of interest are of entirely dynamical nature},
so the corresponding {\it relativistic strain} tensors must take care of this.

According to the preliminary considerations important vector fields in our
approach to describe electromagnetic PhLO are
$\bar{\zeta},\,\bar{A},\,\bar{A^*}$, so, we consider the corresponding three
electromagnetic strain tensors: $ L_{\bar{\zeta}}\,\eta; \,L_{\bar{A}}\,\eta;
\,L_{\bar{A^*}}\,\eta$. \vskip 0.3cm {\bf Proposition 4}. The following
relations hold: \[ L_{\bar{\zeta}}\,\eta=0, \ \ \
(L_{\bar{A}}\,\eta)_{\mu\nu}\equiv D_{\mu\nu}= \begin{Vmatrix}2u_x & u_y+p_x &
u_z & u_{\xi} \\ u_y+p_x & 2p_y & p_z & p_{\xi} \\ u_z & p_z & 0 & 0 \\ u_{\xi}
& p_{\xi} & 0 & 0 \ \ \ \end{Vmatrix} , \] \[
(L_{\bar{A^*}}\,\eta)_{\mu\nu}\equiv D^*_{\mu\nu} =
\begin{Vmatrix}-2\varepsilon p_x & -\varepsilon(p_y+u_x) & -\varepsilon p_z &
-\varepsilon p_{\xi} \\ -\varepsilon(p_y+u_x) & 2\varepsilon u_y & \varepsilon
u_z & \varepsilon u_{\xi} \\ -\varepsilon p_z & \varepsilon u_z & 0 & 0 \\
-\varepsilon p_{\xi} & \varepsilon u_{\xi} & 0 & 0 \end{Vmatrix} . \]

{\bf Proof}. Immediately verified.
\vskip 0.3cm
We give now some important from our viewpoint relations.
\[
D(\bar{\zeta},\bar{\zeta})=D^*(\bar{\zeta},\bar{\zeta})=0,
\]
\[
D(\bar{\zeta})\equiv D(\bar{\zeta})_\mu dx^\mu\equiv D_{\mu\nu}\bar{\zeta}^\nu
dx^\mu =(u_\xi-\varepsilon u_z)dx + (p_\xi-\varepsilon p_z)dy,
\]
\[
D(\bar{\zeta})^\mu\frac{\partial}{\partial x^\mu}\equiv
D^{\mu}_{\nu}\bar{\zeta}^\nu\frac{\partial}{\partial x^\mu}=
-(u_\xi-\varepsilon u_z)\frac{\partial}{\partial x} -
(p_\xi-\varepsilon p_z)\frac{\partial}{\partial y}=-[\bar{A},\bar{\zeta}],\ \
\]
\[
D_{\mu\nu}\bar{A}^\mu\bar{\zeta}^\nu=
-\frac12\Big[(u^2+p^2)_\xi -\varepsilon(u^2+p^2)_z\Big]=
-\frac12L_{\bar{\zeta}}\phi^2 ,
\]
\[
D_{\mu\nu}\bar{A^*}^\mu\bar{\zeta}^\nu=
-\varepsilon\Big[u(p_\xi-\varepsilon p_z)-p(u_\xi-\varepsilon u_z)\Big]=
-\varepsilon\mathbf{R}=-\varepsilon \phi^2\,L_{\bar{\zeta}}\psi.
\]
We also have:
\[
D^*(\bar{\zeta})=\varepsilon\Big[-(p_\xi-\varepsilon p_z)dx+
(u_\xi-\varepsilon u_z)dy\Big] ,
\]
\[
D^*(\bar{\zeta})^\mu\frac{\partial}{\partial x^\mu}\equiv
(D^*)^{\mu}_{\nu}\bar{\zeta}^\nu\frac{\partial}{\partial x^\mu}=
-\varepsilon(p_\xi-\varepsilon p_z)\frac{\partial}{\partial x} +
(u_\xi-\varepsilon u_z)\frac{\partial}{\partial y}=[\bar{A^*},\bar{\zeta}],\ \
\]
\[
D^*_{\mu\nu}\bar{A^*}^\mu\bar{\zeta}^\nu=
-\frac12\Big[(u^2+p^2)_\xi -\varepsilon(u^2+p^2)_z\Big]=
-\frac12L_{\bar{\zeta}}\phi^2 ,
\]
\[
D^*_{\mu\nu}\bar{A}^\mu\bar{\zeta}^\nu=
\varepsilon\Big[u(p_\xi-\varepsilon p_z)-p(u_\xi-\varepsilon u_z)\Big]=
\varepsilon\mathbf{R}=\varepsilon \phi^2\,L_{\bar{\zeta}}\psi.
\]
Clearly, $D(\bar{\zeta})$ and $D^*(\bar{\zeta})$ are linearly independent in
general:
$$
D(\bar{\zeta})\wedge D^*(\bar{\zeta})=\varepsilon
\Big[(u_\xi-\varepsilon u_z)^2+(p_\xi-\varepsilon p_z)^2\Big]dx\wedge dy
=\varepsilon\phi^2(\psi_\xi-\varepsilon \psi_z)^2\,dx\wedge dy\neq 0.
$$
Recall now that every 2-form $F$ defines a linear map $\tilde{F}$ from
1-forms to 3-forms through the exterior product:
$\tilde{F}(\alpha):=\alpha\wedge F$, where $\alpha\in \Lambda^1(M)$.
Moreover, the Hodge $*$-operator, composed now with $\tilde{F}$, gets
$\tilde{F}(\alpha)$ back to $*\tilde{F}(\alpha)\in\Lambda^1(M)$. In the
previous section we introduced two 2-forms $G=A\wedge\zeta$ and
$G^*=A^*\wedge\zeta$ and noticed that $G^*=*G$. We readily obtain now
\[
D(\bar{\zeta})\wedge G=D^*(\bar{\zeta})\wedge G^*= D(\bar{\zeta})\wedge
A\wedge\zeta=D^*(\bar{\zeta})\wedge A^*\wedge\zeta=
\]
\[
=-\varepsilon\Big[u(p_\xi-\varepsilon p_z)- p(u_\xi-\varepsilon
u_z)\Big]dx\wedge dy\wedge dz- \Big[u(p_\xi-\varepsilon p_z)-
p(u_\xi-\varepsilon u_z)\Big]dx\wedge dy\wedge d\xi=
\]
\[
=-\phi^2\,L_{\bar{\zeta}}\psi\,(\varepsilon\,dx\wedge dy\wedge dz+
dx\wedge dy\wedge d\xi)=-\mathbf{R}\,(\varepsilon\,dx\wedge dy\wedge dz+
dx\wedge dy\wedge d\xi),
\]
\[ D(\bar{\zeta})\wedge
G^*=-D^*(\bar{\zeta})\wedge G= D(\bar{\zeta})\wedge
A^*\wedge\zeta=-D^*(\bar{\zeta})\wedge A\wedge\zeta=
\]
\[
=\frac12\Big[(u^2+p^2)_\xi-\varepsilon(u^2+p^2)_z\Big] (dx\wedge dy\wedge
dz+\varepsilon\,dx\wedge dy\wedge d\xi).
\]
Thus, recalling relations (3.16)-(3.20), we get
\begin{equation}
*\Big[D(\bar{\zeta})\wedge
A\wedge\zeta\Big]= *\Big[D^*(\bar{\zeta})\wedge A^*\wedge\zeta\Big]=
-\varepsilon\mathbf{R}\,\zeta =-i(\bar{G^*})\mathbf{d}G=i(\bar{G})\mathbf{d}G^*,
\end{equation}
\begin{equation} *\Big[D(\bar{\zeta})\wedge
A^*\wedge\zeta\Big]= -*\Big[D^*(\bar{\zeta})\wedge A\wedge\zeta\Big]=
\frac12L_{\bar{\zeta}}\phi^2\,\zeta=
i(\bar{G})\mathbf{d}G=i(\bar{G^*})\mathbf{d}G^*.
\end{equation}

The above relations show various dynamical aspects of the energy-momentum
redistribution during evolution of our PhLO. In particular, equations
(3.36-3.37) clearly show that it is possible the translational and rotational
components of the energy-momentum redistribution to be represented in form
depending on the $\zeta$-directed strains $D(\bar{\zeta})$ and
$D^*(\bar{\zeta})$. So, the local translational changes of the energy-momentum
carried by the two vector components $G$ and $G^*$ of our PhLO are given by the
two 1-forms $*\big[D(\bar{\zeta})\wedge A^*\wedge\zeta\big]$ and
$*\big[D^*(\bar{\zeta})\wedge A\wedge\zeta\big])$ and the local rotational ones
- by the 1-forms $*\big[D(\bar{\zeta})\wedge A\wedge\zeta\big]$ and
$*\big[D^*(\bar{\zeta})\wedge A^*\wedge\zeta\big]$.  In fact, the form
$*\big[D(\bar{\zeta})\wedge A\wedge\zeta\big]$ determines the strain that
"leaves" the 2-plane defined by $(A,\zeta)$ and the form
$*\big[D^*(\bar{\zeta})\wedge A^*\wedge\zeta\big]$ determines the strain that
"leaves" the 2-plane defined by $(A^*,\zeta)$. Since the PhLO is free, i.e. no
energy-momentum is lost or gained from outside, this means that the two
(null-field) components $G$ and $G^*$ exchange locally {\it equal}
energy-momentum quantities: $ *\Big[D(\bar{\zeta})\wedge A\wedge\zeta\Big]=
*\Big[D^*(\bar{\zeta})\wedge A^*\wedge\zeta\Big]. $ Now, the local
energy-momentum conservation law
$\nabla_{\nu}\big[G_{\mu\sigma}\bar{G}^{\nu\sigma}+
(G^*)_{\mu\sigma}(\bar{G}^*)^{\nu\sigma}\big]=0$ requires
$L_{\bar{\zeta}}\phi^2=0$, and the corresponding strain-fluxes become
 zero: $*\big[D^*(\bar{\zeta})\wedge A\wedge\zeta\big]=0$,
$*\big[D(\bar{\zeta})\wedge A^*\wedge\zeta\big]=0$.

It seems important to note
that, only dynamical relation between  the local energy-momentum change and
strain fluxes exists, so NO analog of the assumed in elasticity theory
generalized Hooke law, (i.e. linear relation between the stress tensor and the
strain tensor) seems to exist. This clearly goes along with the fully dynamical
nature of PhLO, i.e. linear relations exist between the divergence terms of our
stress tensor $\frac12\big[-G_{\mu\sigma}\bar{G}^{\nu\sigma}-
(G^*)_{\mu\sigma}(\bar{G}^*)^{\nu\sigma}\big]$
and the $\bar{\zeta}$-directed strain fluxes as given by equations
(3.36)-(3.37).

\chapter{Equations of motion for PhLO. Solutions}

{\it In this chapter we show that appropriate solutions for PhLO can be
obtained by solving {\bf linear} equations}.

 \section{The approach based on the notion for PhLO} Every system of
equations describing the time-evolution of some physical system should be
consistent with the very system in the sense that all identification
characteristics of the system described must not change. In the case of
electromagnetic PhLO we assume the couple $(F,\tilde{F})$ to represent the
field, and in accordance with our notion for PhLO one of the identification
characteristics is straight-line translational propagation of the
energy-density with constant velocity "$c$", therefore, with every PhLO we may
associate appropriate direction, i.e. a geodesic null vector field
$\bar{\zeta}, \bar{\zeta}^2=0$ on the Minkowski space-time. On the other hand,
the complex of field functions $(F_{\mu\nu},\tilde{F}_{\mu\nu})$ admits both
translational and rotational components of propagation. We choose further
$\bar{\zeta}=-\varepsilon \frac{\partial}{\partial z}+\frac{\partial}{\partial \xi}$,
which means that we have chosen the
coordinate system in such a way that the translational propagation is parallel to
the plane $(z,\xi)$. For another such parameter we assume that
the finite longitudinal extension of any PhLO is fixed and is given by an
appropriate positive number $\lambda $. In accordance with the "compatible
translational-rotational dynamical structure" of PhLO we shall assume that {\it
no translation is possible without rotation, and no rotation is possible
without translation}, and in view of the constancy of the translational
component of propagation we shall assume that the rotational component of
propagation is periodic, i.e. it is characterized by a constant frequency. The
natural period $T$ suggested is obviously $T=\frac{\lambda}{c}$. An obvious
candidate for "rotational operator" is the linear map $J$ transforming $F$ to
$\tilde{F}$, which map coincides with the reduced to 2-forms Hodge-$*$.
Geometrically, $*$ rotates the 2-frame $(A,A^*)$ to $\frac\pi 2$, so if such
a rotation is associated with a translational advancement of $l_o$, then a full
rotation should correspond to translational advancement of $4l_o=\lambda $. The
simplest and most natural translational change of the field $(F,\tilde{F})$
along $\bar{\zeta}$ should be given by
the Lie derivative of the field along $\bar{\zeta}$. Hence, the simplest and most natural
equations should read
\begin{equation}
\kappa l_o\,L_{\bar{\zeta}}(F)=\varepsilon\tilde{F},
\end{equation}
where
$F$ and $\tilde{F}$ are given in the preceding section,
$\kappa=\pm 1$ is responsible for left/right orientation of the rotational
component of propagation, and $l_o=const$. Vice versa, since $J\circ J=-id$ and
$J^{-1}=-J$ the above equation is equivalent to \[ \kappa
l_o\,L_{{\bar{\zeta}}}(\tilde{F})=-\varepsilon F. \] It is easy to show that these
equations are equivalent to
\begin{equation}
\kappa l_o\,L_{{\bar{\zeta}}}(V-V_o)=\varepsilon(\tilde{V}-V_o),
\end{equation}
where $V$ is given by (40) and
$V_o=dx\otimes\frac{\partial}{\partial x}+
dy\otimes\frac{\partial}{\partial y}$
in our coordinates is the identity map in $Im(V)=Im(\tilde{V})$.
Another equivalent form is given by
$$
\kappa l_oZ_1=\bar{A^*}, \ \ \ \ \text{or} \ \ \ \  \kappa l_oZ_2=-\bar{A},
$$
where
$\bar{A^*}$ and $\bar{A}$ are $\eta$-corresponding vector fields to the 1-forms
$A^*$ and $A$.

\section{The Lagrangian Approach}
Appropriate lagrangian for the above equations ($l_o$=const.) is
\begin{equation}
\mathbb{L}=\frac14\left(
\varepsilon\kappa l_o \bar{\zeta}^\sigma\frac{\partial
F_{\alpha\beta}}{\partial x^\sigma}- \tilde{F}_{\alpha\beta}\right)
\tilde{F}^{\alpha\beta}
-\frac14\left(
\varepsilon\kappa l_o \bar{\zeta}^\sigma                          
\frac{\partial \tilde{F}_{\alpha\beta}}{\partial x^\sigma}+
F_{\alpha\beta}\right)F^{\alpha\beta},
\end{equation}
$F$ and $\tilde{F}$ are considered as independent, and the relations
$F\wedge F=F\wedge \tilde{F}=\tilde{F}\wedge \tilde{F}=0$ lead to
$$
\bar{\zeta}^\sigma\frac{\partial
F_{\alpha\beta}}{\partial x^\sigma}\tilde{F}^{\alpha\beta}=
\bar{\zeta}^\sigma                          
\frac{\partial \tilde{F}_{\alpha\beta}}{\partial x^\sigma}F^{\alpha\beta}=0.
$$
The corresponding Lagrange equations read
\begin{equation}
\varepsilon\kappa l_o \bar{\zeta}^\sigma                          
\frac{\partial \tilde{F}_{\alpha\beta}}{\partial x^\sigma}+
F_{\alpha\beta}=0,\ \
\varepsilon\kappa l_o \bar{\zeta}^\sigma
\frac{\partial F_{\alpha\beta}}{\partial x^\sigma}-
\tilde{F}_{\alpha\beta}=0 ,
\end{equation}
so, on the solutions the lagrangian becomes zero: $\mathbb{L}(solutions)=0$.
The stress-energy-momentum tensor, in view of the null character of $F$ and
$\tilde{F}$, is given by (3.29), where $*F$ has to be replaced by $\tilde{F}$.
It deserves noting that the above null conditions lead to
$F_{\mu\sigma}F^{\nu\sigma}=\tilde{F}_{\mu\sigma}\tilde{F}^{\nu\sigma}$ and to
$F_{\mu\sigma}\tilde{F}^{\nu\sigma}=0$.
Hence, the two
subsystems represented by $F$ and $\tilde{F}$ carry the same
stress-energy-momentum, therefore, $F\rightleftarrows \tilde{F}$
energy-momentum exchange is possible only in equal quantities. In our
coordinates the above equations reduce to
\[ \kappa
l_o(u_{\xi}-\varepsilon\,u_z)=-p,\ \ \ \kappa l_o(p_{\xi}-\varepsilon\,p_z)=u,
\]
it is seen that the constant $l_o$ satisfies the above given relation (3.28).
From these last equations we readily obtain the relations \[
(u^2+p^2)_{\xi}-\varepsilon\,(u^2+p^2)_z=0, \ \
u\,(p_{\xi}-\varepsilon\,p_z)-p\,(u_{\xi}-\varepsilon\,u_z)=
\frac{\kappa}{l_o}(u^2+p^2),
\]
which represent our equations in energy-momentum terms. Now,
the substitution $u=\Phi\cos\,\psi,\ \  p=\Phi\sin\,\psi $, leads to the
relations \[ L_{{\bar{\zeta}}}\Phi=0, \ \
L_{{\bar{\zeta}}}\psi=\frac{\kappa}{l_o}. \] In terms of nonlinear connections
recalling that $\Phi^2=-\frac12tr(V\circ H^*)$ and computing
$\frac12tr(V\circ L_{{\bar{\zeta}}}\tilde{H}^*)=
\varepsilon\big[u\,(p_{\xi}-\varepsilon\,p_z)-p\,(u_{\xi}-\varepsilon\,u_z)\big]=
\Phi^2\varepsilon\,L_{{\bar{\zeta}}}\psi$
the last two relations can be equivalently written as
\[
L_{{\bar{\zeta}}}\big[tr(V\circ H^*)\big]=0,\ \
tr(V\circ L_{{\bar{\zeta}}}\tilde{H}^*)
=-\frac{\varepsilon\,\kappa}{l_o}tr(V\circ H^*).
\]

\section{Equations of motion in terms of translational-rotational \newline
compatability}

In order to look at the translational-rotational compatability as a generating
tool for writing equations of motion we recall first the concept of
local symmetry of a distribution: a vector field $Y$ is a local (or
infinitesimal) symmetry of a p-dimensional distribution $\Delta$ defined by the
vector fields $(Y_1,\dots,Y_p)$ if every Lie bracket $[Y_i,Y]$ is in $\Delta$:
$[Y_i,Y]\in\Delta $. Clearly, if $\Delta$ is completely integrable, then
every $Y_i$ is a symmetry of $\Delta$, and the
flows of these vector fields move the points of each completely
integral manifold of $\Delta$ inside this integral manifold, that's
why they are called sometimes internal symmetries. If $Y$ is outside $\Delta$
then it is called {\it shuffling} symmetry [28], and in such a case the flow of
$Y$ transforms a given completely integral manifold to another one, i.e. the
flow of $Y$ "shuffles" the lists of the corresponding foliation. We are going
to show that our vector field $\bar{\zeta}=-\varepsilon
\frac{\partial}{\partial z}+\frac{\partial}{\partial \xi}$ is a shuffling
symmetry for the distribution $\Delta_o$ defined by the vector fields
$(\bar{A},\bar{A}^*)$. In fact, $\Delta_o$ coincides with our vertical
distribution generated by
$(\frac{\partial}{\partial x}, \frac{\partial}{\partial y})$, so it is
completely integrable and its integral manifolds coincide with the
$(x,y)$-planes. From physical point of view this should be expected in view of
the intrinsically required stability of our PhLO under translational propagation
along null straight lines: this propagation just
transforms the 2-plane $(x,y)$ passing through the point $(z_1,\xi_1)$ to a
parallel to it 2-plane passing through the point $(z_2,\xi_2)$, and these two
points lay on the same trajectory of our field $\bar{\zeta}$.

The corresponding Lie brackets are
\[
[\bar{A},\bar{\zeta}]=(u_\xi-\varepsilon\,u_z)\frac{\partial}{\partial x}+
(p_\xi-\varepsilon\,p_z)\frac{\partial}{\partial y}, \ \ \
[\bar{A}^*,\bar{\zeta}]=-\varepsilon\,(p_\xi-\varepsilon\,p_z)\frac{\partial}{\partial x}+
\varepsilon\,(u_\xi-\varepsilon\,u_z)\frac{\partial}{\partial y}.
\]
We see that $[\bar{A},\bar{\zeta}]$ and $[\bar{A}^*,\bar{\zeta}]$ are generated by
$(\frac{\partial}{\partial x}, \frac{\partial}{\partial y})$,
but $\bar{\zeta}$ is outside $\Delta_o$, so our field $\bar{\zeta}$ is a
shuffling local symmetry of $\Delta_o$.

We notice now that at each point we have
two frames:
$(\bar{A},\bar{A}^*,\partial_z,\partial_{\xi})$ and
$([\bar{A},\bar{\zeta}],[\bar{A^*},\bar{\zeta}],\partial_z,\partial_{\xi})$.
Since physically we have internal
energy-momentum redistribution during propagation, we could interprete the
permanent availability of these two intrinsically connected frames as
corresponding mathematical adequate of this physical process.
Taking into
account that only the first two vectors of these two frames change during
propagation we write down the corresponding linear transformation as follows:
\[
([\bar{A},\bar{\zeta}],[\bar{A^*},\bar{\zeta}])=(\bar{A},\bar{A}^*)
\begin{Vmatrix}\alpha & \beta \\ \gamma & \delta\end{Vmatrix} .
\]
Solving this system with respect to $(\alpha,\beta,\gamma,\delta)$ we obtain
\[
\begin{Vmatrix}\alpha & \beta \\ \gamma & \delta \end{Vmatrix}=
\frac{1}{\phi^2}
\begin{Vmatrix} -\frac12 L_{\bar{\zeta}}\Phi^2 &
\varepsilon \mathbf{R} \\ -\varepsilon \mathbf{R}
& -\frac12 L_{\bar{\zeta}}\Phi^2 \end{Vmatrix}=
-\frac12\frac{L_{\bar{\zeta}}\Phi^2}{\Phi^2}
\begin{Vmatrix} 1 & 0 \\ 0 & 1\end{Vmatrix}+
\varepsilon L_{\bar{\zeta}}\psi\begin{Vmatrix} 0 & 1 \\ -1 & 0\end{Vmatrix} ,
\]
where $\mathbf{R}=u\,(p_\xi-\varepsilon\,p_z)-p\,(u_\xi-\varepsilon\,u_z)$.
If the translational propagation is governed by the conservation law
$L_{\bar{\zeta}}\Phi^2=0$, then we obtain that the rotational component of
propagation is governed by the matrix $\varepsilon L_{\bar{\zeta}}\psi\,J$, where $J$
denotes the canonical complex structure in $\mathbb{R}^2$, and since
$\Phi^2\,L_{\bar{\zeta}}\psi=u\,(p_\xi-\varepsilon\,p_z)-p\,(u_\xi-\varepsilon\,u_z)\neq
0$ we conclude that the rotational component of propagation would be available
if and only if $\mathbf{R}\neq 0$. We may also say that a compatible
translational-rotational dynamical structure is available if the amplitude
$\Phi^2=u^2+p^2$ is a running wave along $\bar{\zeta}$ and the phase
$\psi=\mathrm{arctg}\frac{p}{u}$ is NOT a running wave along $\bar{\zeta} : L_{\bar{\zeta}}\psi\neq
0$. Physically this means that the rotational component of propagation is
entirely determined by the available internal energy-momentum exchange:
$i(\tilde{F})\mathbf{d}F=-i(F)\mathbf{d}\tilde{F}$.

Now, if we have to
{\it guarantee the
conservative and constant character of the rotational aspect} of the PhLO
nature, we can assume
$L_{\bar{\zeta}}\psi=const=\kappa l_o^{-1}, \kappa=\pm 1$.
Thus, the frame rotation
$(\bar{A},\bar{A^*},\partial_z, \partial_\xi)\rightarrow
([\bar{A},\bar{\zeta}],[\bar{A^*},\bar{\zeta}],\partial_z, \partial_\xi)$,
i.e. $[\bar{A},\bar{\zeta}]=-\varepsilon\bar{A^*}\,L_{\bar{\zeta}}\psi$
and $[\bar{A^*},\bar{\zeta}]=\varepsilon\bar{A}\,L_{\bar{\zeta}}\psi $,
gives the following equations for the two functions $(u,p)$:
\[
u_\xi-\varepsilon u_z=-\frac{\kappa}{l_o}\,p, \ \ \
p_\xi-\varepsilon p_z=\frac{\kappa}{l_o}\,u \ .
\]

The quantity
$\mathbf{R}=u\,(p_\xi-\varepsilon\,p_z)-p\,(u_\xi-\varepsilon\,u_z)=
\Phi^2L_{\bar{\zeta}}\psi=\kappa l_o^{-1}\Phi^2$ suggests
to find an integral characteristic of the PhLO rotational nature. In fact, the
two co-distributions $(A,\zeta)$ and $(A^*,\zeta)$ define the two (equal in our
case) Frobenius 4-forms
$\mathbf{d}A\wedge A\wedge \zeta=\mathbf{d}A^*\wedge A^*\wedge \zeta$.
Each of these two 4-forms is equal to $\varepsilon\mathbf{R}\omega_o$.
Now, multiplying by $l_o/c$ any of them we obtain:
\begin{equation}
\frac{l_o}{c}\,\mathbf{d}A\wedge A\wedge \zeta=
\frac{l_o}{c}\,\mathbf{d}A^*\wedge A^*\wedge \zeta=
\frac{l_o}{c}\varepsilon\mathbf{R}\omega_o =
\varepsilon\kappa\frac{\Phi^2}{c}\omega_o\ .
\end{equation}
Integrating over the 4-volume $\mathbb{R}^3\times(\lambda=4l_o)$ (and having in
view the spatially finite nature of PhLO) we obtain the finite quantity
$\mathcal{H}=\varepsilon\kappa ET$, where $E$ is the integral energy of
the PhLO, $T=\frac{\lambda}{c}$, which clearly is the analog of the Planck
formula $E=h\nu$, i.e. $h=ET$. The combination $\varepsilon\kappa$ means that
the two orientations of the rotation, defined by $\kappa=\pm 1$, may be
observed in each of the two spatial directions of translational propagation
of the PhLO along the $z$-axis: from $-\infty$ to  $+\infty$, or from $+\infty$
to $-\infty$.

Finally, recalling relations (3.15), we can easily see that in case of
$L_{\bar{\zeta}}\psi=\kappa l_o^{-1}$ and $L_{\bar{\zeta}}\Phi^2=0$ the 3-form
$\delta F\wedge F=\varepsilon \mathbf{R}*\zeta=
\frac{\varepsilon \kappa}{l_o}\Phi^2*\zeta$
becomes closed: $\mathbf{d}(\delta F\wedge F)=0$, which also gives
an integral conservation law. In fact, the 3-integral of the reduced on
$\mathbb{R}^3$ 3-form $\frac{l_o^2}{c}(\delta F\wedge F)$ gives
$\varepsilon\kappa ET$, where $E$ is the integral energy, so, the Planck
formula holds.

\vskip 0.5cm
\section{Photon-like Solutions}
We consider the equations obtained in terms of the two functions
$\Phi=\sqrt{u^2+p^2}$ and $\psi=\mathrm{arctg}\frac{p}{u}$.
The equation for $\Phi$
in our coordinates is $\Phi_{\xi}-\varepsilon \Phi_{z}=0$, therefore,
$\Phi=\Phi(x,y,\xi+\varepsilon z)$, where $\Phi$ is allowed to be spatially
finite, as assumed further, or spatially localized function. The equation for
$\psi $ is $\psi_{\xi}-\varepsilon\psi_{z}=\frac{\kappa}{l_o}$. Two families of
solutions for $\psi$, depending on an arbitrary function $\varphi$ can be given
by
\[
\psi_1=-\frac{\varepsilon\kappa}{l_o}z+\varphi(x,y,\xi+\varepsilon z),\ \ \
\text{and}\ \ \ \psi_2=\frac{\kappa}{l_o}\xi+\varphi(x,y,\xi+\varepsilon z) .
\]
Since $\Phi^2$ is a spatially finite function representing the energy density
we see that the translational propagation of our PhLO is represented by a
{\it spatially finite running wave} along the $z$-coordinate. Let's assume that
the phase is given by $\psi_1$ and, for simplicity, $\varphi=0$. The form of
this solution suggests to choose the initial condition $u_{t=0}(x,y,\varepsilon
z),p_{t=0}(x,y,\varepsilon z)$ in the following way. Let for $z=0$ the initial
condition be located on a disk $D=D(x,y;a,b;r_o)$ of small radius $r_o$, the
center of the disk to have coordinates $(a,b)$, and the value of
$\Phi_{t=0}(x,y,0)=\sqrt{u_{t=0}^2+p_{t=0}^2}$ to be
proportional to some appropriate for the case bump function $f>0$ on $D$ of the
distance $\sqrt{(x-a)^2+(y-b)^2}$ between the origin of the coordinate system
and the point $(x,y,0)$, such that it is centered at the point $(a,b)$, so,
$f(x,y)=f(\sqrt{(x-a)^2+(y-b)^2}\,)$, $D$ is defined by
$D=\{(x,y)|\sqrt{(x-a)^2+(y-b)^2}\leq r_o\}$, and $f(x,y)$ is zero outside $D$.
Let also the dependence of $\Phi_{t=0}$ on $z$ be given by be the corresponding
bump function $\theta(z;\lambda)>0$ of an interval $(z,z+\lambda)$ of length
$\lambda=4l_o$ on the $z$-axis. If $\gamma>0$ is the proportionality coefficient
we obtain
\begin{align*}
u=\gamma\,f(x,y;a,b)\,
\theta(ct+\varepsilon z;\lambda)\,\cos(\psi_1), \\
p=\gamma\,f(x,y;a,b)\,
\theta(ct+\varepsilon z;\lambda)\,\sin(\psi_1).
\end{align*}
We see that because of the available {\it sine} and {\it cosine} factors in
the solution, the initial condition for the solution will occupy a $3d$-spatial
region of shape that is close to a helical cylinder of height $\lambda$, having
internal radius of $r_o$ and wrapped up around the $z$-axis. Also, its center
will always be $\sqrt{a^2+b^2}$-distant from the $z$-axis. Hence, the solution
will propagate translationally along the coordinate $z$ with the velocity $c$,
and, rotationally, inside the corresponding infinitely long helical cylinder
because of the $z$-dependence of the available periodical multiples.

On the two figures below are given two theoretical examples with $\kappa=-1$
and $\kappa=1$ respectively, amplitude function $\Phi$ located inside a
one-step helical cylinder $\mathcal{D}$ with height of $\lambda$, and phase
$\psi=-\varepsilon\kappa \frac{2\pi z}{\lambda}=
-\varepsilon\kappa\frac{\pi\,z}{2l_o}$. The solutions with
$\varepsilon=-1$ will propagate left-to-right along the coordinate $z$.
\vskip0.5cm
\begin{center}
\begin{figure}[ht!]
\centerline{
{\mbox{\psfig{figure=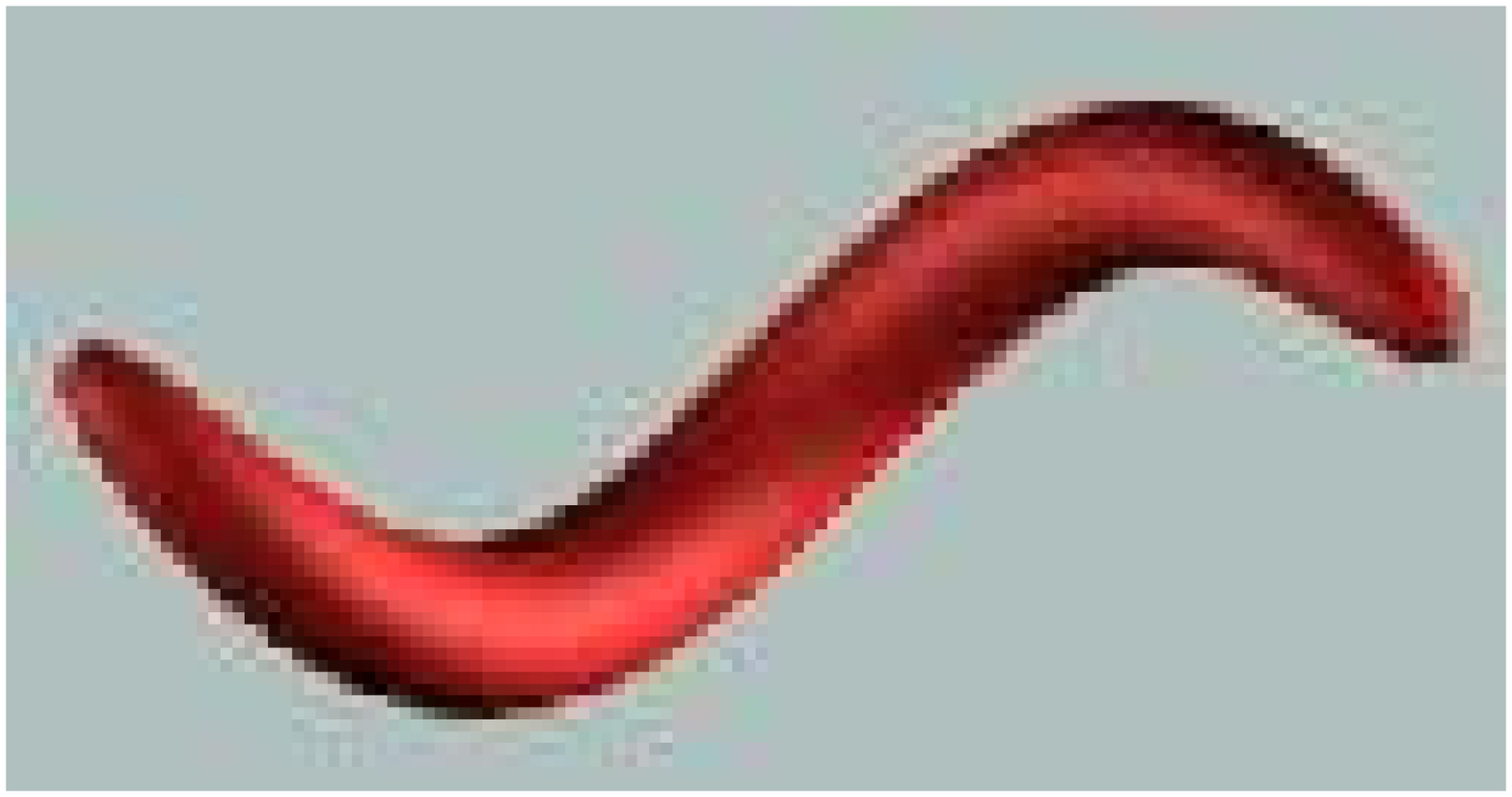,height=1.8cm,width=3.5cm}}
\mbox{\psfig{figure=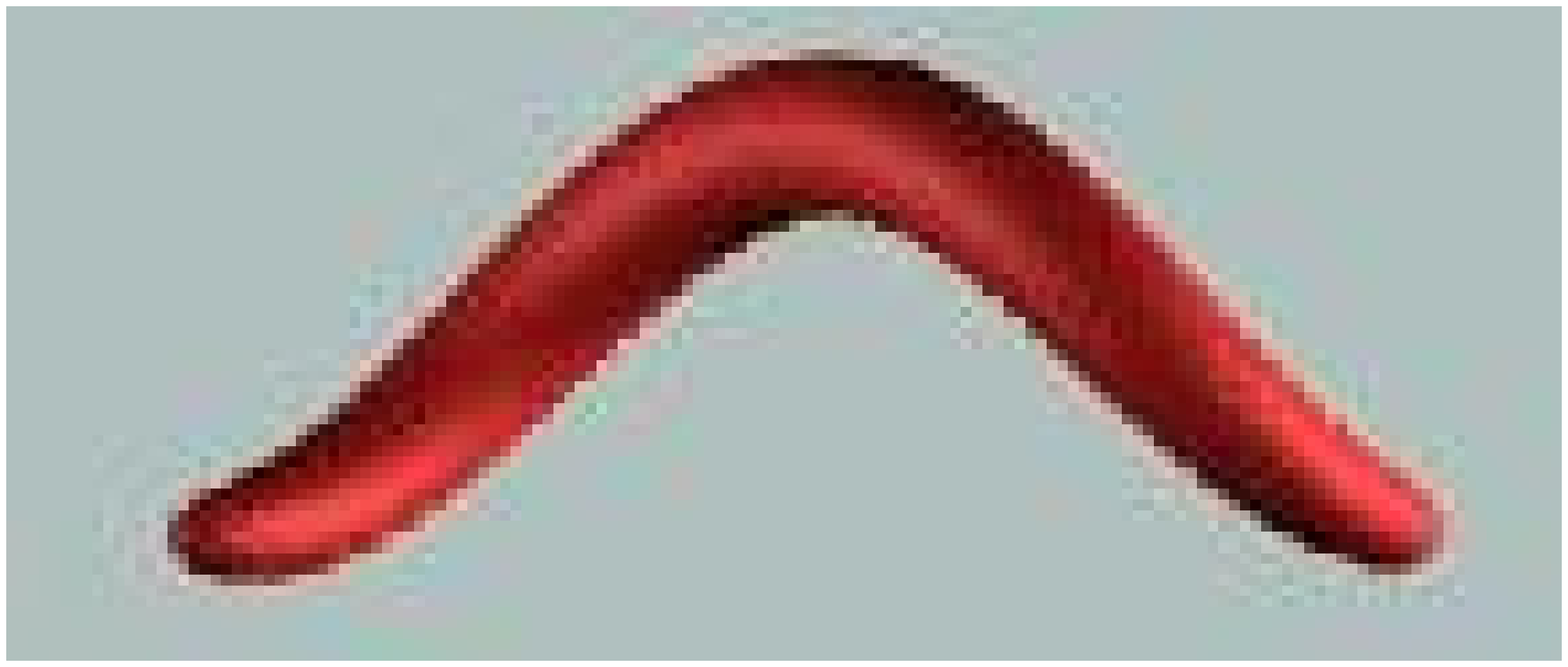,height=1.8cm,width=4.2cm}}
\mbox{\psfig{figure=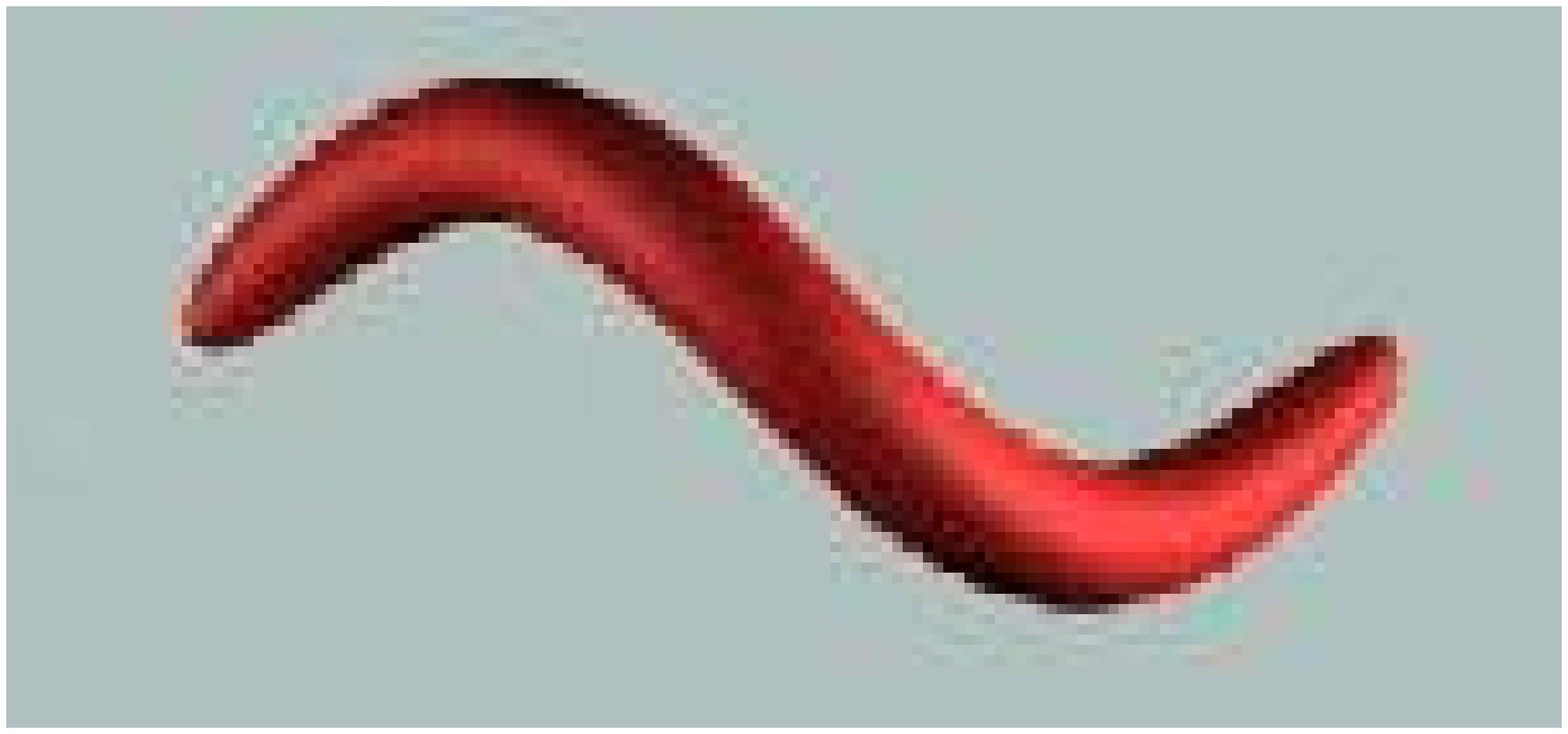,height=1.8cm,width=4.2cm}}}}
\caption{Theoretical example with $\kappa=-1$. The translational propagation is
directed left-to-right.}
\end{figure}
\end{center}
\begin{center}
\begin{figure}[ht!]
\centerline{
{\mbox{\psfig{figure=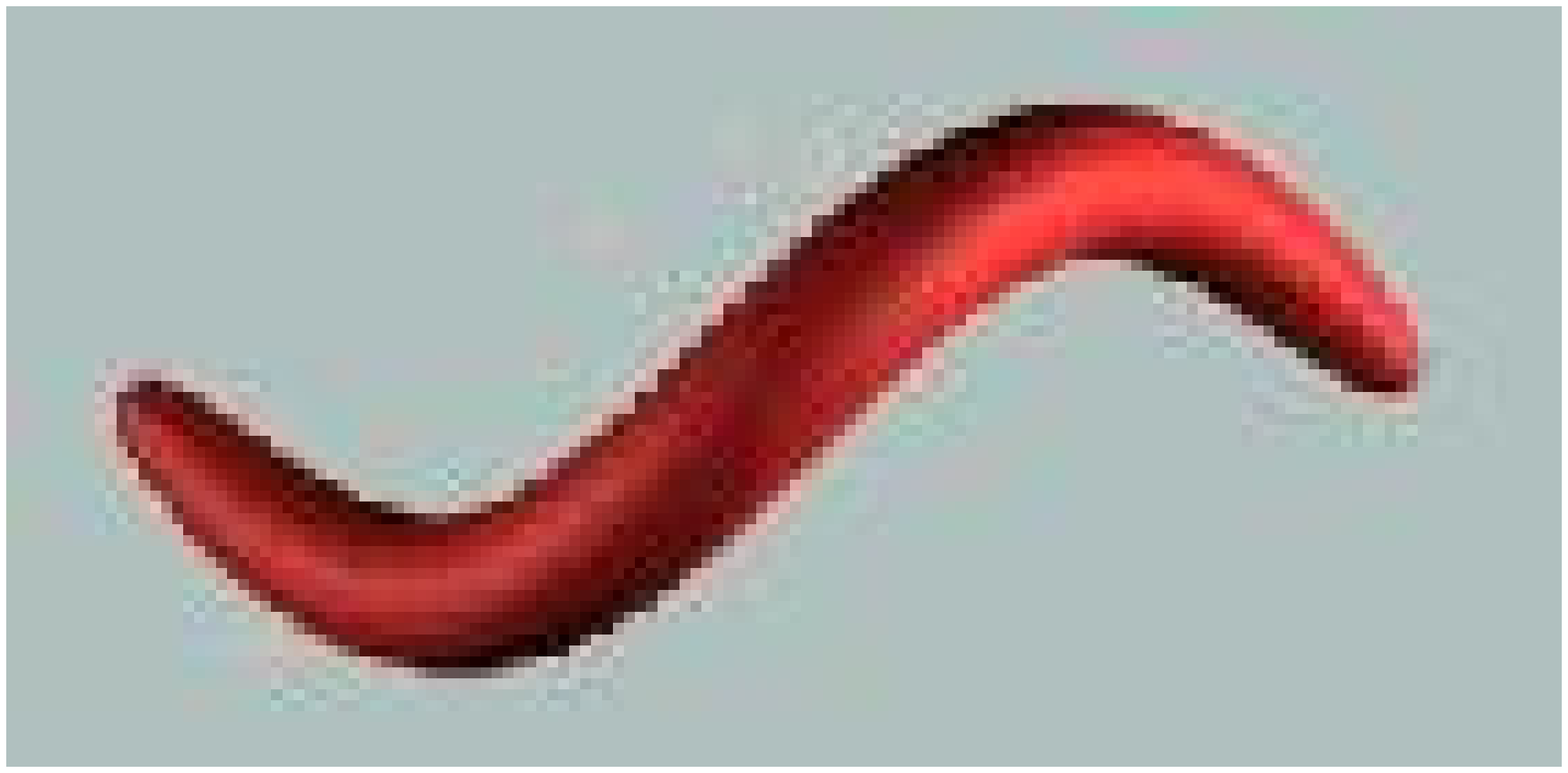,height=1.8cm,width=3.5cm}}
\mbox{\psfig{figure=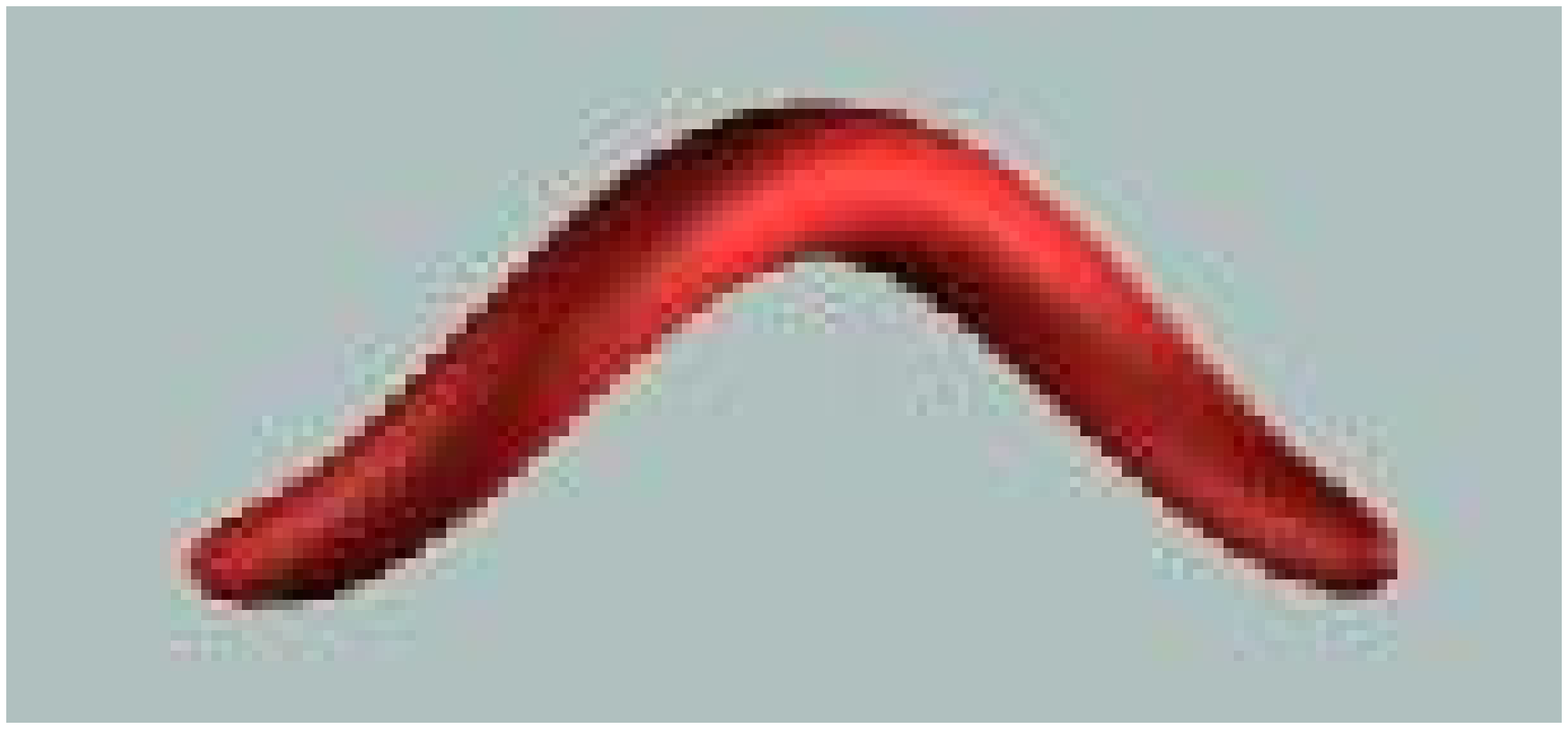,height=1.8cm,width=4.2cm}}
\mbox{\psfig{figure=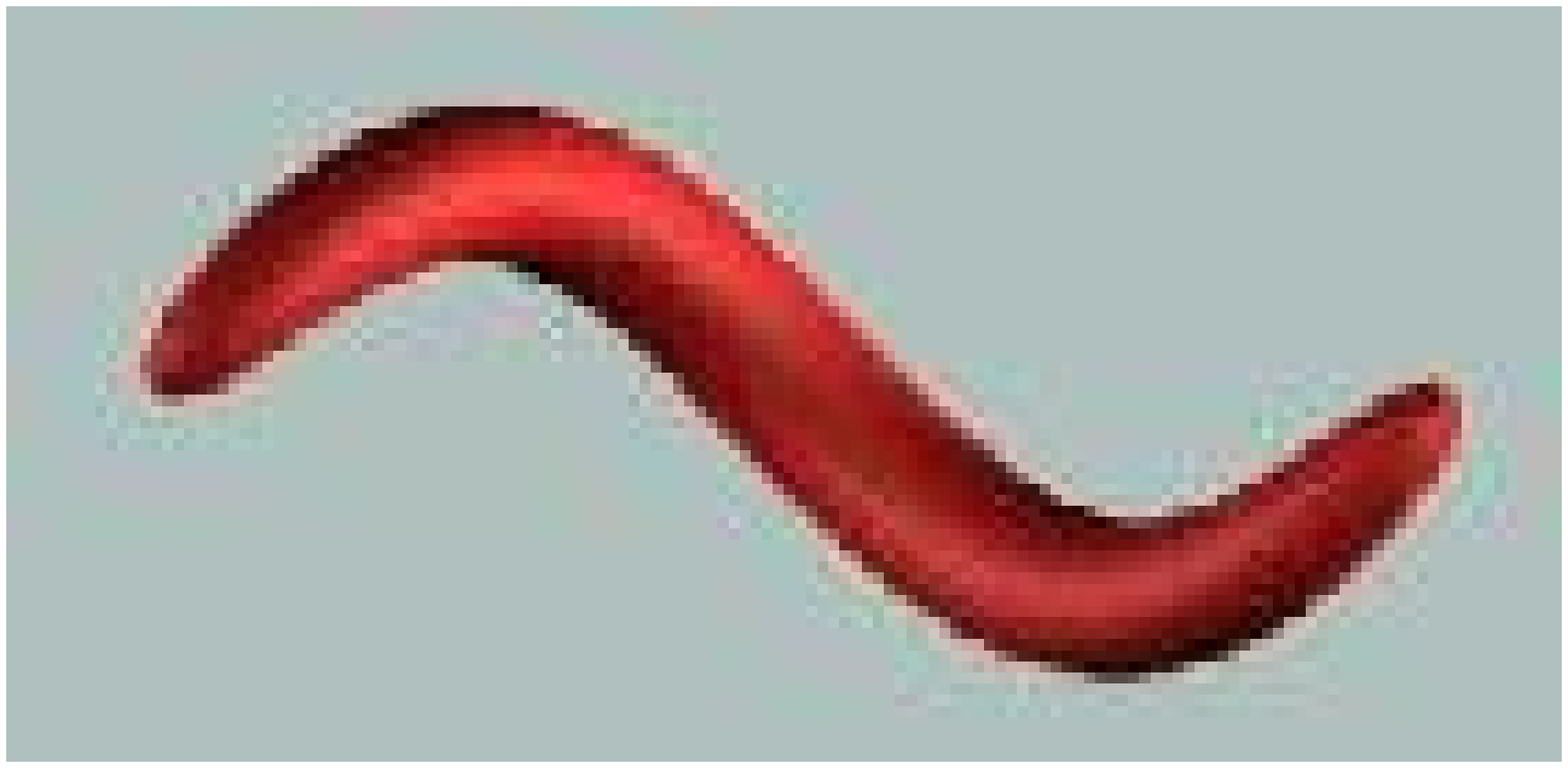,height=1.8cm,width=4.2cm}}}}
\caption{Theoretical example with $\kappa=1$. The translational propagation is
directed left-to-right.}
\end{figure}
\end{center}
The curvature $K$ and the torsion $\tau$ of the screwline inside
$\mathcal{D}$ through the point $(x,y,0)\in D$ will be \[
K=\frac{\gamma\,f\theta}{(\gamma\,f\theta)^2+b^2},\ \ \ \
\tau=\frac{\kappa\,b}{(\gamma\,f\theta)^2+b^2} \ ,
\]
where $b=\lambda/2\pi=2l_o/\pi$. The
rotational frequency $\nu$ will be $\nu=c/\lambda=c/4l_o$, so we can introduce
period $T=1/\nu$ and elementary action $h=E.T$, where $E$ is the
(obviously finite) integral energy of the solution defined as 3d-integral
of the energy density $\Phi^2=(\gamma\,f\theta)^2$.

\chapter{Retrospect}
In trying to understand our observational knowledge of the real world we must
be able to separate the {\it important} structural and behavioral
properties of the real objects from those, the changes of which
during time-evolution do not lead to annihilation of the objects
under consideration. One of the basic in our view lessons that we more or
less have been taught is that the physical objects are {\it spatially finite
entities}, and that for their detection and further study some
energy-momentum exchange is necessarily {\it required}. So, every physical
object necessarily carries energy-momentum and every interaction between two
physical objects has such an energy-momentum exchange aspect. The second
lesson concerning any interaction is that, beyond its {\it universality},
energy-momentum is a {\it conserved} quantity, so NO loss of it is allowed:
it may only pass from one object to another. This means that every {\it
annihilation} process causes {\it creation} process(es), and the full
energy-momentum that has been carried by the annihilated objects, is carried
away by the created ones. Energy-momentum always needs carriers, as well as
every physical object always carries energy-momentum.  Hence, the
energy-momentum exchange abilities of any physical object provide
protection against external influence on one side, and reveal its intrinsic
nature, on the other side. Therefore, our knowledge about the entire
complex of properties of a physical object relies on getting information
about its abilities in this respect and finding corresponding quantities
describing quantitatively these abilities.

The spatially finite nature of a physical object implies spatial structure
and finite quantity of energy-momentum needed  for its creation, so NO
structureless and infinite objects may exist. The approximations for
"point object" and "spatially infinite field", although useful in some respects,
seem theoretically inadequate and should not be considered as basic ones. More
reliable appears to be the approximation "finite continuous object", which we
tried to follow throughout our exposition. This last approximation suggests
that the usually nonlinear local energy-momentum conservation relations should
be the basic suggesting tool for finding appropriate mathematical models of
local nature of such objects. The natural physical sense of the
corresponding model equations is not necessarily supposed to be local
energy-momentum exchange, but must be consistent with it, and closely connected
to it.

Another useful observation is that physical objects are {\it many-aspect
entities}, they have complicated structure and their very existence is
connected with {\it internal} energy-momentum exchange among
the various structural components. So, the mathematical model objects should
be many-component ones, and with appropriate mathematical structure.  Of
basic help in finding appropriate mathematical objects is having knowledge
of the internal symmetry properties of the physical object under
consideration.  This "step by step" process of getting and accumulating
important information about the physical properties of natural objects
reflects in the "step by step" process of refining the corresponding
mathematical models.

The greatest discovery at the very beginning of the last century was that the
notion of electromagnetic field as suggested by Maxwell equations is
inadequate: the time dependent electromagnetic field is not an infinite
smooth perturbation of the aether, on the contrary, it consists of many
individual time-stable objects, called later {\it photons}, which are
created/destroyed mainly during intra-atomic energy-transition processes.
Photons are finite objects, they carry energy-momentum and after they have
been radiated outside their atom-creator, they propagate as a whole
translationally by the speed of light.  Moreover, their propagation is not
just translational, it includes rotational component, which is of {\it
intrinsic} and {\it periodical} nature. The corresponding intrinsic action
for one period $T$ is $h=ET$, where $E$ is the full energy of the photon, and
all photons carry the same elementary intrinsic action $h$. During the entire
20th century physicists have tried to understand the dynamical structure/nature
of photons from various points of view, and this process is still going on
today. The conviction that a new point of view on the dynamical nature of the
field equations is needed is shortly summarized by Ziolkowsi [29]:
"Finite-energy, diffraction-free beams for the linear free-space wave equation
are imposible".

In order to come to a new look at the situation in nonrelativistic terms we made
use of the Newton approach: the identifying features of the object considered
must be kept unchanged during evolution and the admissible changes most
naturally should be expressed by means of specializing the energy-momentum
exchange abilities of the object considered, as well as by paying due respect
to the available translation-rotation interrelation. We showed that this
approach works well in the nonrelativistic $(\mathbf{E},\mathbf{B})$
formalism and concluded that the relativistic
$(F,*F)$ structure is much more adequate to the PhLO dynamical structure than
the $(\mathbf{E},\mathbf{B})$ one.

The basic theoretical idea in the relativistic formalism was to make use of the
Frobenius integrability/nonintegrability theorems as an appropriate
mathematical machinary: the integrability of a distribution we connected with
the time-stability  of the basic identification properties of the object
considered, and the nonintegrability of the available subdistributions was
interpreted physically as internal interaction among the subsystems, where the
corresponding curvatures appeared as natural mathematical tools for generating
appropriate mathematical images of the local energy-momentum exchange
fluxes between any two subsystems.

We introduced a notion of PhLO as a spatially finite physical object with a
compatible translational-rotational dynamical structure and propagating
translationally with the frame independent velocity of light "c". We showed that
Frobenius integrability theory possesses all necessary features to meet the
physical aspects of this notion. From physical viewpoint, two dynamically
interacting subsytems of a PhLO can be individualized, these subsystems carry
the same stress-energy-momentum, and they exchange energy-momentum locally
always in equal quantities, so they exist in a {\it dynamical equilibrium}. The
mathematical realization of the two subsystems of a PhLO was made in two ways:
through a direct choice of two nonintegrable subdistributions, and by means of
a couple of two nonlinear connections $V$ and $\tilde{V}$ with a common image
space. Their inter-communication is carried out and guaranteed by the nonzero
curvature forms $\Omega$ and $\Omega^*$ in the first case, and by the
nonzero curvature forms $\mathcal{R}$ and $\tilde{\mathcal{R}}$ in the second
case. The values of these curvature forms define two 1-dimensional space-like
subspaces, so, the corresponding two exterior products with the null direction of
translational propagation give the mathematical images $F$ and $\tilde{F}$ of
the two interacting subsystems. This approach allows to get some information
concerning the dynamical nature of the PhLO structure not only algebraically,
but also {\it infinitesimally}, i.e. through the curvature forms.

While the energy density $\Phi^2$ of a PhLO propagates only
translationally along straight isotropic lines, the available interaction of the
two subsytems of a PhLO demonstrates itself through a rotational component of
the entire propagational behaviour and is available only if the curvature
forms are not zero. The mutual energy-momentum exchanges are given by the
inner products of the curvature images  with $F$ and
$*F$. The dynamical equilibrium between $F$ and
$*F$ is given by $i(*F)(\mathbf{d}F)=-i(F)(\mathbf{d}*F)$.

Besides the spatially finite nature of PhLO that is allowed by our model and
illustrated with the invariant parameter $l_o$, two basic identifying
properties of PhLO were substantially used: straight-line translational
propagation with constant speed, and constant character of the rotational
component of propagation. The physical characteristics of a PhLO are
represented by an analog of the Maxwell-Minkowski stress-energy-momentum
tensor. An interesting moment is that $F$ and $\tilde{F}$ have zero horizonal
and vertical components with respect to the two nonlinear connections.

It was very interesting to find that some of the basic characteristics of PhLO
could be given in terms of the two strain-tensors, i.e. through the Lie
derivatives of the Minkowski pseudo-metric with respect to spatial direction
generators of the two nonintegrable subdistributions, so, each of the two
nonintegrable subdistributions has its own strain tensor. This gives entirely
new viewpoint on PhLO, namely, the PhLO energy-momentum propagates through
deformations! The values of each strain tensor on the generators of its
subdistribution gives the translational change of the energy density, and its
value on the generators of the other subdistribution gives the Frobenius
curvature. The corresponding dynamical aspects are given by expressions
(3.36)-(3.37).

It seems important to note that the curvature forms are not zero only if the
component-functions of the vector fields defining the distributions (or the
component-functions of the associated nonlinear connections) are NOT running
waves along the translational propagation, e.g. the squares of the curvature
forms are equal to $|\mathbf{d}F|^2=|\mathbf{d}*F|^2=|\delta F|^2=|\delta
*F|^2=(u_{\xi}-\varepsilon u_z)^2+(p_{\xi}-\varepsilon p_z)^2\neq 0$. Also, the
dually invariant longitudinal size parameter $l_{o}$ acquires sense only for
finite nonzero curvatures.

The equations of motion can be viewed from different viewpoints:
as compatability conditions between the rotational and translational components
of propagation, as Lagrange equations for an action principle, as the nonlinear
part of the solutions of the vacuum equations of EED, and also as naturally
defined transformation of 2-dimensional frames. In all these aspects of the
equations of motion the curvature forms play essential role through controlling
the inter-communication between $F$ and $\tilde{F}$. Moreover, the Frobenius
curvature turns out to be proportional to the energy density, which
recalls the main idea of General Relativity from one side, and
allows an analog of the famous Planck formula $E=h\nu$  to be introduced, from
the other side.

The solutions considered illustrate quite well the positive aspects of our
approach. It is interesting to note that the phase terms of these solutions
depend substantially only on spatial variables, so, the spatial
structure of the solutions considered participates directly in the rotational
component of the PhLO dynamical structure. \vskip 0.3cm Our basic conclusion
reads: PhLO are complex objects with dynamical structure of special kind, so
any mathematical model of PhLO shall need corresponding mathematical structure.
According to the results given in this study the basic adequate mathematical
structure in case of electromagnetic PhLO is the 3-dimensional
distribution/codistribution $(\bar{A},\bar{A}^*,\bar{\zeta})/(A,A^*,\zeta)$ on
Minkowski space-time together with the corresponding interconnections
represented by the integrability/nonintegrability properties of its
subdistributions.

\vskip 1cm

This study was partially supported by Contract $\phi\,15\,15/2005$ with the
Bulgarian National Fund "Science Research".
\newpage
\vskip 0.5cm
 {\bf REFERENCES} \vskip 0.3cm

[1] {\bf Poisson, S. D}. {\it Mem. Acad. sci.}, vol.3, p.121 (1818)

[2] {\bf Courant, R., Hilbert, D}., {\it Methoden der mathematischen Physik}, Berlin,
vol.2 \S 6 (1937)

[3] {\bf Farlow, S. J}., {\it Partial Differential equations for Scientists and
Engineers}, John Wiley and Sons, Inc., 1982

[4] {\bf G. Mie}, {\it Ann. der Phys.} Bd.37, 511 (1912); Bd.39, 1 (1912);
Bd.40, 1 (1913)

[5] {\bf M.Born, L.Infeld}, {\it Proc.Roy.Soc.}, A 144 (425), 1934.

[6] {\bf Plebanski, J.}, {\it Lectures on Non-linear Electrodynamics}, Nordita,
1970

[7] {\bf B. Lehnert, S. Roy}, {\it Extended Electromagnetic Theory}, World
Scientific, 1998.

[8] {\bf G.Hunter,R.Wadlinger}, {\it Phys.Essays}, vol.2, 158 (1989).

[9] {\bf D. Funaro}, {\it Electromagnetsm and the
Structure of Matter}, Worldscientific, 2008;

see also arXiv:physics/0505068)

[10] {\bf Maxwell, J. C}., On Physical Lines of Force. Part 1., {\it Phil.
Mag}. vol.XXI (1861), vol. XXIII (1862); also, {\it The Scientific Papers of
James Clerk Maxwell}, vol.I, pp.451-513 (1890)

[11] {\bf Donev, S., Tashkova, M}., {\it Proc. Roy. Soc. of London} A 450, 281
(1995), see also: \newline hep-th/0403244 .

[12] {\bf Michor, P.}, {\it Remarks on the Schouten-Nijenhuis bracket},
available at: \newline http://www.mat.univie.ac.at/~michor/listpubl.html,
No.25.

[13] {\bf N.Bourbaki}, {\it Set Theory (short version)}

[14] {\bf Poynting, J. H}., Phil. Trans. {\bf 175}, 1884, pp.343-361.

[15] {\bf Thomson, J.J.}, Recent Researches in Elect. and Mag., 1893, p.13;
{\bf Poincare, H}., Archives Neerland Sci., vol.{\bf 2}, 1900, pp.252-278; {\bf
Abraham, M.}, Gott.Nach., 1902, p.20; see also the corresponding comments in
Whitakker's {\it History of the theories of Aether and Electricity}, vol.1,
Ch.10.

[16] {\bf Planck, M.}, {\it Ann. d. Phys.}, {\bf 4}, 553 (1901)

[17] {\bf Einstein, A.}, {\it Ann. d. Phys.}, {\bf 17}, 132 (1905)

[18] {\bf De Broglie, L.},  {\it Ondes et quanta}, C. R. {\bf 177}, 507 (1923)

[19] {\bf Lewis, G. N.}, {\it Nature}, {\bf 118}, 874 (1926)

[20] {\bf Speziali, P.}, Ed. {\it Albert Einstein-Michele Besso
Correspondence} (1903-1955), (1972)

[21] {\bf Dainton, J.}, 2000,  {\it Phil. Trans. R. Soc. Lond.
A}, {\bf 359}, 279

[22] {\bf Godbole, R. M.}, arXiv: hep-th/0311188

[23] {\bf Nisius, R.}, arXiv: hep-ex/0110078

[24] {\bf Stumpf, H., Borne, T.}, {\it Annales de la Fond. Louis De
Broglie}, {\bf 26}, No. {\it special}, 429 (2001)

[25] {\bf Godbillon, C.},  {\it Geometrie differentielle et mecanique
analytiqe}, Hermann, Paris (1969)

[26] {\bf Vacaru, S}. et al., arXiv/gr-qc/0508023v2

[27] {\bf Marsden, J., Hughes, T.}, 1994,
{\it Mathematical foundations of Elasticity},
Prentice Hall 1983;
Reprinted by Dover Publications, 1994

[28] {\bf Kushner, A., Lychagin, V., Rubtsov, V.}, {\it Contact Geometry and
Non-linear Differential Equations}, Cambridge University Press 2007

[29] {\bf Zilokowski, R.W.}, {\it Phys.Rev.Lett.} {\bf 66}, No.6, p.839, 1991

\end{document}